\documentclass[12pt]{article}
\usepackage{latexsym}
\usepackage{mathrsfs,stmaryrd}
\usepackage{amssymb,cite}
\usepackage{amsmath}
\usepackage{epsfig,graphicx}
\newcommand{\half}{\frac{\scriptstyle 1}{\scriptstyle 2}}

\newcommand{\C}{\mathbb{C}}

\newcommand{\CP}{\mathbb{CP}}
\newcommand{\RP}{\mathbb{RP}}
\newcommand{\PT}{\mathbb{PT}}
\newcommand{\R}{\mathbb{R}}

\newcommand{\cH}{\mathcal{H}}

\newcommand{\M}{\mathbb{M}}

\newcommand{\Z}{\mathbb{Z}}
\newcommand{\del}{\partial}
\newcommand{\e}{\mathrm{e}}
\newcommand{\im}{\mathrm{i}}

\newcommand{\rd}{\mathrm{d}}
\newcommand{\D}{\mathrm{D}}
\newcommand{\rW}{{\rm W}}
\newcommand{\rZ}{{\rm Z}}

\newcommand{\cG}{\mathcal{G}}
\newcommand{\cN}{\mathcal{N}}
\newcommand{\cO}{\mathcal{O}}
\newcommand{\cP}{\mathcal{P}}

\newcommand{\SL}{\mathrm{SL}}
\newcommand{\sgn}{\mathrm{sgn}}

\newcommand{\SU}{\, \mathrm{SU}}

\newcommand{\MHVbar}{\overline{\rm MHV}}

\newcommand{\cA}{{\mathcal{A}}}
\newcommand{\cM}{{\mathcal{M}}}
\newcommand{\be}{\begin{equation}\label}
\newcommand{\ee}{\end{equation}}
\newcommand{\bea}{\begin{eqnarray}\label}
\newcommand{\eea}{\end{eqnarray}}

\newcommand{\la}{\langle}
\newcommand{\ra}{\rangle}

\def\mathbox#1{\vbox{\hrule\hbox{\vrule\kern8pt\vbox{\kern8pt\begin{centering}
\hbox{$\displaystyle #1$}\end{centering}\kern8pt}\kern8pt\vrule}\hrule}}

\newtheorem{defn}{Definition}[section]

\begin{document}

\title{Scattering Amplitudes and BCFW Recursion in Twistor Space\bigskip}

\author{Lionel Mason and David Skinner 
\\ \small The Mathematical Institute, University of Oxford,\\
\small 24-29 St.~Giles, Oxford OX1 3LP, United Kingdom 
\cr
\small \&\cr
\small Institut des Hautes {\'E}tudes Scientifiques,
Le Bois Marie 35, 
\\ \small Route de Chartres,
91440 Bures-sur-Yvette, France}


\maketitle

\begin{abstract}

  Twistor ideas have led to a number of recent advances in our
  understanding of scattering amplitudes.  Much of this work has been
  indirect, determining the twistor space support of scattering
  amplitudes by examining the amplitudes in momentum space. In this
  paper, we construct the actual twistor scattering amplitudes
  themselves.  We show that the recursion relations of Britto,
  Cachazo, Feng and Witten have a natural twistor formulation that,
  together with the three-point seed amplitudes, allows us to
  recursively construct general tree amplitudes in twistor space. We
  obtain explicit formulae for $n$-particle MHV and NMHV
  super-amplitudes, their CPT conjugates (whose representations are
  distinct in our chiral framework), and the eight particle N$^2$MHV
  super-amplitude.  We also give simple closed form formulae for the
  $\cN=8$ supergravity recursion and the MHV and $\overline{\mbox{MHV}}$
  amplitudes.  This gives a formulation of scattering amplitudes in
  maximally supersymmetric theories in which superconformal symmetry
  and its breaking is manifest.

  For N$^k$MHV, the amplitudes are given by $2n-4$ integrals in the
  form of Hilbert transforms of a product of $n-k-2$ purely
  geometric, superconformally invariant twistor delta functions,
  dressed by certain sign operators.  These sign operators subtly
  violate conformal invariance, even for tree-level amplitudes in
  $\cN=4$ super Yang-Mills, and we trace their origin to a
  topological property of split signature space-time. We develop the 
  twistor transform to relate our work to the ambidextrous twistor diagram
  approach of Hodges and of Arkani-Hamed, Cachazo, Cheung and Kaplan.

\end{abstract}

\section{Introduction}
\label{sec:intro}

In his development of twistor-string theory~\cite{Witten:2003nn},
Witten showed that gauge theory scattering amplitudes have remarkable
properties when analyzed in twistor space.  The subsequent activity
led to substantial progress in our understanding of amplitudes
including the construction of the tree-level Yang-Mills S-matrix via
the connected prescription of twistor-strings~\cite{Roiban:2004yf},
the MHV diagram formalism \cite{Cachazo:2004kj}, the BCFW recursion
relations~\cite{Britto:2004ap,Britto:2005fq} and the generalized
unitarity and leading singularity
methods~\cite{Britto:2004nc,Bern:2005iz,Bern:2006ew,Bern:2007ct,Buchbinder:2005wp,Cachazo:2008dx,Cachazo:2008vp,Cachazo:2008hp,Drummond:2008bq}. Despite
there having been much work examining the support of amplitudes in
twistor space (see {\it
 e.g.}~\cite{Witten:2003nn,Cachazo:2004kj,Cachazo:2004zb,Cachazo:2004by,Bena:2004ry,Bena:2004xu,Bedford:2004py,Britto:2004tx,Bidder:2004vx}),
until this point there has been no systematic study of explicit
formulae for the actual twistor space amplitudes themselves.

There are many reasons why it is interesting to investigate the
twistor representation of scattering amplitudes more closely. Firstly,
such an analysis should make the conformal properties of scattering
amplitudes manifest; these can be difficult to see on momentum
space.  Secondly, twistor theory brings out the complete integrability
of the anti self-dual sectors of both Yang-Mills and General
Relativity~\cite{Newman:1976gc,Penrose:1976js,Ward:1977ta,MasonWoodhouse:1996};
in the language of twistor
actions~\cite{Mason:2005zm,Boels:2006ir,Mason:2007ct,Mason:2008jy},
there exist gauges for which the anti-selfdual sector is
\emph{free}~\cite{Boels:2007qn}.  Furthermore, Drummond, Henn \&
Plefka~\cite{Drummond:2009fd} have recently shown that the generators
of the dual superconformal
algebra~\cite{Drummond:2007au,Drummond:2008vq} -- one of the main
tools in the construction of
multi-particle~\cite{Drummond:2008cr,Brandhuber:2008pf} and
multi-loop~\cite{Bern:2005iz,Bern:2006ew,Bern:2007ct,Bern:2008ap,Drummond:2008aq,Drummond:2008bq}
scattering amplitudes in $\cN=4$ SYM -- have a simple (though
second-order) representation on twistor space.

Additional motivation comes from Penrose's twistor
programme~\cite{Penrose:1999cw}, which seeks to reformulate
fundamental physics on twistor space as a route towards quantum
gravity. Indeed, there has been a long-standing programme to
understand scattering amplitudes in twistor theory via twistor
diagrams~\cite{Penrose:1972ia,Hodges:1980hn}, but this has proved
technically difficult for two reasons. Firstly, the standard form of
the Penrose transform of on-shell states requires the use of
cohomology: this builds in extra gauge freedom, and requires one to
understand multi-dimensional contour integrals in large complex
manifolds. In this paper, cohomology will be sidestepped at the
expense of working in $(++--)$ space-time signature. With this
signature, twistor space has a real slice $\RP^3$ and massless fields
on split signature space-time correspond to homogenous {\it functions}
on real twistor space, rather than cohomology classes. The Penrose
transform then becomes the closely related `X-ray transform' of Fritz
John~\cite{John:1938} (see also~\cite{Atiyah:1979}) and, as we show in
appendix~\ref{app:halfFT}, the Fourier transform of the X-ray
transform is precisely Witten's half Fourier
transform~\cite{Witten:2003nn} to on-shell momentum space. We will
base our analysis on this half Fourier transform.  The expectation is
that the twistorial structures underlying amplitudes in all signatures
will be visible in this signature also, although we will see that
complications that seem to be specific to split signature also arise
in this approach.

A second technical problem one encounters when trying to describe
twistor scattering amplitudes is that \emph{off-shell} states (arising
in the internal legs of a Feynman diagram) cannot be encoded
holomorphically on
twistor space alone. In momentum space, such off-shell states are
easily incorporated by relaxing the condition that the wavefunction is
supported only on the mass shell. In the twistor diagram
approach~\cite{Penrose:1972ia,Hodges:1980hn}, one instead relaxes the
condition that the wavefunction be holomorphic, and then proceeds to
complexify $f(Z,\bar Z)\to f(Z,W)$. This procedure leads to a
description in terms of both\footnote{Twistor diagrams are
 ambidextrous, and one must choose arbitrarily whether to represent a
 given external state on twistor or dual twistor space. While there
 is no problem with working with these two representations
 simultaneously (analogous to describing some particles on momentum
 space and others on space-time), for a basic description in terms of
 an action, one needs to specify the basic fields and the space on
 which they live.} twistors and dual twistors and hence provides a
six-dimensional description of four-dimensional objects, implying
still more redundancy.  However, the recursion relations discovered by
Britto, Cachazo \& Feng~\cite{Britto:2004ap} and proved by Britto,
Cachazo, Feng \& Witten~\cite{Britto:2005fq} involve only
\emph{on-shell, gauge invariant} objects, thereby eliminating much of
the redundancy (even on momentum space) of the Feynman diagram
approach to scattering amplitudes. The BCFW relations generate the
full perturbative content of Yang-Mills and gravity starting from only
the three-particle amplitudes for $\la++-\ra$ and $\la--+\ra$ helicity
configurations. Basing a twistor scattering theory on the BCFW
relations allows us to avoid this second difficulty.

The first twistor formulation of the BCFW recursion procedure was
given in terms of twistor diagrams by
Hodges~\cite{Hodges:2005bf,Hodges:2005aj,Hodges:2006tw}.  Hodges'
construction uses both twistors and dual twistors, and has recently
been re-derived by Arkani-Hamed, Cachazo, Cheung \&
Kaplan~\cite{AHCCK} in work that is parallel to this paper, and
similarly uses Witten's half Fourier transform. By contrast, in this
article we will work with a chiral formulation in which all external
states are represented on (dual\footnote{For better agreement with the
 perturbative scattering theory literature, we work on \emph{dual
   twistor space}, where Yang-Mills amplitudes with 2 \emph{negative}
 and $n-2$ \emph{positive} helicity gluons are supported on a line. A
 full summary of our conventions may be found in
 appendix~\ref{app:conventions}.}) twistor space. The twistor diagram
representation
of~\cite{Hodges:2005bf,Hodges:2005aj,Hodges:2006tw,AHCCK} is related
to ours by a (complete) Fourier transform on non-projective twistor
space. In section~\ref{sec:ambi} we discuss the relation of this
Fourier transform to the twistor transform between projective twistor
and dual twistor spaces, and use this to explain the detailed relation
between the work of Arkani-Hamed {\it et al.}~\cite{AHCCK} and the
current paper.

If only dual twistor variables $\rW$ are used, the (super-)momentum shift that is the first step in the BCFW procedure corresponds to the simple shift  
$$
	\cA(\rW_1,\ldots,\rW_n)\to \cA(\rW_1,\dots,\rW_n-t\rW_1)
$$
in the twistor amplitude $\cA$, where $t$ is 
the shift parameter. This formula is proved in section~\ref{sec:mom-shift} and makes manifest the (super-)conformal invariance of the BCFW shift.  The original use of the shift was to introduce a complex parameter in which propagators within the amplitude generate poles, so that contour integration yields the recursion relation as a residue
formula.  Here, our aim is not to prove the recursion relation (for
which see~\cite{Britto:2005fq,Benincasa:2007qj,Brandhuber:2008pf,ArkaniHamed:2008gz}). Instead,  $t$ will be a real parameter that provides the one degree of freedom required to go off-shell in twistor space. This is the key idea from the point of view of a twistor theorist seeking to study perturbative scattering theory: rather than representing an off-shell state by both a twistor and a dual twistor as in twistor diagrams, one can describe it using a single twistor together with a BCFW shift. (See also~\cite{Boels:2007gv} for a hybrid formulation in which off-shell states are treated on momentum space while on-shell states are treated on twistor space.)

To begin the recursion procedure, one must seed the BCFW relations
with the three-particle amplitudes. In $(++--)$ space-time, these can
be obtained directly by taking the half Fourier transform of the
momentum space expressions. Doing so leads in the
first instance to formulae\footnote{See appendix~\ref{app:amplitudes} for a detailed derivation of the half Fourier transforms of various momentum space amplitudes} whose superconformal properties remain somewhat obscure, in contrast to our aim of making such behaviour transparent.  To remedy this, in
section~\ref{sec:distributions} we introduce distributions on twistor
space that are manifestly conformally invariant. These are the basic
objects out of which we construct the seed amplitudes in
section~\ref{sec:seedamp}.

One of the surprises of our analysis is that these basic
three-particle amplitudes are \emph{not} conformally invariant, even
in $\cN=4$ SYM. The failure of conformal invariance is rather subtle
and is discussed in section~\ref{sec:confbreaking} and further in the
conclusions. In some sense, it is merely the problem of choosing an
overall sign for the amplitude; nevertheless, there is a topological
obstruction to doing this in a way that is consistent with conformal
invariance.  The obstruction is analogous to the impossibility of
choosing a holomorphic branch for $\sqrt{z}$ on the punctured complex
plane; one must choose a cut across which the function will not be
holomorphic.  Similarly, to make the sign factor in the amplitude well-defined requires the choice of a light-cone at infinity.  Moreover, although the BCFW shift is superconformally invariant, the recursion
relations themselves are not. Once again, the violation of conformal
invariance is rather subtle; for example we explicitly show that when
$n$ is odd, the $n$-particle MHV and googly MHV 
super-amplitudes in $\cN=4$ SYM break conformal invariance -- even at tree-level -- 
in the same way as the three-particle amplitudes. However, when $n$ is even the conformal
breaking of the seed amplitudes and the recursion relations cancel
each other out, so that these MHV (and hence googly MHV) amplitudes are
genuinely conformally invariant.  We argue in the conclusion that the
origin of the failure of conformal invariance in our formalism is
likely to be associated with our resorting to split signature in order
to side-step cohomology.  The topological obstruction is only present
in split signature. Furthermore, in the  twistor actions for gauge
theory~\cite{Mason:2005zm,Boels:2006ir, Witten:2003nn} the relevant
sign factors are essentially determined by the differential forms used
in the Dolbeault cohomology description, but these signs are lost when
reducing the forms to functions so as to reach a split signature description.

In section~\ref{sec:recursion} we translate the BCFW recursion
relation itself onto twistor space, obtaining a recursion formula that
decomposes arbitrary tree-level twistor amplitudes into more
elementary ones. We focus on maximally supersymmetric Yang-Mills
and gravity, and so only discuss the twistor form of the
supersymmetric BCFW
relations~\cite{Brandhuber:2008pf,ArkaniHamed:2008gz}. Despite their
extra field content, these theories are much simpler than their
non-supersymmetric counterparts, even at tree-level (where the
non-supersymmetric theories are contained as a subset). This is
because there are $2^n$ $n$-particle scattering amplitudes in
non-supersymmetric Yang-Mills or gravity, corresponding to the
different choices of helicity for the external particles, whereas
there are only $(n+1)$-distinct $n$-particle tree amplitudes in
$\cN=4$ SYM or $\cN=8$ SG -- the N$^k$MHV amplitudes, associated with
a polynomial of degree $(k+2)\cN$ in the on-shell Grassmann momenta.
This exponential simplification was a crucial ingredient in the recent
solutions of the classical Yang-Mills S-matrix by Drummond \&
Henn~\cite{Drummond:2008cr}, and the classical gravitational S-matrix
by Drummond, Spradlin, Volovich \& Wen~\cite{Drummond:2009ge} using
the supersymmetric version of BCFW recursion.

Armed with the twistor form of both the seed amplitudes and the BCFW
recursion relation, in section~\ref{sec:using-recursion} we proceed to
construct twistor space versions of various tree amplitudes in $\cN=4$
SYM. A simple consequence of the twistor representation of the
recursion relations and the tree amplitudes is that, modulo some sign
factors, the N$^k$MHV components of the complete classical S-matrix of
$\cN=4$ SYM can written as an integral over $2n-4$ `shift' parameters
of $n-k+2$ super-conformally invariant delta functions
$\delta^{4|4}(W)$.
In more detail, when an amplitude is expressed via recursion as a
combination of two subamplitudes with one being a
three-point MHV or $\MHVbar$ amplitude, the integrals can be performed
explicitly leading to an action of  operators
$\cH^i_{i-1,i+1}\delta^{4|4}(W_i)$ or $\widetilde\cH^{i-1,i+1}_i$
respectively.  These insert a new particle at point $i$ in between particles
$i-1$ and $i+1$ in the other subamplitude, the first preserving MHV
degree and the second raising it.  The $\cH$ and $\widetilde
\cH$ operators are each a pair of integral operators (Hilbert transforms)
integrating shift parameters in the amplitudes on which they act.  
These Hilbert transforms 
can formally be expressed as the sign functions of certain
first order differential operators.
These two operations seem to be sufficient to
generate the general amplitude (at least as far as the examples we
have calculated are concerned) as a sum of terms consisting of $n-3$
such operators acting on a three point amplitude.  These two operators
can be identified with the inverse soft limits of~\cite{AHCCK}.
We explicitly perform the recursion for the $n$-particle MHV and
NMHV super-amplitudes, their CPT conjugates, and the 8-particle
N$^2$MHV super-amplitude. 
We also give an algorithm for computing some
more general tree amplitude from the BCFW recursion relations.

We consider $\cN=8$ supergravity in section~\ref{sec:sg}.  The
structure of the BCFW recursion relation is unchanged compared to
Yang-Mills except that the sum is over all partitions, the seed
amplitudes are different and we work with $\cN=8$.  We solve the
recurrence in this case to give the MHV and $\overline{\mbox{MHV}}$
amplitudes.  For gravity the general shape of the result is
essentially the same as for Yang-Mills (i.e., $2n-4$ bosonic integrals
of $n-k-2$ delta functions) except that certain derivatives and
additional factors are introduced at each stage.  A partial solution
of the BCFW recursion for gravity has recently been constructed
in~\cite{Drummond:2009ge}, using the interplay of the KLT
relations~\cite{Kawai:1985xq} (which motivate a particular form for
the MHV amplitude~\cite{Elvang:2007sg}) with the properties of the
earlier solution of the $\cN=4$ SYM classical
S-matrix~\cite{Drummond:2008cr}.  It is clear that there is a very
close relationship between the twistor $\cN=8$ SG and $\cN=4$ SYM
amplitudes for the twistor amplitudes also in the sense that the
underlying structure of the Hilbert transforms and corresponding
support of the amplitudes is the same in both cases.  The distinction
between the two sets of amplitudes that there are extra factors and
derivatives in the Gravitational case.

We go on to give a preliminary discussion of loops.  We first give the
half Fourier transform of the dimensionally regularised 4 particle
1-loop amplitude.  The analytic continuation of such an amplitude to
split signature is ambiguous as, for example, we are taking logs or
fractional powers of quantities that have a definite sign in Lorentz
signature, but which have no fixed sign in split signature. We can
nevertheless find formulae for both the finite and IR divergent parts
of the amplitude by choosing an analytic continuation to split
signature and performing the half-Fourier transform.  We show that
gives a straightforward answer although of course it depends on the
choice of analytic continuation we started with.  We discover that the finite
part is superconformally invariant.  The
non-supersymmetric (finite) all plus loop amplitude is rather easier
as it is a rational function on momentum space with unique analytic
continuation, and we give this also.
It is also to be hoped that
the generalised unitarity and leading singularity
methods~\cite{Britto:2004nc,Bern:2005iz,Bern:2006ew,Bern:2007ct,Buchbinder:2005wp,Cachazo:2008dx,Cachazo:2008vp,Cachazo:2008hp,Drummond:2008bq}
have a natural formulation on twistor
space\footnote{See~\cite{ArkaniHamed:2009dn,Mason:2009qx,ArkaniHamed:2009vw,
    Bullimore:2009cb,Kaplan:2009mh} for 
  subsequent work that realizes this aim.}; there is no analytic
continuation issue for leading singularities.

We do of course intend that eventually there will be a systematic
method for obtaining loop amplitudes on twistor space.  We give a
further discussion on the prosepects for this in \S\ref{sec:signs} in
the conclusions.  The main conclusion is that in order to translate from
the formalism obtained here to one appropriate to Lorentz signature,
we must re-interpret our twistor functions as representatives of Cech
cohomology classes, and re-interpret all our integrals as
contour integrals.  In that re-interpretation, the sign factors can be
dropped from the formulae, but must
then be incorporated as part of the definition of the Cech cocycles
dictating which sets in the cover the Cech cocycles should be
attached to.

The twistor space support of amplitudes has previously been analysed
and fruitfully exploited by a number of
authors~\cite{Witten:2003nn,Cachazo:2004kj,Cachazo:2004zb,Cachazo:2004by,Bena:2004ry,Bena:2004xu,Bedford:2004py,Britto:2004tx,Bidder:2004vx}. However,
this was done implicitly, {\it e.g.} by use of differential operators
acting on the momentum space amplitudes, or by integral
representations. Our explicit representation of the twistor amplitudes
is in fact smeared out by certain non-local operators that also break
conformal symmetry. It is reasonable to regard these as an artifact of
the use of split signature\footnote{These operators are signs on
  momentum space, and so are not visible to the differential operators
  that were used there to determine the twistor support of the
  amplitudes. Similarly, they are not visible when momentum space
  conformal generators are used to test for conformal invariance,
  unless the detailed behaviour of the amplitudes across its
  singularities is examined.} and it is therefore reasonable to ignore
them if one is interested in the structures valid in the complex or
other signatures; this is also the conclusion of the discussion in
\S\ref{sec:signs}.  If the non-local sign operators are ignored, we
find that the MHV amplitudes are indeed localized along lines (indeed
they are simply products of delta-functions that restrict the
corresponding twistors to lines).  Howver, the NMHV amplitudes are a
sum of terms that are supported on three lines in contradiction to the
expectations raised by the MHV
formalism~\cite{Witten:2003nn,Cachazo:2004kj,Cachazo:2004zb,Cachazo:2004by,
  Bena:2004ry,Bena:2004xu,
  Bedford:2004py,Britto:2004tx,Bidder:2004vx}.  It is clear that some
decomposition and re-summation is needed for agreement with the
expectations of the MHV formalism.  However, this picture is in
agreement with the NMHV 1-loop formulae~\cite{Bern:2004bt}.  At higher
N$^k$MHV degree, the Hilbert transforms lead to an additional two
lines for each $k$.  Subsequent to version one of this paper, this
picture has been re-expressed more concretely and much
extended~\cite{Korchemsky:2009jv, Bullimore:2009cb}.

At present, although the twistor form of the BCFW recursion relation
and scattering amplitudes have many remarkable properties, this work
does not constitute a complete {\it theory} in twistor space, because
both the BCFW relation itself and the three-point seed amplitudes
currently need to be imported from momentum space by half Fourier
transform. We conclude in section~\ref{sec:conclusions} with a
discussion of a possible way to relate our results to the twistor
action of~\cite{Boels:2006ir} (and its ambitwistor
cousin~\cite{Mason:2005kn}), which goes some way towards making the
twistor theory self-contained.  Our approach is complementary to the
momentum space picture Drummond \& Henn~\cite{Drummond:2008cr}, and we
discuss the relation of the dual superconformal invariants
of~\cite{Drummond:2008vq,Drummond:2008bq,Drummond:2008cr} to the
twistor space structures we find\footnote{Subsequent to version 1 of
  this paper, the relationships betweeen twistor amplitudes and dual
  conformal invariance are now much better
  understood~\cite{Hodges:2009hk, Mason:2009qx, ArkaniHamed:2009vw}}.

Note that one can similarly
transform Risager's momentum shift~\cite{Risager:2005vk} and its
multiline extensions~\cite{Kiermaier:2008vz,Kiermaier:2009yu} into
twistor space, obtaining a twistor space version of the MHV diagram
formalism. 

In appendix \ref{app:conventions} we summarise our conventions and the
basic background structures and formulae.  In appendix
\ref{app:halfFT} we derive the half Fourier transform from the X-ray
transform.  In appendix \ref{app:amplitudes} we give the basic
calculations of the half Fourier transform of the seed amplitudes.

\section{The Momentum Shift on Twistor Space}
\label{sec:mom-shift}

The amplitude $A(1,\ldots,n)$ for a process with $n$ massless
particles is a function of $n$ null momenta 
$p_1,\ldots ,p_n$.  Decomposing these null momenta into
their spinor factors $p_i=|i]\la i|$ (where $|i]$ and $|i\ra$ are
spinor-helicity notation for left and right spinors
$\tilde\lambda_{iA'}$ and $\lambda_{iA}$, respectively) the BCFW
procedure starts by shifting two of them:
\be{shift} |i] \to
|\hat{i}]:=|i]+ t|j]\ ,\qquad\qquad|j\ra \to|\hat{j}\ra:=|j\ra-t|i\ra
\, .  
\ee 
This shift apparently treats left- and right-handed spinors
symmetrically. However, there is some chirality in the BCFW shift~\eqref{shift} because the
`permissible shifts' -- whether one should translate a given state's
primed spinor or unprimed spinor -- are correlated with the helicities
of the states being shifted~\cite{Britto:2005fq}. This chirality is
more apparent in the maximally supersymmetric extensions of the BCFW
procedure~\cite{Brandhuber:2008pf,ArkaniHamed:2008gz}: any
(irreducible) representation of an $\cN=4$ YM supermultiplet or an
$\cN=8$ gravity supermultiplet with maximal on-shell supersymmetry is
necessarily chiral, as either the positive or negative helicity state
must be chosen to sit at the top of the supermultiplet. In particular,
all external supermultiplets have the same helicity, so~\eqref{shift}
together with the `permissible shift' rule are replaced by the chiral
super-shift 
\be{supershift} 
	\|i\rrbracket \to \|\hat{i}\rrbracket:=\|i\rrbracket + t\|j\rrbracket\ ,\qquad\qquad |j\ra \to|\hat{j}\ra:=|j\ra - t|i\ra\ , 
\ee 
where $\|i\rrbracket =(\tilde\lambda_i,\eta_i)$ denotes both the primed spinor momenta and
the Grassmann co-ordinate of the on-shell momentum superspace of the
$i^{\rm th}$ state; $\eta_j$ itself is not shifted.

In $(++--)$ space-time signature, Witten showed in~\cite{Witten:2003nn}
that the (dual) twistor and on-shell momentum space superfields are related by the \emph{half Fourier transforms} 
\be{halfFT}
    f(\lambda,\mu,\chi) = \int\rd^{2|\cN}\tilde\lambda\ \e^{\im\llbracket\mu\tilde\lambda\rrbracket}\,
    \Phi(\lambda,\tilde\lambda,\eta)\ ;\qquad
    \Phi(\lambda,\tilde\lambda,\eta) = \frac{1}{(2\pi)^2}\int\rd^2\mu\ 
    \e^{-\im\llbracket\mu\tilde\lambda\rrbracket}\,f(\mu,\lambda,\chi)\ ,
\ee
where $(\lambda_A,\mu^{A'})$ and $\chi_a$ are the bosonic and
fermionic components of a (dual) supertwistor $\rW$, and
\begin{equation}
	\llbracket\mu\tilde\lambda\rrbracket:=\mu^{A'}\tilde\lambda_{A'}+\chi_a\eta^a
\end{equation}
is the natural pairing between $(\mu,\chi)$ and the on-shell momentum space co-ordinates
$(\tilde\lambda,\eta)$. Under the momentum
supershift~\eqref{supershift} (choosing $i=1$ and $j=n$ for
simplicity), the twistor super-amplitude transforms as 
\be{BCFWshift2}
\begin{aligned}
    \widehat \cA(\rW_1,\ldots,\rW_n)&=\int
    \prod_{i=1}^n\rd^{2|\cN}\tilde\lambda_i\  
    \e^{\im\llbracket\mu_i\tilde\lambda_i\rrbracket}\ A\left(\hat{1},\ldots,\hat{n}\right)\\
    &=\int \rd^{2|\cN}\hat{\tilde\lambda}_1\,\rd^{2|\cN}\tilde\lambda_n\ 
    \e^{\im\llbracket\mu_1\hat{\tilde\lambda}_1\rrbracket}\ \e^{\im\llbracket(\mu_n-t\mu_1)\tilde\lambda_n\rrbracket}
    \prod_{j=2}^{n-1}\rd^{2|\cN}\tilde\lambda_j\,\e^{\im\llbracket\mu_j\tilde\lambda_j\rrbracket} 
    A\left(\hat{1},\ldots, \hat{n}\right)\\
    &=\cA(\rW_1,\ldots,\rW_n-t\rW_1)\ ;
\end{aligned}
\ee
{\it i.e.} the half Fourier transform combines with the shift $|n\ra\to|n\ra-t|1\ra$ in the unprimed spinor to result in a shift
of the entire (super)twistor\footnote{That only W$_n$ is shifted
should not be surprising: \eqref{supershift} is generated by 
$\tilde\lambda_{n}\del/\del\tilde\lambda_1$, $\eta_n\del/\del\eta_1$
and $-\lambda_1\del/\del\lambda_n$. Under the half Fourier
transform~\eqref{halfFT} one replaces $\tilde\lambda\to\del/\del\mu$,
$\del/\del\tilde\lambda\to-\mu$, $\eta\to\del/\del\chi$ and
$\del/\del\eta\to-\chi$, so these generators combine to form
$-\rW_1\del_{\rW_n}$.} W$_n$ along the line joining it to
W$_1$. Equation~\eqref{BCFWshift2} provides a key motivation for the rest of this
paper. It shows that the BCFW shift is (super)conformally invariant
and may be simply expressed on twistor space.

\section{Twistor Amplitudes and Conformal Invariance}
\label{sec:amplitudes}

The BCFW recursion procedure is seeded by the three-point MHV and $\MHVbar$ amplitudes. For $\cN=4$ SYM, twistor space versions of these may be obtained by directly taking the half Fourier transform of the momentum space expressions
\begin{equation}
\begin{aligned}
	A_{\rm MHV}(1,2,3) &= \frac{\delta^{(4|8)}\!\left(\sum_{i=1}^3 |i\ra\llbracket i\|\right)}{\la12\ra\la23\ra\la31\ra}\\
	A_{\MHVbar}(1,2,3) &=\frac{\delta^{(4)}\!\left(\sum_{i=1}^3|i\ra[i|\right)\,
	\delta^{(0|4)}\left(\eta_1[23]+\eta_2[31]+\eta_3[12]\right)}{[12][23][31]}\ .
\end{aligned}
\ee
For example, for the 3-particle MHV amplitude one finds  in the first instance\footnote{See appendix~\ref{app:amplitudes} for a detailed derivation of the half Fourier transforms of various momentum space amplitudes.} 
\be{MHV-twist-explicit}
    \cA_{\rm MHV}(\rW_1,\rW_2,\rW_3)=
    \frac{\delta^{(2|4)}\!\left(\mu_1\la23\ra+\mu_2\la31\ra+\mu_3\la12\ra\right)}{\la12\ra\la23\ra\la31\ra}\ , 
\ee 
where the $\delta$-functions run over the supertwistor components
$(\mu^{A'},\chi_a)$. $\cA_{\rm MHV}(\rW_1,\rW_2,\rW_3)$ has support precisely where W$_1$, W$_2$ and W$_3$ are collinear, and has the standard `current correlator' denominator~\cite{Nair:1988bq} that inspired twistor-string theory~\cite{Witten:2003nn}. 

While~\eqref{MHV-twist-explicit} is manifestly (super)Poincar\'e invariant, its conformal properties are still not transparent, since $(\mu^{A'}, \chi_a)$ appear in~\eqref{MHV-twist-explicit} on a rather different footing to $\lambda_A$.  
As indicated in the introduction, the conformal properties of scattering amplitudes are exhibited most clearly by writing them in terms of manifestly $\SL(4|\cN;\R)$ invariant\footnote{The superconformal group in $(++--)$ signature space-time is $\SL(4|\cN;\R)$ or ${\rm PSL}(4|4;\R)$ when $\cN=4$.} distributions on real projective twistor space. In section~\ref{sec:distributions} we discuss the twistor distributions that in section~\ref{sec:seedamp} turn out to be relevant for describing twistor space scattering amplitudes.


\subsection{Distributions on projective twistor space}
\label{sec:distributions}

The most elementary distribution is the delta function supported at a
point $Y\in\RP^3$, We write this as $\delta^{(3)}_{-n-4}(W,Y)$ and it has the
defining property 
\be{delta-fn} 
    f(Y)=\int_{\RP^3} f(W)\delta^{(3)}_{-n-4}(W,Y) \,D^3W
\ee 
for $f$ a function of homogeneity degree $n$ and where $D^3W :=
\frac{1}{4\!}\epsilon^{\alpha\beta\gamma\delta}W_\alpha\rd
W_\beta\wedge\rd W_\gamma\wedge\rd W_\delta$ is the canonical top form
of homogeneity $+4$. This $\RP^3$ delta function can be described
using an elementary integral of the standard, non-projective $\delta$-function on $\R^4$:
\be{delta-int} 
    \delta^{(3)}_{-n-4}(W,Y)=\int_{-\infty}^\infty \frac{\rd t}{t^{1+n}}\,\sgn(t)\,\delta^{(4)}(W-tY)\ . 
\ee 
Equation~\eqref{delta-int} has the right support because $W_\alpha-tY_\alpha=0$ only if $W$ and $Y$ lie on the same line through the origin in $\R^4$ and hence define the same point in the projective space. It is also easy to check that under the scalings
$W_\alpha\to a W_\alpha$ and $Y_\alpha \to b Y_\alpha$, we have (at least for $a/b>0$)
\be{;lkjas;lfk;}
    \delta^{(3)}_{-n-4}(aW, bY) = \frac{b^n}{a^{n+4}}\delta^{(3)}_{-n-4}(W,Y)\ ,
\ee
so that the subscript labels the homogeneity of the first entry. As $\R-\{0\}$ is not connected, the behaviour under sign reversal must be considered separately.  Scaling $(W_\alpha,Y_\alpha)\to(aW_\alpha ,bY_\alpha)$ with $a/b<0$ induces the scaling
\be{kjghkjg}
    \frac{\rd t}{t^{1+n}}\,\sgn(t) \to (-1)\,\times\,\frac{b^n}{a^n}\frac{\rd t}{t^{1+n}}\,\sgn(t)
\ee
where the extra sign change comes from the factor of $\sgn(t)$. However, under this scaling the limits of the $t$ integral also change sign, so that $\delta_{-n-4}^{(3)}(W,Y)$  itself has no extra signs. These properties ensure that~\eqref{delta-fn} is well-defined projectively whenever $f(W)$ is.

Perhaps surprisingly, we actually do want to consider twistor functions and distributions that have {\em wrong sign} behaviour, in the sense that 
\be{wrongscaling}
    \tilde f(aW)=
    \begin{cases}
            a^n \tilde f(W) & a\in \R^+\\
            -a^n \tilde f(W)& a\in \R^-
    \end{cases}
\ee
so that they scale with an extra minus sign when the scaling parameter is negative\footnote{Just as ordinary homogeneous functions can be thought of as sections of a line bundle $\cO(n)$ on the projective
space, such `wrong sign' functions correspond to sections of another
family of invariantly defined line bundles $\widetilde\cO(n)$ on the
projective space.}.  For these we can define a tilded $\delta$-function
$\tilde\delta^{(3)}_{-n-4}(W,Y)$ such that
\be{tilde-delta}
    \tilde f(Y)=\int_{\RP^3} \tilde f(W)\tilde\delta^{(3)}_{-n-4}(W,Y)\, D^3W \ .
\ee
For~\eqref{tilde-delta} to be well-defined, $\tilde\delta^{(3)}_{-n-4}(W,Y)$ must have `wrong sign' behaviour in both
$W$ and $Y$, so it is related to~\eqref{delta-int} by
\be{tilde-untilde-relate}
    \tilde\delta^{(3)}_{-n-4}(W,Y) = \sgn\left(\frac{W}{Y}\right)\delta^{(3)}_{-n-4}(W,Y) 
    =\int_{-\infty}^\infty \frac{\rd t}{t^{1+n}}\,\delta^{(4)}(W-tY)\ ,
\ee
where the second equality follows because $t=W/Y$ on the support of the $\delta$-function.

In concrete calculations, the $\rd t$ integrals are easily performed
explicitly using one of the 
$\delta$-functions, say the component of  the $|\lambda\ra$ spinor in
the direction of some fixed spinor $|\alpha\ra$.  On the support of
$\delta^{(4)}(W_1-tW_2)$, $t= \la \alpha 
1\ra/\la\alpha 2\ra$, so we can reduce to three $\delta$-functions. However
this breaks conformal invariance. Both the $\delta^{(3)}(W,Y)$ and
$\tilde\delta^{(3)}(W,Y)$ enforce the conformally invariant condition
that $W$ and $Y$ coincide projectively, but the only ways for us to
express this condition invariantly is via the formal definition~\eqref{delta-fn} or the integral
formulae~\eqref{delta-int} \&~\eqref{tilde-untilde-relate}.  This is because
it is not possible to impose $W_\alpha\propto Y_\alpha$ with an invariant set of {\it irreducible} equations:
The three conditions $W_\alpha\propto Y_{\alpha}$ are given by the six equations
\begin{equation}
    W_\alpha Y_{\beta}-W_\beta Y_{\alpha}=0\ ,
\end{equation}
but are subject to three relations.
Choosing any three of the equations breaks conformal invariance and
will also admit spurious solutions for which $W_\alpha$ is not
proportional to $Y_\alpha$.  The integrals over $\rd t$ above are the
easiest way to express  the full projective invariance.

\smallskip

The projective delta functions combine naturally to form the supersymmetric delta functions. For example, consider  the `wrong sign' $\tilde\delta$-function
\be{proj-susy-delta}
\begin{aligned}
    \tilde\delta^{(3|\cN)}_{\cN-4-n}(\rW,{\rm Y})
    &=\int_{-\infty}^\infty \frac{\rd t}{t^{1+n}}\,\delta^{(4|\cN)}(\rW-t{\rm Y})\\
    &:=\int_{-\infty}^\infty\frac{\rd t}{t^{1+n}}\,\delta^{(4)}(W-tY)\prod_{i=1}^\cN (\chi_i-t\psi_i)\ ,
\end{aligned}
\ee 
where $\rW=(W_\alpha,\chi_i)$ and ${\rm Y}=(Y_\beta,\psi_j)$.  By including a factor of $\sgn(t)$ in the measure, we can
also define a supersymmetric $\delta$-function with the correct sign
behaviour, but because of the twistor structure of the BCFW recursion relations, we will find that~\eqref{proj-susy-delta} is more directly useful. When $\cN=4$ and $n=0$, \eqref{proj-susy-delta} has
homogeneity zero in both its arguments (as appropriate for $\cN=4$ twistor supermultiplets), whereas for $\cN=8$ SG we will most frequently set $n=1$ so that $\tilde\delta^{(3|8)}(\rW,{\rm Y})$ has homogeneities $(3,1)$ in $(\rW,{\rm Y})$. In each of these cases, we omit the subscript.

One can also define a family of descendant $\delta$-functions and
$\tilde\delta$-functions that enforce collinearity and coplanarity
{\it etc.} of more twistors, rather than just coincidence. In
particular, we will make use of the $\cN=4$ and $\cN=8$ collinear $\tilde\delta$-functions 
\be{collinear-tilde-delta}
\begin{aligned}
   \tilde\delta^{(2|4)}(\rW_1;\rW_2,\rW_3) 
   &:=\int_{\R^2}\frac{\rd s}{s}\frac{\rd t}{t}\,\delta^{(4|4)}(\rW_1-s \rW_2-t \rW_3)\\
   \tilde\delta^{(2|8)}(\rW_1;\rW_2,\rW_3) 
   &:=\int_{\R^2}\frac{\rd s}{s^2}\frac{\rd t}{t^2}\,\delta^{(4|8)}(\rW_1-s\rW_2-t\rW_3)\ .
\end{aligned}
\ee
which are again superconformally invariant by construction. These collinear $\tilde\delta$-functions have correct sign behaviour for $\rW_1$, but wrong sign behaviour for  $\rW_2$ and $\rW_3$. The completely right sign untilded
collinear $\delta$-function is (for $\cN=4$)
\be{collinear-delta}
\begin{aligned}
    \delta^{(2|4)}(\rW_1,\rW_2,\rW_3) 
    &:= \int_{\R^2}\frac{\rd s}{|s|}\frac{\rd t}{|t|}\,\delta^{(4|4)}(\rW_1-s \rW_2-t \rW_3)\\
    &= \sgn(\la\rW_1\rW_2\ra\la\rW_3\rW_1\ra)\,\tilde\delta^{(2|4)}(\rW_1;\rW_2,\rW_3)\ ,
\end{aligned}
\ee
where the second line follows from using the delta functions in the $\lambda_A$ co-ordinates
to evaluate $s= \la 13\ra/\la 23\ra$ and $t=\la 12\ra/\la 32\ra$ so that 
\be{signs}
    \sgn (st)=\sgn(\la\rW_1\rW_2\ra\la\rW_3\rW_1\ra)\, .
\ee
For $\cN=4$, \eqref{collinear-delta} is invariant under scaling of each of its arguments, and performing
elementary substitutions shows that in fact it is totally symmetric
under exchange.

Non-projectively $\rW_1-s \rW_2-t\rW_3$ can only vanish
when $\rW_1$ lies in the two-plane containing the origin that is spanned
by $\rW_2$ and $\rW_3$. Therefore
$\tilde\delta^{(2|\cN)}(\rW_1;\rW_2,\rW_3)$ and
$\delta^{(2|\cN)}(\rW_1,\rW_2,\rW_3)$ have co-dimension $2|\cN$ support
on the set where $\rW_1,\,\rW_2$ and $\rW_3$ are collinear. Again, in
order to explicitly perform the $s, t$ integrals we must break
conformal invariance: The invariant condition for collinearity (in the
bosonic twistor space) is
\begin{equation}
    \varepsilon^{\alpha\beta\gamma\delta}W_{1\beta}W_{2\gamma}W_{3\delta}=0\ , 
\ee
but there is no conformally invariant or global way to take just two of these four equations\footnote{There is a Poincar\'e invariant choice in this case, which leads to the explicit form~\eqref{MHV-twist-explicit} of the three-point MHV amplitude.}.

A natural extension to the coplanarity $\tilde\delta$-function is
\be{coplanar-delta}
\begin{aligned}
    \tilde\delta^{(1|\cN)}(\rW_1,\rW_2,\rW_3,\rW_4)&:=\int_{\R^3}\frac{\rd r}{r}\frac{\rd s}{s}\frac{\rd t}{t}\,
    \delta^{(4|\cN)}\!\left(\rW_1 -r\rW_2 -s\rW_3 -t\rW_4\right)\, .
\end{aligned}
\ee and similarly for $\delta^{(1|\cN)}$.  Differently weighted
versions are obtained by including various powers of the
$r,s,t,\ldots$ variables in the measures, together with signs to change
the right/wrong sign behaviours. (For example, replacing $\rd r/r\to\rd r/|r|$ produces a version that is right sign in 
$\rW_1$ and $\rW_2$ and wrong sign in $\rW_3$ and $\rW_4$)


\subsubsection{The Hilbert transform and the $\sgn$ function}
\label{sec:hilbert}

The original Hilbert transform is a complex structure on functions on the
real line that fall off as $O(1/x)$ as $x\to\infty$. It is given by
the formula 
\be{orig-Hilb}
    H[f](x)={\rm p.v. } \frac{1}{\pi}\int_\R \frac{\rd
      y}{y}\,f(x-y)\, , \quad H[H[f]]=-f \  .
\ee
One way to view the Hilbert transform is to say that if $f={\rm Re}(F)$
where $F$ is a complex function that is holomorphic on the upper half
plane, then $H[f]={\rm Im}(F)$. This follows by expressing the 
principal value regularisation for the distribution $1/y$ as
\begin{equation}
    \frac1y=\frac12 \lim_{\epsilon\to 0}\left(\frac1{y+i\epsilon}+\frac1{y-i\epsilon}\right)\ ;
\ee
the right hand side give rise to the Cauchy integral formula for $\im F$ and its complex conjugate   
in terms of an integral of $f$ along the real axis.

We can extend the idea of the Hilbert transform to distributions
on twistor space (or more general projective spaces) as follows: choose a line in twistor space by fixing a point $A$ and then perform a Hilbert transform along the line joining $W$ to $A$. That is, we make the following

\begin{defn}
The \emph{Hilbert transform} of a function (or distribution) $f(W)$ in the direction $A$ is 
\be{hilb-trans-int}
    {\rm H}_A[f](W):=\frac{1}{\pi}\int_{-\infty}^\infty \frac{\rd t}t f(W_\alpha-tA_\alpha)\ , 
\ee
where the integral is understood by the principal value prescription. 
\end{defn}

\noindent Thus, the basic wrong sign $\tilde\delta$-functions may
be viewed as Hilbert transforms of the non-projective
$\delta$-function, for example 
\begin{equation}
\begin{aligned}
    \tilde\delta^{(3|\cN)}(\rW_1,\rW_2) &= \pi{\rm H}_{\rW_2}[\delta^{(4|\cN)}(\rW_1)]\\
    \tilde\delta^{(2|\cN)}(\rW_1;\rW_2,\rW_3) &= \pi^2{\rm H}_{\rW_2}\left[{\rm H}_{\rW_3}[\delta^{(4|\cN)}(\rW_1)]\right]
\end{aligned}
\ee
and so on.

The Hilbert transform has a useful interplay with the Fourier transform. Representing a (non-projective) distribution $f(W)$ by its Fourier transform $F(Z)$ we can write
\be{hilb-fourier}
\begin{aligned}
    \int_{-\infty}^{\infty}\frac{\rd t}{t} f(W_\alpha-tA_\alpha) &= \int\frac{\rd t}{t}\,\rd^4Z\ \e^{\im (W-tA)\cdot Z} F(Z)\\
    &= -\im\pi\int\rd^4Z\ \sgn(A\!\cdot\!Z)\,\e^{\im W\cdot Z} \, F(Z)\\
    &=\im\pi\,\sgn\left(\im A\!\cdot\!\frac{\del}{\del W}\right) f(W)
\end{aligned}
\ee
where in the second line we use the fact that 
\be{sgn-int}
    {\rm p.v.}\int_{-\infty}^\infty \frac{\rd t}t\, \e^{-\im at}=-\im \pi\, \sgn \, a\ ,
\ee
and in the last line follows by setting\footnote{Note that $\im
A\!\cdot\!\frac{\del}{\del W}$ is Hermitian.}
$Z^\alpha=-\im\del/\del W_\alpha$. At least formally, this allows us
to express the weighted projective $\delta$-functions in terms of
pseudo-differential operators. In particular, when $\cN=4$ we
have\footnote{We can similarly write formally 
\be{log-transform}
    \log\left(\im A\!\cdot\!\frac{\del}{\del W}\right) f(W) =\int \frac{\rd t}{|t|} f(W+tA)
\nonumber
\ee
although this integral needs to be regularised more carefully~\cite{Gelfand-Shilov}. So similarly,
\be{delta-log} 
    \delta^{(3|4)}(W_1,W_2)=\log\left(\im W_2\!\cdot\!\frac{\del}{\del W_1}\right)\delta^{(4|4)}(W_1)\, .
\nonumber
\ee
We will not have so much use for this formula however.} 
\begin{equation}
\begin{aligned}
    \tilde\delta^{(3|4)}(\rW_1,\rW_2)
    &=\int\frac{\rd t}{t}\,\delta^{(4|4)}(\rW_1-t\rW_2)\\
    &=\im\pi\,\sgn\left(\im\rW_2\!\cdot\!\frac{\del}{\del\rW}\right)\delta^{(4|4)}(\rW_1)
\end{aligned}
\ee
and
\be{N=4sgn}
\begin{aligned}
     \tilde\delta^{(2|4)}(\rW_1;\rW_2,\rW_3) 
    &=\int\frac{\rd s}{s}\frac{\rd t}{t}\,\delta^{(4|4)}(\rW_1-s\rW_2-t\rW_3)\\
    &=(\im\pi)^2\,
    \sgn\left(\im\rW_2\!\cdot\!\frac{\del}{\del\rW_1}\,\im\rW_3\!\cdot\!\frac{\del}{\del\rW_1}\right)\,\delta^{(4|4)}(\rW_1)\ ,
\end{aligned}
\ee
whereas when $\cN=8$ the principal value integral (see {\it e.g.}~\cite{Gelfand-Shilov})
\begin{equation}
{\rm p.v.}\int_{-\infty}^\infty\frac{\rd t}{t^2}\,\e^{-\im at} = -\pi|a|
\end{equation}
gives
\begin{equation}
\begin{aligned}
    \tilde\delta^{(3|8)}(\rW_1,\rW_2)
    &=\int\frac{\rd t}{t^2}\,\delta^{(4|8)}(\rW_1-t\rW_2)\\
     &=-\pi\,\left|\im\rW_2\!\cdot\!\frac{\del}{\del\rW_1}\right|\delta^{(4|8)}(\rW_1)
\end{aligned}
\ee
and
\be{N=8sgn}
\begin{aligned}
     \tilde\delta^{(2|8)}(\rW_1;\rW_2,\rW_3)
     &=\int\frac{\rd s}{s^2}\frac{\rd t}{t^2}\,\delta^{(4|8)}(\rW_1-s\rW_2-t\rW_3)\\
     &=\pi^2\left|\im\rW_2\!\cdot\!\frac{\del}{\del\rW_1}\,\im\rW_3\!\cdot\!\frac{\del}{\del\rW_1}\right|\,\delta^{(4|8)}(\rW_1)
\end{aligned}
\ee
This notation helpfully encodes the scaling behaviour, and it will often be convenient to write both the basic seed amplitudes and the recursion relations themselves in terms of these sign operators.


\subsection{Seed amplitudes in $\cN=4$ SYM}
\label{sec:seedamp}

We now have all the necessary ingredients to state the twistor space
form of the basic three-point amplitudes in a way that clarifies their
conformal properties. In this section, we confine ourselves to a
discussion of $\cN=4$ SYM, postponing the (largely parallel) case of
$\cN=8$ SG until section~\ref{sec:sg}.  

The twistor form of the 3-particle MHV super-amplitude in $\cN=4$ SYM may be written as
\be{MHV3-twist}
\mathbox{ \begin{aligned}
    \cA_{\rm MHV}(\rW_1,\rW_2,\rW_3)&=\sgn(\la\rW_2\rW_3\ra)\,
                    \tilde\delta^{(2|4)}(\rW_1;\rW_2,\rW_3) \\
                    &= \sgn(\la\rW_1\rW_2\ra\la\rW_2\rW_3\ra\la\rW_3\rW_1\ra)\,\delta^{(2|4)}(\rW_1,\rW_2,\rW_3)\ .
    \end{aligned}  }
\ee
where $\la \rW_i\rW_j\ra=\la\lambda_i\lambda_j\ra$ is the usual spinor
product of the $\lambda$-parts of the spinor.  Thus the complete
super-amplitude is a {\it superconformally invariant
$\delta$-function} imposing collinearity of W$_1, \rW_2, \rW_3$, times a
sign factor. The sign factor ensures that~\eqref{MHV3-twist} is
antisymmetric under the exchange of any two points, compensating the
antisymmetry of the colour factor ${\rm Tr}(T_1[T_2,T_3])$. Since the
sign only depends on the ordering of the three twistors, we see that
the twistor amplitude is completely geometric: it depends on the three
ordered points on a line. 

Remarkably, the extension of this amplitude to the $n$-point MHV
amplitude is the product 
\be{MHVn}
    \mathbox{
    \cA_{\rm MHV}(\rW_1,\ldots,\rW_n)
    = (-1)^{n-3}\prod_{i=3}^n\cA_{\rm MHV}(\rW_1, \rW_{i-1}, \rW_i)\ ,} 
\ee
as shown by half Fourier transform in appendix~\ref{app:amplitudes} and obtained from twistor BCFW recursion
below.  Each three-particle MHV amplitude enforces collinearity of W$_1$, W$_{i-1}$ and W$_i$, so the product of such
three-particle amplitudes has the well-known collinear support in twistor space. Again, the amplitude is purely geometric; there are no extra spinor or twistor products. The cyclic symmetry of the MHV amplitude
is not explicit, but follows from~\eqref{MHVn} and the cyclic symmetry of the 4-point amplitude 
\be{4cyclic}
    \cA_{\rm MHV}(1,2,3)\cA_{\rm MHV}(1,3,4) = \cA_{\rm MHV}(2,3,4)\cA_{\rm MHV}(2,4,1)\ .
\ee
This four-point identity is easily proved using the three-point amplitude in the form~\eqref{MHV-twist-explicit}.

\smallskip

The three-point $\MHVbar$ amplitude, given in on-shell momentum superspace by~\cite{Brandhuber:2008pf}
\be{MHV-bar3-mom}
    A_{\MHVbar}(p_1,p_2,p_3)
    =\frac{\delta^{(4)}\!\left(p_1+p_2+p_3\right)\delta^{(4)}\!\left(\eta_1[23]+\eta_2[31]+\eta_3[12]\right)}{[12][23][31]}\ , 
\ee 
has the twistor space representation
\be{MHV3-bar-twist}
    \mathbox{
            \cA_{\MHVbar}(\rW_1,\rW_2,\rW_3) 
            = \sgn\left(\left[\frac{\del}{\del\rW_2}\frac{\del}{\del\rW_3}\right]\right)
            \tilde\delta^{(3|4)}(\rW_1,\rW_2)\,\tilde\delta^{(3|4)}(\rW_1,\rW_3)\ .
    }
\end{equation}
The operator $\sgn[\del_2\,\del_3])$ is a pseudo-differential operator that is closely related to the Hilbert transform. Like the Hilbert transform, it can be easily understood in terms of its Fourier transform (whence it arose). $\cA_{\MHVbar}$ may also be written explicitly as
\be{MHV-bar-explicit}
\begin{aligned}
    \cA_{\overline{\rm MHV}}(\rW_1,\rW_2,\rW_3)
    &= \frac{\lambda_1}{\lambda_2}\,\frac{\lambda_1}{\lambda_3}\,\delta(\la12\ra)\, \delta(\la13\ra)\ \times\\
    &\hspace{-2cm}\delta^\prime\!\left(\!\!\left(\mu_2^{A'} - \frac{\lambda_2}{\lambda_1}\mu_1^{A'}\right)  
    \!\!\left(\mu_{3A'}-\frac{\lambda_3}{\lambda_1}\mu_{1A'}\right)\!\!\right)
    \delta^{(0|4)}\left(\chi_2-\frac{\lambda_2}{\lambda_1}\chi_1\right)
    \delta^{(0|4)}\left(\chi_3-\frac{\lambda_3}{\lambda_1}\chi_1\right)
\end{aligned}
\ee
which follows directly from the half Fourier transform. Again, this explicit representation obscures the conformal properties and in practice the implicit form~\eqref{MHV3-bar-twist} will actually be more useful. It is easy to show that~\eqref{MHV3-bar-twist} is antisymmetric under exchange of any two external states, again compensating the antisymmetry of the colour factor.

\smallskip

Using the sign-function representations of the delta functions, we can also write the three-point amplitudes as 
\be{MHV3formal}
    \mathbox{\begin{aligned}
            \cA_{\rm MHV}(\rW_1,\rW_2,\rW_3) 
            &= (\im\pi)^2\sgn\left(\la23\ra\,\im\rW_2\!\cdot\!\del_{\rW_1}\,
            \im\rW_3\!\cdot\!\del_{\rW_1}\right)\delta^{(4|4)}(\rW_1)\\
            \cA_{\MHVbar}(\rW_1,\rW_2,\rW_3)
            &=(\im\pi)^2\sgn\left(\left[\del_{\rW_2}\del_{\rW_3}\right]\,\im\rW_1\!\cdot\!\del_{\rW_2}\,
            \im\rW_1\!\cdot\!\del_{\rW_3}\right)\,\delta^{(4|4)}(\rW_2)\,\delta^{(4|4)}(\rW_3)\ .
   \end{aligned} }
\end{equation}
This representation may seem rather formal, but it is well-defined as a distribution and in any case is completely equivalent to the concrete forms~\eqref{MHV-twist-explicit} \&~\eqref{MHV-bar-explicit}. Alternatively, one can obtain an ambidextrous representation of the amplitudes by writing the $\delta^{(4|4)}$-functions in terms of Fourier transforms, as is done in~\cite{AHCCK} and as we discuss further in section~\ref{sec:ambi}. The differential operators inside the sign functions in~\eqref{MHV3formal} play a prominent role in what follows. For future reference, we therefore define
\be{C-def}
    \cH^i_{jk} :=
    (\im\pi)^2\sgn\left(\la\rW_j\rW_k\ra\,\im\rW_j\!\cdot\!\frac{\del}{\del\rW_i}\,\im\rW_k\!\cdot\!\frac{\del}{\del\rW_i} \right)
\ee
and similarly
\be{tildeC-def}
    \widetilde \cH^{jk}_i := (\im\pi)^2\sgn \left(
\left[\frac{\del}{\del\rW_j}\frac{\del}{\del\rW_k}\right]\, \im\rW_i\!\cdot\!\frac{\del}{\del\rW_j}\,
    \im\rW_i\!\cdot\!\frac{\del}{\del\rW_k} \right) \ .
\ee
These $\cH^i_{jk}$ and $\widetilde \cH^{ij}_k$ operators are conjugate, in the sense that they are related by making the replacements $\rW_i\leftrightarrow\del/\del\rW_i$ and exchanging the infinity twistor and its dual $I^{\alpha\beta}\leftrightarrow I_{\alpha\beta}$.  Each of these operators will play an important role in the construction of general tree amplitudes. In this representation, the cyclic symmetry of the three-point amplitude is the identity
\be{3cyclic-Cop}
    \cH^1_{23}\delta^{(4|4)}(\rW_1)=\cH^2_{31}\delta^{(4|4)}(\rW_2)\ ,
\ee
while the cyclic symmetry~\eqref{4cyclic} of the four-point amplitude is
\be{4cyclic-Cop} 
\cH^2_{13} \cH^4_{13} \delta^{(4|4)}(\rW_2) \delta^{(4|4)}(\rW_4)= 
\cH^3_{24} \cH^1_{24} \delta^{(4|4)}(\rW_3) \delta^{(4|4)}(\rW_1)\, .
\ee


\subsection{On conformal invariance}
\label{sec:confbreaking}

The results of the previous subsection showed that seed amplitudes in
$\cN=4$ SYM are one or two superconformally invariant
$\delta$-functions, dressed by certain signs (which may not be locally
defined). Although the delta functions in~\eqref{MHV3-twist}
\&~\eqref{MHV3-bar-twist} are manifestly superconformally invariant,
the factors of $\sgn (\la\rW_i \rW_j\ra)$ and the operator
$\sgn([\del_2\,\del_3])$ are not. No choice of tilded or 
untilded  $\delta$-function removes all of these signs, which are
necessary for the amplitudes to have the correct cyclic and exchange properties.

Is this a failure of conformal invariance, or merely a failure to make invariance manifest? Consider the three-point MHV amplitude, written in the second form of~\eqref{MHV3-twist} which makes the exchange properties transparent. On the support of the delta function, the three twistors are collinear and the sign factor just depends on
the ordering of the points along this line. Thus, if we could consistently orientate all the lines in twistor space, we would be able to replace the factor $\sgn(\la\rW_1\rW_2\ra\la\rW_2\rW_3\ra\la\rW_3\rW_1\ra)$ by the prescription that the collinear delta-function is to be multiplied by $+1$ if the ordering of the three twistors agrees with the chosen orientation, and by $-1$  if not. The amplitude would then be conformally invariant. However, there is a topological obstruction to doing this: an oriented line in $\RP^3$ can always be continuously deformed so that it comes back to itself with the opposite orientation ({\it e.g.} by rotating it through $\pi$ about an axis that is perpendicular to the line, thinking of it in affine $\R^3$). Globally, the space of $\RP^1$s inside $\RP^3$ is conformally compactified split signature space-time with topology $(S^2\times S^2)/\Z_2$. This space has fundamental group $\Z_2$, and this fundamental group precisely corresponds to the possible orientations of the twistor line\footnote{A related fact is that the integral in the X-ray transform requires an orientation, and so gives rise to massless  fields that have wrong-sign behaviour with respect to this $\Z_2$, {\it i.e.}, they are wrong-sign sections of the line bundle of functions of homogeneity $-1$ on the 4-quadric with signature $(3,3)$ in $\RP^5$.  This `wrong-sign' behaviour is not correlated with that of the amplitude.}.

After removing a line $I$ `at infinity', one \emph{can} orientate all the $\RP^1$s that do not meet this line. The remaining twistor space fibres over $\RP^1$ and we can fix the orientation on any line that does not meet $I$ by pulling back the orientation of this $\RP^1$. Equivalently, removing a line $I$ from twistor space removes a point $i$ from conformally compactified space-time. The space of twistor lines that do not intersect $I$ corresponds to the region of conformally compactified space-time that is not null separated from the point $i$, in other words affine space-time $\R^{2,2}$. Thus, provided one stays within a single copy of split signature affine space-time, the sign factors in~\eqref{MHV3-twist} just amount to an overall sign that may be consistently chosen. However, there is no consistent way to extend this over the whole of $\RP^3$, and conformal invariance is genuinely broken.

How do we reconcile this with the fact that momentum space amplitudes
are annihilated by all the superconformal generators (see {\it e.g.}~\cite{Witten:2003nn})?  Let us examine in detail how conformal invariance is broken. Acting on the three-point MHV amplitude (in the form of the second equation of \eqref{MHV3-twist}) with the conformal generators $J^{\ \beta}_\alpha=\sum_{i=1}^3\rW_{i \alpha}\del/\del\rW_{i \beta}$, the only possible contribution comes from the sign function. We obtain
\be{3pt-conf-invce}
    J_\alpha^{\ \beta} \cA_{\rm MHV}(\rW_1,\rW_2,\rW_3)
    =(\rW_{2\alpha}I^{\beta\gamma}\rW_{3\gamma}-\rW_{3\alpha}I^{\beta\gamma}\rW_{2\gamma})\,
    \delta(\la \rW_2\, \rW_3\ra)\, \tilde \delta^{(2|4)}(\rW_1;\rW_2,\rW_3)\ .
\ee
On the support of the delta functions in this expression, the $\rW_i$ are all collinear and  $\la \rW_2\rW_3\ra=0$. Geometrically, the condition  $\la \rW_2\rW_3\ra=0$ means that W$_2$ and W$_3$ lie in the same plane through the line `at infinity' ($I^{\alpha\beta}\rW_\alpha=0$, or $\lambda_A=0$).  So the delta functions in~\eqref{3pt-conf-invce} give support only when the $\rW_i$ all lie on a line that intersects the line at infinity. On such lines, the coefficient $\rW_{2\alpha}I^{\beta\gamma}\rW_{3\gamma}-\rW_{3\alpha}I^{\beta\gamma}\rW_{2\gamma}$ does not vanish (unless $\rW_2$ and $\rW_3$ actually coincide). Thus, the failure of conformal invariance occurs where all three particles' $\lambda_A$ spinors are proportional. This is the most singular part of the momentum space amplitude -- a momentum space calculation (or one based on the explicit twistor form~\eqref{MHV-twist-explicit}) could only uncover the failure of conformal invariance with a careful analysis of anomalous terms in the action of the conformal generators in the triple-collinear limit.  

However, although the failure of conformal invariance is associated with the collinear singularities of the momentum space amplitudes, note that nothing singular is happening in twistor space. Given a line in `affine' twistor space, a collinear singularity occurs when two or more marked points on this line collide -- this process is conformally invariant.
By contrast, the violation of conformal invariance above is associated with support on lines that intersect $I$. No twistors need collide, and from the point of view of the conformally compactified space, this line is on the same footing as any other. Moreover, the collinear delta function $\delta^{(2|4)}(\rW_1,\rW_2,\rW_3)$ corresponds to
\be{genuine-superconformal}
     \frac{\delta^{(4|4)}\left(\sum_{i=1}^3 |i\ra\llbracket i\|\right)}{\left|\la12\ra\la23\ra\la31\ra\right|}
\ee
on momentum space, so (at least away from singularities) it equals\footnote{Recall that in split signature, the momentum space spinors are real.} $\pm A_{\rm MHV}(1,2,3)$. Thus, on the open region of momentum space with collinear singularities removed, \eqref{genuine-superconformal} is likewise annihilated by all the generators of the superconformal algebra. However, under a finite conformal transformation, \eqref{genuine-superconformal} fits together across the singularities in a way which is conformally invariant, while the amplitude itself does not.

\bigskip

The conformal properties of the three-point $\MHVbar$ amplitude
follows similarly from a CPT transformation (or the Fourier transform
of section~\ref{sec:ambi}). Although the $n$-particle MHV amplitudes might at first sight appear
worse, many of the sign factors cancel: Arrange the 3-point factors
in~\eqref{MHVn} pair-wise, and use cyclic symmetry and the first line
of~\eqref{MHV3-twist} to ensure that the only sign factors are $\sgn
\la 1i\ra$, occurring in both the $i^{\rm th}$ and $(i+1)^{\rm th}$
term. These signs then cancel. With this cancellation, the even-point
MHV amplitudes are manifestly conformally invariant. The odd-point
amplitudes still end up with the one three-point sub-amplitude in the
product~\eqref{MHVn} whose conformal breaking sign factor cannot be
made to cancel. This is consistent with the topological argument
because the relation  
\be{inversion}
   \cA(1,2,\ldots,n-1,n)=(-1)^n\cA(n,n-1,\ldots,2, 1)\, ,
\ee
requires that odd-point amplitudes, but not even-point amplitudes,
change sign under a reversal of the  
orientation of points along the line.


\section{BCFW Recursion in Twistor Space}
\label{sec:recursion}

We now use the half Fourier transform to translate the supersymmetric BCFW recursion relation into twistor space. See~\cite{Britto:2005fq} for a proof of the original BCFW rule in Yang-Mills,~\cite{Benincasa:2007qj} for gravity and~\cite{Brandhuber:2008pf,ArkaniHamed:2008gz} for the supersymmetric extension. 

In both $\cN=8$ SG and ${\cN=4}$ SYM, the (super-)BCFW recursion rule states that
\be{BCFW} 
    \widetilde A(1,\ldots,n) = \sum\int\rd^\cN\eta\  \widetilde A_L\!\left(\hat 1,2,\ldots,i,\{-\hat p,\eta\}\right)
    \frac{1}{p^2}\,\widetilde A_R\!\left(\{\hat p,\eta\},{i+1},\ldots,{n-1},\hat n\right) 
\ee
where $\widetilde A$ denotes a tree-level super-amplitude that has been stripped of
its overall momentum conserving $\delta$-function (and, in Yang-Mills,
also of its colour factor).  The sum is taken over all possible ways
of splitting the external states among the two sub-amplitudes, subject to the requirement that the shifted momenta
reside in separate sub-amplitudes (and subject to cyclic symmetry in
SYM). The integral over the Grassmann variables $\eta$ of the internal
supermultiplet accounts for the possible helicity states of the
internal particle. The propagator momentum $p$ is defined as usual,
{\it i.e.} 
\begin{equation}
    p := \sum_{j\in L}{p_j}\ ,
\end{equation}
where $L$ is the set of external particles attached to the left
sub-amplitude. The supermomenta in the sub-amplitudes are shifted
compared to the external momenta according to the general 
prescription~\eqref{supershift}. Similarly, in the sub-amplitudes
$\widetilde A_{L,R}$, the propagator momentum $p$ is shifted as
$p\to\hat p:= p - t|1\ra[n|$. For a given term in the sum
in~\eqref{BCFW} ({\it i.e.}, a given decomposition into 
sub-amplitudes) the shift parameter $t$ is fixed to the value $t_*$
that ensures $\hat p^2(t_*)=0$. Consequently, all the momenta in
$\widetilde A_{L,R}$ are null, so these are fully on-shell
sub-amplitudes. Note that the $t_*$ are real in $(++--)$ signature
space-time.

As a preliminary step towards transforming the BCFW relation to
twistor space,  first restore the momentum-conserving 
$\delta$-functions to~\eqref{BCFW}. One finds
\be{momdelta1}
    A(1,\ldots,n) =\sum \int\rd^4p\,\rd^\cN\eta\ \delta^{(4)}\left(-p+\sum_{j\in L}p_j\right)\widetilde
    A_L(t_*)\frac{1}{p^2} \delta^{(4)}\left(p+\sum_{k\in R}p_k\right)\widetilde A_R(t_*)\ , 
\ee
where now $p$ is {\it a priori} unconstrained (and in particular is generically off-shell).  We can
always obtain a null momentum by projecting the arbitrary momentum $p$ along some fixed null
momentum direction, so we can always set
\be{2null} 
    p = \ell - t |1\rangle[n|\ ,
\ee 
where $\ell=|\lambda\rangle[\tilde\lambda|$ is a null but otherwise
arbitrary momentum, and $t$ is an arbitrary parameter. 

In terms of the
$(\ell,t)$ variables, the integration measure and propagator combine
to become 
\be{Nair} 
\frac{\rd^4p}{p^2} = \sgn (\langle 1
|\ell|n])\frac{\rd t}{t}\rd^3\ell = \sgn (\langle 1 \lambda\rangle
[\tilde\lambda n]) \frac{\rd
t}{t}\left(\langle\lambda\rd\lambda\rangle\rd^2\tilde\lambda
-[\tilde\lambda\rd\tilde\lambda]\rd^2\lambda\right)
\ee 
as in~\cite{Brandhuber:2004yw}. The sign factor $\sgn (\langle 1 |\ell|n])$ arises because the orientation on
the $\rd t$ factor changes sign with $\langle 1|\ell|n]$.  This can be
seen from the fact that the momentum light-cone is naturally oriented
by the orientation of momentum space, together with the choice of
outward normal going from $p^2<0$ to $p^2>0$.  The direction $\rd t$
is essentially that of $|1\rangle[n|$, and is aligned or anti-aligned
with this outgoing normal according to the sign of $\langle
1\lambda\rangle[\tilde\lambda n]$. Therefore we must incorporate this
sign in order to have agreement with the given orientation on momentum
space. The measure
\begin{equation}
    \rd^3\ell =
    \langle{\lambda\rd\lambda}\rangle\rd^2\tilde\lambda- [\tilde\lambda\rd\tilde\lambda]\rd^2\lambda
\end{equation}
on the null cone in momentum space is invariant under the scaling
$(\lambda,\tilde\lambda)\to (r^{-1}\lambda,r\tilde\lambda)$ where $r$
is an arbitrary function of the projective spinors $[\lambda_A]$ and
$[\tilde\lambda_{A'}]$.  We can represent this null cone as a rank two
bundle over the $\RP^1$ factor, co-ordinatised by the $[\lambda_A]$ on
the base and $\tilde\lambda_{A'}$ up the fibre; doing so amounts to
restricting $r$ to be a function of $[\lambda_A]$ alone. The measure then reduces to
$\rd^3\ell=\la\lambda\rd\lambda\ra\,\rd^2\tilde\lambda$, which combines with the
integral over the internal $\eta$s to give
\be{nullsupercone} 
    \frac{\rd^4p}{p^2}\rd^\cN\eta =\sgn(\langle 1\lambda\rangle[\tilde\lambda n]) 
    \frac{\rd t}{t}\la\lambda\rd\lambda\ra\,\rd^{2|\cN}\tilde\lambda \ .
\ee 
Thus the integral and propagator in the BCFW recursion may together be interpreted as an integral over the on-shell
momentum superspace of the internal state, together with an integral over the BCFW shift parameter.

In the $(\lambda,\tilde\lambda,t)$ variables, the momentum $\delta$-functions in the integral in~\eqref{momdelta1} become 
\be{shiftdelta}
    \delta^{(4)}\left(-\lambda\tilde\lambda+ 
    \sum_{j\in L}\hat{p}_j(t)\right) \delta^{(4)}\left(\lambda\tilde\lambda
    +\sum_{k\in R}\hat{p}_k(t)\right)\ , 
\ee 
which are the $\delta$-functions associated with the sub-amplitudes $A_{L,R}(t)$ for
{\it arbitrary} values of the shift parameter $t$. However, on the
support of these $\delta$-functions, $t$ is fixed to be precisely
$t=t_*$ and then $\ell=\hat p(t_*)$. Hence the $\delta$-functions
allow us to replace $\widetilde{A}_{L,R}(t_i)$ by $A_{L,R}(t)$ inside
the integral~\eqref{momdelta1}. Thus we have a form of the BCFW
recursion relation in which the propagator has been absorbed into the
measure and all the ingredients are manifestly on-shell: 
\begin{multline}
\label{BCFW2}
    A(1,\ldots,n) = \sum \int\frac{\rd t}{t}\la\lambda\rd\lambda\ra\,
    \rd^{2|\cN}\tilde\lambda\, \sgn(\langle1\lambda\rangle[\tilde\lambda n])\\
    \hspace{-5cm}\times\ 
    A_L(\hat{1},\ldots,\{-\lambda,\tilde\lambda,\eta\})\, A_R(\{\lambda,\tilde\lambda,\eta\},\ldots,\hat{n})\ .  
\end{multline} 
This form of the BCFW recursion relation is somewhat similar to a completeness relation: One decomposes the amplitude by inserting a complete set of on-shell states. However, such an interpretation does not account for the shift of the external states 1 and $n$, nor the integral over the shift parameter $t$.


\subsection{The $\im\epsilon$-prescription on $\R^{2,2}$}
\label{sec:ieps}

There is a subtlety\footnote{We  thank N. Arkani-Hamed for discussions of this
point.} in the definition of the propagator measure
in~\eqref{Nair} \&~\eqref{nullsupercone} because the $1/t$ factor
means that the $t$-integral is singular.  We have seen that for tree
amplitudes, the integral is performed by integration against a
delta-function, and so the regularisation is not urgent, but it
nevertheless should in general be regularised, particularly if one
wishes to apply these ideas to loop processes. 

In quantum field theory in Minkowski space, stability of the vacuum requires
that only positive energy states be allowed to propagate towards the
future. This is achieved by using the Feynman
propagator $\Delta_F(x-y) = \la0|T\phi(x)\phi(y)|0\ra$ which is time-ordered.
Using the Fourier transform of the time-ordering step
functions 
\be{Fourier-step}
    \theta(x^0-y^0) = -\frac{1}{2\pi\im}\int_{-\infty}^{\infty} \rd E\,\frac{\e^{-\im E(x^0-y^0)}}{E+\im\epsilon}
\ee
one arrives at the $\im\epsilon$-prescription $1/p^2\to1/(p^2+\im\epsilon)$ for the momentum space propagator.

However, in this paper we are tied to split signature
space-time, $\R^{2,2}$, which makes no distinction between past and future.
The light-cone is connected as are the `time-like' vectors which are now
on the same footing as space-like ones.  There is no past or future so
it does not make sense to ask that positive
energy particles propagate `forwards' in time and negative energy ones
`backwards'. Correspondingly, in split signature momentum space, the
natural choice of $\im\epsilon$ prescription is 
\be{principal-value-prop}
    \frac{1}{p^2}\to\frac{1}{2}\left(\frac{1}{p^2+\im\epsilon} +
      \frac{1}{p^2-\im\epsilon}\right) \, .
\ee
We will therefore adopt this prescription when we need to. Thus, the
measure in~\eqref{Nair} can be written as
\be{principal-value-meas}
\begin{aligned}
    \frac{\rd^4p}{p^2}&\to\frac{1}{2}\left(\frac{\rd^4p}{p^2+\im\epsilon}+\frac{\rd^4p}{p^2-\im\epsilon}\right)\\
    &=\frac{1}{2}\left(\frac{\rd t}{t+\im\epsilon}+\frac{\rd t}{t-\im\epsilon}\right)\sgn (\langle 1 |\ell|n])\rd^3\ell\ ,
\end{aligned}
\ee
and this amounts to treating the $\rd t$ integral via a Cauchy principal
value integral. 
Although we will often write the abbreviated
form~\eqref{Nair}, we will take the $\rd t$ integral to be a
principal value integral. (This is in contrast to the proof~\cite{Britto:2005fq} of the
BCFW relations which treats it as an $S^1$ contour integral.)
This wont make any difference to tree
level calculations where the integral is determined by
delta-functions, but this will make a difference for loops.  


\subsection{Transform to twistor space}
\label{sec:twistorBCFW}

We will now take the half Fourier transform of equation~\eqref{BCFW2} with
respect to the $\tilde\lambda$ variables of the external states and
substitute in the inverse half Fourier transform from twistor space for
the sub-amplitudes. On
the lhs, this is just the definition of the 
twistor super-amplitude: 
\be{twistamp} 
    \cA(\rW_1,\ldots,\rW_n) :=\int\prod_{i=1}^n\rd^{2|\cN}\tilde\lambda_i\ 
            \e^{\im\llbracket\mu_i\tilde\lambda_i\rrbracket}\ A(1,\ldots,n)\ .  
\ee 
Whereas on the right hand side we obtain
\be{left-right-transform}
\begin{aligned}
    A_L(t)
    &=\int\frac{\rd^{2|\cN}\mu}{(2\pi)^2}\,\e^{\im \llbracket\mu\tilde\lambda\rrbracket}
    \prod_{j\in L}\frac{\rd^{2|\cN}\mu'_j}{(2\pi)^2}\, \e^{-\im\llbracket\mu'_j\hat{\tilde\lambda}_j(t)\rrbracket}\,
    \cA_L(\rW'_1,\ldots,\rW)\\
    A_R(t)
    &=\int\frac{\rd^{2|\cN}\mu'}{(2\pi)^2}\,\e^{-\im\llbracket\mu'\tilde\lambda\rrbracket}
    \prod_{k\in R}\frac{\rd^{2|\cN}\mu'_k}{(2\pi)^2}\, \e^{-\im\llbracket\mu'_k\tilde\lambda\rrbracket}\,
    \cA_R(\{\lambda,\mu',\chi'\},,\ldots,\{\lambda_n-t\lambda_1,\mu'_n,\chi'_n\}))
\end{aligned}
\end{equation}
where we use the hatted variables 
\begin{equation}
    \|\hat{\tilde\lambda}_1(t)\rrbracket = \|\tilde\lambda_1\rrbracket + t\|\tilde\lambda_n\rrbracket
    \qquad
    \|\hat{\tilde\lambda}_j(t)\rrbracket = \|\tilde\lambda_j\rrbracket \quad\hbox{for}\ j\neq1
\end{equation}
in the transformation of the external states in $A_L$.  It
makes no difference whether we write
$\|\hat{\tilde\lambda}_k\rrbracket$ or simply
$\|\tilde\lambda_k\rrbracket$ for the variables in $A_R$, since the
shifts in $A_R$ do not involve these variables. However, as indicated
in~\eqref{left-right-transform},  we must account for the shift
$|n\ra\to|n\ra-t|1\ra$ explicitly. The change in sign between the
Fourier transform of the internal state in $A_L$ compared to $A_R$
accounts for the fact that $\ell$ is the momentum flowing {\it out} of
$A_L$ and {\it in} to $A_R$. Note that the supertwistors associated to
the internal state have the same unprimed spinor part $\lambda_A$ in
both $\cA_L$ and $\cA_R$.

We now insert these sub-amplitudes into~\eqref{BCFW2} and transform the
whole expression back to twistor space. For all the external states
except 1 and $n$, this is trivial.  Changing variables
$\tilde\lambda_1\to\hat{\tilde\lambda}_1$ also allows us to perform
the $\rd^{2|\cN}\hat{\tilde\lambda}_1\,\rd^{2|\cN}\mu'_1$ integrals
directly. The remaining integrals are 
\be{BCFW-twist1}
\begin{aligned}
    &\frac{1}{(2\pi)^6}\int\frac{\rd t}{t}\la\lambda\rd\lambda\ra\rd^{2|\cN}\mu\,
    \rd^{2|\cN}\tilde\lambda\,\rd^{2|\cN}\tilde\lambda_n\,
    \rd^{2|\cN}\mu'\,\rd^{2|\cN}\mu'_n\ \e^{\im\llbracket(\mu-\mu')\,\tilde\lambda\rrbracket}\,
    \e^{\im\llbracket(\mu_n-\mu'_n-t\mu_1)\,\tilde\lambda_n\rrbracket}\\
    &\hspace{2cm}\times\ \sgn(\la1\lambda\ra[\tilde\lambda n])\ 
    \cA_L(\rW_1,\ldots,\rW)\,\cA_R(\{\lambda,\mu',\chi'\},\ldots,\{\lambda_n-t\lambda_1,\mu'_n,\chi'_n\})\\
    =&\frac{1}{(2\pi)^6}\int\frac{\rd t}{t}D^{3|\cN}\rW\,\rd^{2|\cN}\tilde\lambda\,\rd^{2|\cN}\tilde\lambda_n\,
    \rd^{2|\cN}\mu'\,\rd^{2|\cN}\mu'_n\ \e^{\im\llbracket(\mu-\mu')\,\tilde\lambda\rrbracket}\,
    \e^{\im\llbracket(\mu_n-\mu'_n)\,\tilde\lambda_n\rrbracket}\\
    &\hspace{3cm}\times\ \sgn(\la1\lambda\ra[\tilde\lambda n])\ 
    \cA_L(\rW_1,\ldots,\rW)\,\cA_R(\rW',\ldots,\rW'_n-t\rW_1)\ ,
\end{aligned}
\ee
where in the second line we have translated $\|\mu'_n\rrbracket$ to $\|\mu'_n\rrbracket - t\|\mu_1\rrbracket$ and defined
\begin{equation}
    \rW':=(\lambda,\mu',\chi')\qquad\hbox{and}\qquad \rW'_n:=(\lambda_n,\mu'_n,\chi'_n)\ .
\end{equation}
We also combined $\la\lambda\rd\lambda\ra\rd^{2|\cN}\mu$ into the measure $D^{3|\cN}\rW$ on the supertwistor space of the internal state. 

To proceed, we somewhat formally write 
\be{formal}
    \sgn[\tilde\lambda\tilde\lambda_n] = \sgn\left[\frac{\del}{\del\mu'}\frac{\del}{\del\mu'_n}\right]
\ee
inside the integrals\footnote{The definition of this operator will
always be via the Fourier transform.  In particular, this makes
transparent that $\sgn ([\del_i\,\del_j][\del_i\,\del_j])=1$, which will be a key property in manipulating the recursion relations}. The operator $\sgn
[\frac{\del}{\del\mu'}\frac{\del}{\del\mu'_n}]$ then acts on $\cA_R$
(as a distribution), whereupon the remaining integrals (except those
over $t$ and the internal supertwistor W) become straightforward. We
are left with 
\begin{multline}
\label{BCFW-twist2}
    \cA(\rW_1,\ldots,\rW_n) = \sum \int \frac{\rd t}{t}D^{3|\cN}\rW\,\cA_L(\rW_1,\ldots,\rW)\\
    \times\ \sgn\left(\la1\lambda\ra\left[\frac{\del}{\del\mu}\frac{\del}{\del\mu_n}\right]\right)
    \cA_R(\rW,\ldots,\rW_n-t\rW_1)\ .
\end{multline}
The only $t$-dependence is inside $\cA_R$. Since our split signature $\im\epsilon$-prescription means that the $\rd t/t$ integral to be understood as a principal value integral, from section~\ref{sec:hilbert} we can write
\begin{equation}
    \int\frac{\rd t}{t} \cA_R(\rW,\ldots,\rW_n-t\rW_1) 
    = \im\pi \, \sgn\left(\im
    \rW_1\!\cdot\!\frac{\del}{\del\rW_n}\right)\cA_R(\rW,\ldots,\rW_n)\ . 
\ee
Combining this with~\eqref{BCFW-twist2}, we arrive at our final form of the BCFW recursion relation in (dual) supertwistor space: 
\be{BCFWtwist}
    \mathbox{
    \begin{aligned}
            \cA(\rW_1,\ldots,\rW_n) &= \sum\int D^{3|\cN}\rW\,\cA_L(\rW_1,\ldots,\rW)\\
            &\qquad\times\ \im\pi\,\sgn\left(\la\rW_1\rW\ra\, \im\rW_1\!\cdot\!\frac{\del}{\del\rW_n}\,
            \left[\frac{\del}{\del\rW}\frac{\del}{\del\rW_n}\right]\right)\ \cA_R(\rW,\ldots,\rW_n)
    \end{aligned}}\ ,
\ee
where $\la\rW_1\rW_2\ra=\la12\ra$ and
$\left[\del_{\rW}\,\del_{\rW_n}\right]
=\del^2/\del\mu^{A'}\del\mu_{A'n}$. Thus, in twistor space, BCFW
recursion recursion involves gluing two sub-amplitudes together using
the operator 
\be{non-local}
    \sgn\left(\la\rW_1\rW\ra\, \rW_1\!\cdot\!\frac{\del}{\del\rW_n}\,
    \left[\frac{\del}{\del\rW}\frac{\del}{\del\rW_n}\right]\right)
\ee
and then integrating over the location of the intermediate
supertwistor. Based on~\eqref{BCFWtwist} and the forms of the
three-point amplitudes given in~\eqref{MHV3formal}, we immediately
conclude that the complete classical S-matrix of $\cN=4$ SYM can be
written as a sum of products of sign operators acting on basic 
$\delta$-functions, with these products then integrated over some
number of copies of supertwistor space.  

Although the non-local operator~\eqref{non-local} and integrals
over internal twistors seem rather awkward and may initially seem
disappointing, just as the momentum 
$\delta$-functions allow us to perform the integrals in~\eqref{BCFW2}
and  
return to the unintegrated form~\eqref{BCFW}, we will see in the
examples below that the recursion relation~\eqref{BCFWtwist} is
quite tractable and the integrals and operators can often be
evaluated explicitly, although not yet so systematically as in the
momentum space representation.   


\section{SYM Twistor Amplitudes from BCFW Recursion}
\label{sec:using-recursion}

In this section we use the BCFW  recursion relations to calculate the
twistor form of various scattering amplitudes in $\cN=4$ SYM. We
denote the $n$-particle N$^k$MHV twistor super-amplitude by
$\cA_k^n(1,\ldots,n)$, although we occasionally omit the super- or subscript
when the context makes it clear.


\subsection{On the general structure of $\cN=4$ Amplitudes}
\label{sec:general-structure}

Scattering amplitudes in a theory with unbroken supersymmetry such as $\cN=4$
SYM can only depend on combinations 
of Grassmann variables that are invariant under the 
R-symmetry group. In split signature space-time this is $\SL(\cN;\R)$, so invariants can only be constructed by complete
contractions with the $\cN$-dimensional Levi-Civita symbol. Thus, decomposing momentum space amplitudes into their homogeneity in the $\eta$s, only multiples of $\cN$ will appear for the homogeneities, with N$^k$MHV amplitudes being homogeneous polynomials of degree $(k+2)\cN$ in the $\eta$s.  The $n$-fold half
Fourier transform for an $n$-particle amplitude sends such a homogeneous polynomial to a homogeneous
polynomial of degree $(n-k-2)\cN$ in the $\chi$s, so $n$-particle
N$^k$MHV amplitudes on (dual) supertwistor space have homogeneity $(n-k-2)\cN$ in the anti-commuting
variables.  According to this counting, the 3-particle $\MHVbar$ amplitude should be taken as having $k=-1$, but for all other tree amplitudes, $k\geq 0$.

We can use the recursion relations to show that, ignoring the
conformal breaking sign factors,  a general $n$-particle
N$^k$MHV super-amplitude is obtained by acting on $(n-2-k)$
$\delta^{(4|4)}$-functions with $2(n-2)$ Hilbert transforms.  To start
the induction,  recall from section~\ref{sec:seedamp} the $3$-point MHV and $\MHVbar$ amplitudes 
\be{3-point-recall}
\begin{aligned}
    \cA_0^3(1,2,3) &= \sgn\la23\ra\int\frac{\rd s}{s}\frac{\rd t}{t}\,\delta^{(4|4)}(\rW_1-s\rW_2-t\rW_3)\\
    \cA_{-1}^3(1,2,3) & = \sgn[\del_2\del_3]\int\frac{\rd s}{s}\frac{\rd t}{t}\,
    \delta^{(4|4)}(\rW_2-\rW_1)\,\delta^{(4|4)}(\rW_3-t\rW_1)\, ;
\end{aligned}
\ee
these are constructed from two ($=2(n-2)$) Hilbert transforms acting on a
product of $(1-k)$  $\delta^{(4|4)}$-functions, where $k=0,-1$ for the
MHV and $\MHVbar$ amplitudes, respectively. Now proceed by induction on $n$ and
$k$: Suppose that a given term in the BCFW recursion decomposes an
$n$-point N$^k$MHV amplitude $\cA^n_k$ into an $r$-point N$^l$MHV amplitude 
$\cA^r_l$ and a $s$-point N$^m$MHV amplitude $\cA^s_m$. Then 
\begin{equation}
    n=r+s-2\qquad\hbox{and}\qquad k=l+m+1\ .
\end{equation}
The $\D^{3|4}W$ integration in the recursion removes one projective $\delta^{(3|4)}$-function, and this $\delta^{(3|4)}$-function is a single Hilbert transform of a $\delta^{(4|4)}$-function. Thus the total number of constituent $\delta^{(4|4)}$-functions in $\cA^n_k$ is one less than the sum of the numbers in $\cA^r_l$ and $\cA^2_m$, {\it i.e.}
\be{kjghjkli8}
    \#\left(\delta^{(4|4)}\hbox{-functions}\right) = (r-2-l)+(s-2-m) = (n-2-k)
\ee
by induction from the three-point amplitudes. On the other hand, the
gluing operator~\eqref{non-local} itself involves the Hilbert
transform $\sgn\rW_1\!\cdot\!\del_n$ (which cannot cancel with one in
the right hand amplitude as W$_1$ is not a variable in that
amplitude), so the net number of Hilbert 
transforms remains the same and we inductively find 
\begin{equation}
    \#\left(\hbox{Hilbert}\ \hbox{transforms}\right) = 2(r-2)+2(s-2)=2(n-2)
\ee
as was to be proved. 

The other constituents in the 3-point amplitudes and the recursion relations
are the local and non-local sign factors $\sgn\la ij\ra$ and
$\sgn [\del_i\del_j]$.  It is clear from the form of the discussion of
conformal properties of the $n$-point 
MHV amplitudes in section~\ref{sec:confbreaking} that there is ample
scope for the cancellation of 
these factors so we can make no uniform statement about how many of
these survive in a final formula for an amplitude.


\subsection{Solving the recursion relations}
\label{sec:homog-cojhomog}

There are two terms in the BCFW decomposition of a generic amplitude
$\cA^n_k$ that play a somewhat distinguished role -- when one or other
of the two sub-amplitudes is a three-point amplitude. In these cases,
momentum space 
kinematics ensure that with the $[1n\ra$ shift we have chosen, only
the right sub-amplitude can be $\MHVbar$ (in which case the left
sub-amplitude is $\cA^{n-1}_k$), while only the left sub-amplitude can
be MHV (in which case the right sub-amplitude is
$\cA^{n-1}_{k-1}$). We call the first case the `homogeneous
contribution' following~\cite{Drummond:2008cr}, and the second case
the `conjugate homogeneous contribution'. We now explain how to
perform the integral in the twistor BCFW recursion explicitly in these
two cases. This will form the basis of our strategy for solving the
recursion relations in general.  The outcome of these recursions simply leads to
the action of operators  $\cH^i_{i-1,i+1} \delta^{4|4}(W_i)$ and
$\widetilde\cH^{i-1,i+1}_i$ on 
the remaining sub-amplitudes that insert particle $i$ inbetween $i-1$
and $i+1$ in the remaining subamplitude.  These operators can be
identified with the inverse 
soft limits of \cite{AHCCK}.


\subsubsection{The homogeneous term and MHV amplitudes}
\label{sec:homogeneous}

The homogeneous contribution to the twistor BCFW decomposition of $\cA^k_n$ is
\be{hgs-rec0}
    \int D^{3|4}\rW\, \cA_k(1,\ldots,{n-2},\rW)\,
    \sgn\left(\la 1\rW\ra\,\im\rW_1\!\cdot{\del}_n\,\left[\del_\rW\,\del_n\right]\right)\,
    \cA_{-1}(\rW,{n-1},n)\  .
\ee
Recalling the form~\eqref{3-point-recall} of the $\MHVbar$ amplitude we see that the $\sgn[\del_\rW\del_n]$
operators in the sub-amplitude and recursion cancel up to a constant factor of $-1$, coming from the different ordering of W and W$_n$ in the two terms. Since 
\begin{equation}
    \im\pi\,\sgn\left(\im\rW\!\cdot\del_{n-1}\right)\delta^{(4|4)}(\rW_{n-1})=\tilde\delta^{(3|4)}(\rW_{n-1},\rW)
\ee
we can perform the $D^{3|4}\rW$ integration trivially, yielding the contribution 
\be{hgs-rec1}
    \cA_k(1,\ldots,{n-2},n-1)\ \times\ 
    \sgn\left(\la 1\, n-1\ra\,\im\rW_1\!\cdot {\del}_n\,\im\rW_{n-1}\!\cdot\del_n \right)\, \delta^{(4|4)} (\rW_n)\  .
\ee
Recognising the 3-point MHV amplitude, we obtain the final form
\be{homogcontrib}
    -\cA_k(1,\ldots,{n-2},{n-1})\times\cA_{0}(1,n, {n-1})
\ee
for the contribution to $\cA_{k}^n$ from the homogeneous term. Thus the
homogeneous term simply tacks on a 3-point MHV amplitude to the
$(n-1)$-particle N$^k$MHV amplitude.  This has the effect of inserting
the dual twistor $\rW_n$ in between $\rW_{n-1}$ and
$\rW_1$, that were adjacent in the sub-amplitude.    

\begin{figure}
\begin{center}
\includegraphics[width=70mm]{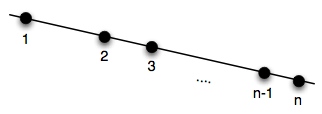}
\caption{The MHV amplitude is supported on a line in twistor space.}
\label{fig:MHV}
\end{center}
\end{figure}

\bigskip

For MHV amplitudes ($k=0$), the homogeneous term is the complete BCFW
decomposition and we immediately obtain
\be{MHVn-2}
\begin{aligned}
    \cA^n_0(1,\ldots,n) &= (-1)^{n-3}\prod_{i=3}^{n}\cA_0(1,{i-1},i)\\
    &= \prod_{i=3}^n  \cH^i_{i-1,i-2}\,\delta^{(4|4)}(\rW_i)
\end{aligned}
\ee
in agreement with equation~\eqref{MHVn}.  The basic three point MHV
amplitude is proportional to a collinear delta function, so the
$n$-particle MHV amplitude requires that all the points are collinear
in twistor space, as is well-known. (See figure~\ref{fig:MHV}.)


\subsubsection{The conjugate homogeneous term and the $\MHVbar$ amplitudes}
\label{sec:conjhomog}

The conjugate homogeneous contribution to the decomposition of $\cA^n_k$ is
\begin{equation}
    \int \D^{3|4}\rW \cA_{0}(1,2,\rW) \,
    \sgn\left(\la 1\rW\ra\,\im\rW_1\!\cdot\del_{n}\,\left[\del_\rW\,\del_{n}\right]\right)\,
    \cA^{n-1}_{k-1}(\rW,3,\ldots,n)\, . 
\end{equation}
From~\eqref{3-point-recall} we have
\begin{equation}
    \cA_{0}(1,2,\rW) = \im\pi\,\sgn\left(\la\rW1\ra\,\im\rW_1\!\cdot\!\frac{\del}{\del\rW_2}\right)\,
    \tilde\delta^{(3|4)}(\rW_2-s\rW_1,\rW)
\end{equation}
and the $\tilde\delta^{(3|4)}$-function again allows us to perform the $D^{3|4}\rW$ integral directly.
We obtain
\be{googly-rec0}
    -\pi^2\sgn\left(\left[\del_2\,\del_n\right]\,\im\rW_1\!\cdot\del_2\,\im\rW_1\!\cdot\del_{n}\right)\,
    \cA^{n-1}_{k-1}(2,3,\ldots,n) 
    = -\widetilde\cH^{2n}_1\cA^{n-1}_{k-1}(2,3,\ldots,n)\, . 
\ee
Applying $\widetilde\cH^{2n}_1$ to $\cA^{n-1}_{k-1}(2,\ldots,n)$ inserts the point $\rW_1$ in between $\rW_2$
and $\rW_n$, which are adjacent in the colour ordering of the sub-amplitude.

\bigskip

Just as the homogeneous term is the only contribution to the BCFW decomposition of an MHV amplitude, so too this conjugate homogeneous term is the only contribution to the `googly MHV' amplitude $\cA^n_{n-4}$ -- the CPT conjugates of the MHV amplitudes.   To see this, first note that for $\cA^n_k\neq0$, generically $k\leq n-4$ (with the equality holding for the googly MHV amplitudes) except that the three-point MHV amplitude has $k=n-3$. This CPT conjugate to the statement that, with the exception of the 3-point $\MHVbar$ amplitude $\cA^3_{-1}$, amplitudes with $k<0$ vanish.
Now, if we decompose a googly MHV amplitude $\cA^n_{n-4}$ into $\cA^r_l$ and $\cA^s_m$ sub-amplitudes, since $r+s=n+2$ and $l+m=n-4+1$ we must have $(r-l) + (s-m) = 7$. Consequently, one of the sub-amplitudes must be a three-point MHV and momentum kinematics dictates that it is the left sub-amplitude. The other sub-amplitude is then $\cA^{n-1}_{n-5}$; the googly MHV amplitude with one fewer leg. Thus~\eqref{googly-rec0} is the only contribution to the BCFW decomposition of a googly MHV amplitude. Continuing recursively we have
\be{googly-MHV-amps}
    \mathbox{
\begin{aligned}
    \cA^n_{n-4}(1,\ldots,n)&= (-1)^n\left(\prod_{i=2}^{n-1}\widetilde \cH_{i-1}^{in} \right)
    \delta^{(4|4)}(\rW_{n-1})\,\delta^{(4|4)}(\rW_n)\
\\   &=   \left(\prod_{i=2}^{n-3}
\widetilde
  -\cH_{i-1}^{in}\right)\cA^4_0(n-3,n-2,n-1,n)  
    \ .
\end{aligned} }
\ee
In this expression, the $\widetilde \cH^{in}_{i-1}$ do not commute and are ordered with increasing $i$ to the right. To perform the last step of the induction we used the specific form of the 3-point $\MHVbar$ amplitude. Thus the googly MHV amplitudes are built up from a product of $\widetilde\cH$ operators acting on two $\delta^{(4|4)}$-functions.

Cyclic symmetry of the googly MHV amplitudes implies many identities in these formul\ae\ that are not
manifest, but which will be useful in the following. In particular, there is an obvious relation from the cyclic symmetry of the 3-point amplitude, while that of the four-point amplitude yields
\be{tHtH-reln}
    \widetilde \cH_1^{24}\widetilde \cH_2^{34}\,\delta^{(4|4)}(\rW_3)\,\delta^{(4|4)}(\rW_4) =
    \widetilde \cH_2^{31}\widetilde \cH_3^{41}\,\delta^{(4|4)}(\rW_4)\,\delta^{(4|4)}(\rW_1) 
\ee
which is the CPT conjugate of the relation~\eqref{4cyclic-Cop} among the $\cH^i_{jk}$ operators. Finally, since the four-point amplitude may be represented either as an MHV or a googly MHV amplitude, we have
\be{H-reln}
    \cH^2_{13} \cH^4_{13}\, \delta^{(4|4)}(\rW_2)\, \delta^{(4|4)}(\rW_4)= 
    \widetilde \cH_1^{24}\widetilde \cH_2^{34}\,\delta^{(4|4)}(\rW_3)\,\delta^{(4|4)}(\rW_4) \ .
\ee
As with the MHV amplitudes, we can use the cyclic identities to ensure that the $\sgn[\del_i\del_j]$ factors cancel pairwise, leaving us with at most one such factor. Thus, following the discussion of section~\ref{sec:confbreaking}, the $\MHVbar$ amplitudes with an odd number of external particles violate conformal invariance -- they do not extend to the conformal compactification of affine spacetime, but rather to its double cover.


\subsection{NMHV amplitudes}
\label{sec:NMHV-recursion}

We now compute the twistor form of the NMHV amplitude $\cA^n_1$. For $n=5$ this is a googly MHV and~\eqref{googly-MHV-amps} gives
\be{NMHV5}
\begin{aligned}
    \cA_1^5(1,2,3,4,5) &= \widetilde \cH_1^{25}\,\widetilde \cH_2^{35}\,\widetilde \cH_3^{45}\,
    \delta^{(4|4)}(\rW_4)\,\delta^{(4|4)}(\rW_5)\\
    &=-\widetilde \cH_1^{25}\cA_{0}^4(2,3,4,5)\ .
\end{aligned}
\ee
In fact, we will be able to write the general $n$-particle NMHV amplitude in terms of this 5-point amplitude and $m$-point MHV amplitudes. To see this, note that the contribution to an $n$-point NMHV amplitude from all but the homogeneous term is
\be{NMHV0}
    \sum_{i=2}^{n-3}\int D^{3|4}\rW\,\cA_{0}^{i+1}(1,\ldots,i,\rW)\,
    \sgn\left(\la1\rW\ra\,\im\rW_1\!\cdot\del_n\,\left[\del_\rW\,\del_n\right]\right)
    \cA_{0}^{n-i+1}(\rW,i+1,\ldots,n)\ .
\ee
Using~\eqref{MHVn} (or~\eqref{MHVn-2}) we can split the MHV sub-amplitudes as\footnote{When $i=2$ or $n-3$ no splitting is necessary.}
\be{split-MHV}
\begin{aligned}
    \cA_{0}^{i+1}(1,\ldots,i,\rW) &= -\cA^i_{0}(1,\ldots,i)\,\cA_{0}(1,i,\rW)\\
    \cA^{n-i+1}_{0}(\rW,{i+1},\ldots,n) &= \cA_{0}^{n-i-1}({i+1},\ldots,{n-1})\,
    \cA_{0}^4({i+1},{n-1},n,\rW)\ .
\end{aligned}
\ee
The first terms on the rhs of these equations are independent of both $\rW$ and $\rW_n$, so may be
brought outside both the $D^{3|4}\rW$ integral and the non-local sign
operators in~\eqref{NMHV0}. Thus we only need consider the expression 
\be{5-pt1}
    \int D^{3|4}\rW\,\cA_0(1,i,\rW)
    \times\ \sgn\left(\la1\rW\ra\,\im\rW_1\!\cdot\del_n\,[\del_\rW\,\del_n]\right)\cA_0^4(i+1,n-1,n,\rW)\ .
\ee
But this is conjugate homogeneous and is just the five-point NMHV super-amplitude $\cA_1^5(1,i,i+1,n-1,n)$! Therefore, the sum of contributions~\eqref{NMHV0} reduces to\footnote{When $i=2$ or $i=n-3$, the `two-point' MHV amplitudes in this sum should be replaced by unity.}
\begin{equation}
    \sum_{i=2}^{n-3}\cA_0^i(1,\ldots,i)\, \cA_{1}^5(1,i,{i+1},{n-1},n)\,\cA_0^{n-i-1}({i+1},\ldots,{n-1})\ ,
\end{equation}
while the homogeneous term is $\cA_1^{n-1}(1,\ldots,{n-1})\,\cA_0^3(1,n,{n-1})$.

Working by induction, one can show that this recursive formula is
solved by the double sum 
\be{NMHV-total}
    \mathbox{
    \begin{aligned}
    \cA_1(1,\ldots,n)&= \sum_{j=5}^n\sum_{i=2}^{j-3} \cA_1^5(1,i,i+1,j-1,j)\\
    &\hspace{2cm}\times\ \cA_0(1,\ldots,i) \cA_{0}(i+1,\ldots,j-1)
    \cA_0(j,\ldots,n,1) 
    \end{aligned}}
\ee
where all `two-point $\cA_0$ amplitudes' should be replaced by unity.
For example, the twistor form of the 6-particle NMHV amplitude equals
\be{6-pt-NMHV}
    \cA_1(1,2,3,4,5)\cA_0(1,5,6)+\cA_1(1,2,3,5,6)\cA_0(3,4,5) + \cA_{1}(1,3,4,5,6)\cA_0(1,2,3)\ .
\ee
Notice that the $n$-particle MHV amplitude may be decomposed as
\be{sdfghjm}
    \cA_0(1,i,i+1,j-1,j)\ \times\ \cA_0(1,\ldots,i)\cA_0(i+1,\ldots,j-1)\cA_0(j,\ldots,n,1)
\ee
whenever $i$ and $j$ lie in the ranges permitted by the double sum
in~\eqref{NMHV-total}. Thus, to obtain an NMHV amplitude from the MHV
amplitude, one chooses two points $(\rW_i,\rW_j)$ with $i-j>2$, and
replaces the five-point MHV amplitude  
\begin{equation}
\begin{aligned}
     \cA_0(1,i,i+1,j-1,j) &= -\cA_0(1,i,j)\,\cA_0(i,i+1,j-1,j)
\\        &=-\cH^1_{ij}\delta^{(4|4)}(\rW_1)\,\cA_0(i,i+1,j-1,j)
\end{aligned}
\end{equation}
by the five-point NMHV amplitude
\begin{equation}
     \cA_1(1,i,i+1,j-1,j) = -\widetilde\cH^{ij}_1\,\cA_0(i,i+1,j-1,j)\ .
\end{equation}
The twistor $\rW_1$ is distinguished here purely through its role in
the BCFW recursion relation. Put simply, one organises the amplitude
so that $\cA_0(1,i,j)$ is explicit, and replaces it with an $\widetilde\cH_1^{ij}$.  This replacement is perhaps most analogous to multiplication of an MHV amplitude by the first of the dual superconformal invariants $R$ (or ${\mathcal P}$) in the work of Drummond {\it et al.}~\cite{Drummond:2008vq,Drummond:2008bq,Drummond:2008cr}. Although these invariants are local on momentum space, they becomes non-local on twistor space via the
half Fourier transform.  (We discuss the analogues of the higher $R$ invariants of~\cite{Drummond:2008cr} later.)

\begin{figure}[t]
\begin{center}
\includegraphics[width=70mm]{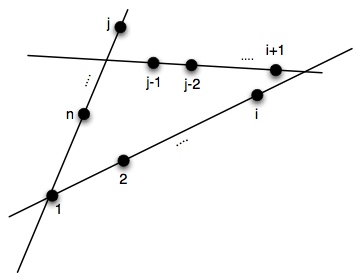}
\caption{The NMHV amplitude is supported on three coplanar lines in twistor space. Point 1 (distinguished by its role in the BCFW recursion relations) is located at the intersection of two of these lines.}
\label{fig:NMHV}
\end{center}
\end{figure}

Geometrically, the twistor support of the NMHV amplitude can be 
understood, provided we ignore the effect of the nonlocal operator
$\sgn([\del_i\,\del_j])$ in $\widetilde\cH^{ij}_1$. (This operator is $\pm1$ on momentum space, so goes unnoticed if one tries to determine the localisation properties by acting on the momentum space amplitudes with certain differential  operators as in~\cite{Witten:2003nn}.) Replacing this operator by 1, we find 
\begin{equation}
\begin{aligned}
&\cA_1(1,i,i+1,j-1,j) \rightarrow 
(\im\pi)^2\sgn\left(\im\rW_1\!\cdot\!\del_i\ \im\rW_1\!\cdot\!\del_j\right)
     \cA_0(i,i+1,j-1)\cA_0(i,j-1,j)\\
&\ \ =\int\frac{\rd s}{s}\frac{\rd
t}{t}\,\cA_0(\rW_i-s\rW_1,\rW_{i+1},\rW_{j-1})\,
\cA_0(\rW_i-s\rW_1,\rW_{j-1},\rW_j-t\rW_1)   
\end{aligned}
\end{equation}
This is a product of coplanar delta functions that altogether impose
the condition that $\rW_1$, $\rW_i$, $\rW_{i+1}$,  
$\rW_{j-1}$ and $\rW_j$ are all coplanar. The remaining MHV amplitudes
in~\eqref{NMHV-total} then require that the MHV amplitudes
$\cA_0(1,\ldots,i)$, $\cA_0(i+1,\ldots,j-1)$ and $\cA_0(j,\ldots,n,1)$
lie in this plane. Thus, ignoring the non-local $\sgn[\del_i\,\del_j]$
operator, the overall picture of an NMHV amplitude is shown in
figure~\ref{fig:NMHV}.  This is initially at variance with the picture
from twistor-string theory, in which the NMHV contribution should be
based on a degree-two curve in the connected prescription, or a pair
of skew lines in the disconnected prescription, rather than the three
coplanar lines of figure~\ref{fig:NMHV}. However, the five-point NMHV
amplitude can be presented in various forms; in particular it may be
represented as a sum of contributions that are supported on two
intersecting lines by half Fourier transforming the MHV
formalism. This then decomposes the 
above formula into a rather larger sum of terms supported on two lines. 
What is perhaps more interesting is that this configuration is
precisely that discovered in~\cite{Bern:2004bt} for the twistor space support of the 1-loop NMHV box coefficients, 
showing that the BCFW decomposition of a tree amplitude encodes its origin in terms of the soft limit of a one loop
amplitude.


\subsection{Conjugate NMHV amplitudes}
\label{sec:conjnmhv}

The recursion for the $\cA^n_{n-5}$ amplitudes is essentially
conjugate to that of the NMHV amplitudes. The conjugate homogeneous
term involves a 3-point MHV sub-amplitude and a conjugate NMHV
sub-amplitude with one fewer points; its contribution is
$$
	-\widetilde\cH^{2\,n}_1\cA^{n-1}_{n-6}(2,\ldots , n)\ .
$$
In the remaining terms, each sub-amplitude is googly MHV. These terms are
given by 
\be{inhom-NMHVbar}
     \int \D^{3|4}\rW \left(\prod_{j=2}^{i-2}\widetilde\cH_{j+1}^{ji}\right)
     \cA_0(1,2,i)\cA_0(1,\rW,i)\, 
     \sgn\!\left(\la 1\rW\ra\, \rW_1\!\cdot\!\del_n\, [\del_\rW\del_n]\right)
     \cA^{n-i}_{n-i-4} (\rW, i+1,\ldots,n)\, ,
\ee
summed over $i$ for $i=2,\ldots ,n-2$ and with the $\widetilde
\cH$-operators ordered with increasing $j$ to the left. (We have used the cyclic
property of the $\MHVbar$ amplitude to   
ensure that 1 and W appear together in a 3-point MHV amplitude and
have furthermore used a reverse ordered form of
\eqref{googly-MHV-amps} to represent the left hand google-MHV amplitude.) To
perform the integral, use the delta-function in $\cA_0(i,1,\rW)$ (as
in the conjugate homogeneous term). We obtain 
\begin{multline}
    -\left( \prod_{j=2}^{i-2}\widetilde\cH_{j+1}^{ji}\right)  \cA_0(1,2,i)\widetilde\cH_1^{i\,n}\cA^{n-i}_{n-i-4} (i,i+1,\ldots,n)\\
     =\ \left(\prod_{j=2}^{i-2}\widetilde\cH_{j+1}^{ji}\right)  \cA_0(1,2,i)
     \cA^{n-i+1}_{n-i-3} (1,i, \ldots,n)\ , \label{inhom-NMHVbar1}
\end{multline}
where we have used \eqref{googly-MHV-amps} to identify $-\widetilde\cH_1^{i\,n}\cA^{n-i}_{n-i-4} (i,i+1,\ldots,n)$ as
an $\MHVbar$ amplitude.  

Adding up the homogeneous and the inhomogeneous terms, one finds
\be{NMHVbar}
\cA^n_{n-5}=\sum_{2\leq i < j\leq n-2}\left(\prod_{l=2}^{i-1} -\widetilde
\cH_{l-1}^{l n}\right)\left(\prod_{m=i}^{j-2}\widetilde\cH_{m+1}^{mj}\right)
  \cA_0(i-1,i,j)  \left(\prod_{r=j}^{n-2}
  \widetilde\cH^{r,n-1}_{r+1}\right) 
  \cA_0(i,j,n-1,n) \ ,
\ee
where the $\widetilde\cH$-operators are ordered with increasing $m,r$ indices to the left, but $l$ increasing to the right and the product should be taken to be 1 if the top limit is smaller than the lower one.  Crudely, this formula differs from the conjugate MHV amplitude by replacing an $\widetilde\cH$-operator by a three-point $\cA_0$ amplitude.


\subsection{Further tree examples}
\label{sec:further-trees}

Rather than give more closed-form formulae, we just describe the
strategy for integrating the recursion relation and outline the
structure of the terms that arise. The homogeneous or conjugate
homogeneous term of any amplitude may be regarded as understood, at
least inductively, via the discussion of
section~\ref{sec:homog-cojhomog}.  

In a generic N$^2$MHV amplitude, the remaining inhomogeneous terms
involve one NMHV and one MHV sub-amplitude. If the MHV sub-amplitude is
on the left, we can perform the integral over the internal twistor
using the same strategy as for the conjugate homogeneous term. This
yields a product of an MHV sub-amplitude and an $\widetilde\cH$
operator acting on the NMHV sub-amplitude on the right. Conversely,
suppose that the MHV sub-amplitude is on the right inside the integral,
so that the NMHV sub-amplitude $\cA_1^r(1,\ldots,r-1,\rW)$ is on the
left. The form~\eqref{NMHV-total} of the NMHV amplitude shows that
either $\rW$ appears only in an MHV sub-amplitude -- in which case we
can again proceed as in the conjugate homogeneous term -- or else must
appear in a five-point NMHV amplitude $\cA_1^5(1,i,i+1,r-1,\rW)$ for
some $i$. Even in this case, the conjugate homogeneous strategy can be
used: The cyclic symmetry of the 5-point NMHV amplitude ensures that
it can always be written so that W is not acted on by $\widetilde\cH$,
{\it e.g.} 
\begin{equation}
     \cA_1(1,i,i+1,r-1,\rW) = \widetilde\cH^{i+1,r-1}_i\cA_0(1,i+1,r-1,\rW).
\end{equation}
Since $\widetilde\cH$ is independent of W it may be brought outside to
act on the result of the remaining integral, and this integral is of
the same type as contributed to the NMHV amplitudes. In all cases we
obtain two $\widetilde \cH$-operators acting on a sequence of $\cA_0$
factors, again demonstrating the similarity between the role of the
$\widetilde\cH$ operators and the dual superconformal invariants $R$
of~\cite{Drummond:2008vq,Drummond:2008bq,Drummond:2008cr}. 

The 7 point N$^2$MHV is the first non-trivial example of an $\overline{\rm NMHV}$ amplitude, and here~\eqref{NMHVbar} 
runs over six terms. The next case is the 8 point N$^2$MHV with 20 terms:
\be{N2MHV-8}
\begin{aligned}
	&\widetilde\cH_1^{28}\left(\cA_0(2678)\widetilde\cH_2^{36}\cA_0(3456)
	+\cA_0(278)\widetilde\cH_2^{37}\cA_0(34567)\phantom{\int}\right.\\
	&\hspace{2.5cm}+\widetilde\cH_2^{38}\cA_0(345678) +\cA_0(234)\cA_0(278)\widetilde\cH_2^{47}\cA_0(4567)\\  
	&\hspace{2.5cm}+\cA_0(234)\widetilde\cH_2^{48}\cA_0(45678)
	+ \cA_0(2345)\widetilde\cH_2^{58}\cA_0(5678)\left.\phantom{\int}\hspace{-0.5cm}\right)\\
	+&\ \cA_0(178)\left(\widetilde\cH_1^{27}\cA_0(267)\widetilde\cH_2^{36}\cA_0(3456)
	+\widetilde\cH_1^{27}\cA_0(456)\widetilde\cH_2^{37}\cA_0(3467)\phantom{\int}\right.\\
	&\hspace{2.5cm}+\widetilde\cH_1^{27}\cA_0(234)\widetilde\cH_2^{47}\cA_0(4567)
	+\cA_0(167)\widetilde\cH_1^{26}\widetilde\cH_2^{36}\cA_0(3456)\\
	&\hspace{2.5cm}+\cA_0(123)\widetilde\cH_1^{37}\widetilde\cH_3^{47}\cA_0(4567)
	+\widetilde\cH_3^{24}\cA_0(124)\widetilde\cH_1^{47}\cA_0(4567)\left.\phantom{\int}\hspace{-0.5cm}\right)\\
	+&\ \cA_0(123)\widetilde\cH_1^{38}\cA_0(378)\widetilde\cH_3^{47}\cA_0(4567)
	+\cA_0(123)\widetilde\cH_1^{38}\cA_0(567)\widetilde\cH_3^{48}\cA_0(4578)\\
	+&\ \cA_0(123)\widetilde\cH_1^{38}\cA_0(345)\widetilde\cH_2^{58}\cA_0(5678)
	+\widetilde\cH_1^{25}\cA_0(2345)\widetilde\cH_1^{58}\cA_0(5678)\\
	+&\ \cA_0(345)\widetilde\cH_3^{25}\cA_0(125)\widetilde\cH_1^{58}\cA_0(5678)
	+\cA_0(123)\widetilde\cH_4^{35}\cA_0(135)\widetilde\cH_1^{58}\cA_0(5678)\\
	+&\ \cA_0(1234)\widetilde\cH_1^{48}\widetilde\cH_4^{58}\cA_0(5678)
	+\widetilde\cH_2^{13}\cA_0(134)\widetilde\cH_1^{48}\cA_0(45678)\, .
\end{aligned}
\ee
In this equation all operators act on everything to their right within
a term.  

For  $\overline{{\rm N}^2{\rm MHV}}$ amplitudes, the inhomogeneous terms involve
sub-amplitudes that are either $\MHVbar$ on the left and $\overline{\rm NMHV}$
on the right, or the other way around.  The construction of the $\overline{\rm NMHV}$ amplitudes already showed how to perform the integration when the $\MHVbar$ is on the left, {\it i.e.} via the conjugate homogeneous strategy.  When the $\MHVbar$ sub-amplitude $\cA_{r-4}^r(\rW,n-r+2,\ldots ,n)$ is on the right,
use~~\eqref{googly-MHV-amps} to express it as a sequence of $(r-2)$
$\widetilde\cH$ operators acting on $\cA_{-1}(n-1,\rW,n)$.  Since
these operators do not depend on W or $\rW_n$, and can be chosen so as
not to act on 
$\rW$,  they can be taken out of the integral. The remaining integral
is then straightforward to perform using the standard, homogeneous
strategy. 


\subsection{General tree amplitudes}
\label{sec:generaltrees}

In the examples above, it was convenient to express an $n$-point
N$^k$MHV amplitude in terms of $k$ $\widetilde \cH$-operators acting
on $(n-2-k)$ three-point  MHV sub-amplitudes in some order and it seems
likely that it will be possible to 
express all tree amplitudes in this way.  It is natural to ask for the
correspondence with the dual superconformal invariants that arise in the scheme of Drummond {\it et
al.}~\cite{Drummond:2008vq,Drummond:2008bq,Drummond:2008cr}.
While we have no definitive answer, it seems likely that multiplication by the simplest dual superconformal invariant $R_{r;st}$ corresponds to a single insertion of an $\widetilde\cH^{ij}_k$-operator.  More complicated $R$-invariants arise when more than one $\widetilde\cH$-operator is inserted ({\it e.g.} in N$^2$MHV amplitudes) and we expect that their additional indices encode the ordering and nesting of the $\widetilde\cH$-operators.  

Each of the $k$ $\widetilde \cH$-operators 
contains two Hilbert transforms and each of the $(n-2-k)$ three-point
MHV amplitudes is   an $\cH$ operator (two Hilbert
transforms) acting on a $\delta^{(4|4)}$-function. This gives a total
of $2(n-2)$ Hilbert transforms acting on $(n-2-k)$
$\delta^{(4|4)}$-functions, decorated by an unspecified number of
$\sgn(\la \rW_i\,\rW_j\ra)$ factors or  $\sgn ([\del_i\, \del_j])$
operators.  These factors are constructed using the infinity twistor
and, as with the three-point seed amplitudes, their presence indicates
a violation of conformal invariance. We know that tree amplitudes in
momentum space for $\cN=4$ SYM are annihilated by the conformal
algebra (at least away 
from their singularities), so this violation must again be rather
subtle and indeed on momentum space these become locally constant
functions and so their derivatives with respect to the conformal
algebra vanish away from the singularities. A further subtlety is that
there is ample opportunity for cancellation of
the conformal-breaking signs, so the question as to whether their
presence is essential in a given amplitude is not always clear.

In general, the existence of at least two $\delta^{(4|4)}$-functions
in each amplitude means that there will always be more than one way to
perform the $\rW$-integration in the recursion, and the two strategies
that we relied upon in our examples are no doubt not exhaustive. The
arguments so far show that the internal state can in practice be
integrated out to leave expressions in terms of $\tilde\cH$ operators
and three point MHV amplitudes, at least up to the N$^3$MHV and
$\overline{{\rm N}^3{\rm MHV}}$ amplitudes ({\it i.e.}, for any amplitude with up to 10
external states).  As in momentum space, our formalism obscures the
underlying cyclic symmetry of the amplitudes, though this is encoded in the
algebraic relations we have written down. Obtaining and making
sense of the formulae for arbitrary tree amplitudes is likely to require a better command of these
symmetries than we currently possess.


\subsection{Some elementary loop amplitudes}
\label{sec:loops}

The structure of loop amplitudes in twistor space has already been much discussed in~\cite{Cachazo:2004zb,Cachazo:2004by,Bena:2004xu,Britto:2004tx}).  As with the twistor structure of tree amplitudes, these articles typically only identified the support of the amplitudes, rather than giving explicit twistor formulae. There is no problem in principle in obtaining twistor loop amplitudes by taking the half Fourier transform of the known momentum space expressions. More interesting would be to translate the generalised unitarity methods -- currently the definitive way of  constructing supersymmetric loop amplitudes.   In this subsection, we content ourselves with calculating the half Fourier transform explicitly for the simplest case of the 4-particle, 1 loop amplitude.   Via the BDS ans{\" a}tz~\cite{Bern:2005iz} (verified up to three loops and consistent with both the dual conformal anomaly equation of Drummond {\it et al.}~\cite{Drummond:2007cf} and the strong-coupling limit computed by Alday \& Maldacena in~\cite{Alday:2007hr}), this one loop amplitude forms the basis of the 4-particle amplitude to all orders. At the end of the subsection, we make some simple comments about some non-supersymmetric 1 loop amplitudes, in particular calculating the non-supersymmetric all-plus 1 loop amplitude.

In momentum space, the 4-particle, 1 loop amplitude is a multiple of the tree amplitude given by~\cite{Green:1982sw,Bern:1994zx} 
\be{1loop-mom}
     A^{1\,{\rm loop}}(1,2,3,4) = \left[-\frac{1}{\epsilon^2}\left(\frac{\mu_{\rm IR}^2}{-s}\right)^\epsilon
     +\frac{1}{\epsilon^2}\left(\frac{\mu_{\rm IR}^2}{-t}\right)^\epsilon
     + \frac{1}{2}\ln^2\frac{s}{t}\right]A^{\rm tree}(1,2,3,4) + {\mathcal O}(\epsilon)
\ee
at some renormalisation scale $\mu_{\rm IR}$, and where $s=(p_1+p_2)^2$ and
$t=(p_2+p_3)^2$.  This naturally divides into the infrared divergent part (the first two terms in the square
bracket) and a finite part (the $3^{\rm rd}$ term in the square bracket).  On the support of the 4-particle momentum $\delta$-function, $s/t$ can be expressed entirely in terms of the unprimed spinors:
\be{oiuhpo}
    \frac{s}{t} = \frac{\la12\ra\la34\ra}{\la14\ra\la23\ra}\ .
\ee
Since these unprimed spinors do not participate in the half Fourier transform, the finite part of the twistor amplitude is simply  
\be{1loop-twist-finite-explicit}
 \cA^{1\,{\rm loop}}_{\rm finite}(1,2,3,4) =  \frac{1}{2}\ln^2\,\left(\frac{\la12\ra\la34\ra}{\la14\ra\la23\ra}\right)\,
 \cA^{\rm tree}(1,2,3,4)\ .
\ee
To study the superconformal properties of this amplitude, recall that the tree level amplitude can be written as
\be{tree-remind}
    \cA^{\rm tree}(1,2,3,4) = \int\frac{\rd a}{a}\frac{\rd b}{b}\frac{\rd c}{c}\frac{\rd g}{g} \,
    \delta^{(4|4)}(\rW_2-a\rW_3-b\rW_1)\,\delta^{(4|4)}(\rW_4-c\rW_1-g\rW_3)\, .
\ee
On the support of the delta functions in~\eqref{tree-remind} one has
\begin{equation}
    \frac{\la12\ra\la34\ra}{\la14\ra\la23\ra} = -\frac{ac}{bg}
\end{equation}
by using the $\lambda_A$ components of each of the delta functions to
perform the integrals. Hence 
\begin{multline}
\label{1loop-twist-finite}
    \cA^{1\,{\rm loop}}_{\rm finite}(1,2,3,4) = 
    \int\frac{\rd a}{a}\frac{\rd b}{b}\frac{\rd c}{c}\frac{\rd g}{g} \,\frac{1}{2}\ln^2\left(-\frac{ac}{bg}\right)\\
    \times\ \delta^{(4|4)}(\rW_2-a\rW_3-b\rW_1)\,\delta^{(4|4)}(\rW_4-c\rW_1-g\rW_3)
\end{multline}
which makes manifest the superconformal invariance of the IR-finite part of the 1 loop amplitude.

The transform of the IR divergent part is more delicate. A distinction
between our split signature context, and Lorentz signature is that $s$
and $t$ can change sign and there is some ambiguity as to how the
functions $(-s)^\epsilon$ and $(-t)^\epsilon$ should be continued
across the zero set of $s$ and $t$.  Once such a choice is made, the
integral for the half Fourier transform can be reduced to a standard
known one as follows: Consider the half Fourier transform of $(-s)^{-epsilon}A^{\rm tree}$ (the second
term is identical).  We can formally replace $(-s)^{-\epsilon}$ by the pseudo-differential
operator  $(\la12\ra[\del_1\,\del_2])^{-\epsilon}$, which then acts
on the tree amplitude
\be{1loop-twistor-diverge}
   \la12\ra^{-\epsilon}[\del_1\,\del_2]^{-\epsilon} 
    \cA^{{\rm tree}}(1,2,3,4)\ .
\ee
From \eqref{MHVn-twis}, the tree amplitude $\cA^{\rm tree}(1,2,3,4)$
can be expressed in the Poincar\'e invariant form 
\be{4-tree-poinc}
     \cA^{\rm tree}=\frac{\la 34\ra^2}{\la 12\ra\la 23\ra\la 34\ra\la 41\ra} 
     \delta^{(2|4)}\left(\mu_1 +\mu_3 \frac{\la 41\ra}{\la 34\ra} + \mu_4\frac{\la 13\ra}{\la 34\ra}\right)
     \delta^{(2|4)}\left(\mu_2 +\mu_3 \frac{\la 42\ra}{\la 34\ra} + \mu_4\frac{\la 23\ra}{\la 34\ra}\right)
\ee
The action of $(\la12\ra[\del_1\,\del_2])^{-\epsilon}$ on these delta
functions can be understood by translation of its action on
$\delta^{(2)}(\mu_1)\delta^{(2)}(\mu_2)$, which in turn can be understood  
from its Fourier transform
\be{Four-IR-div}
     [\del_1\,\del_2]^{-\epsilon}\delta^{(2)}(\mu_1)\delta^{(2)}(\mu_2)
     =\int\rd^2\tilde\lambda_1\, \rd^2\tilde\lambda_2 \ [\tilde\lambda_1\tilde\lambda_2]^{-\epsilon}\,
     \e^{\im[\mu_1\tilde\lambda_1] + \im[\mu_2\tilde\lambda_2]}\ .
\ee
By thinking the pair $q=(\tilde\lambda_1,\tilde\lambda_2)$ as a 4-vector, dual to the 4-vector $y=(\mu_1,\mu_2)$, this can be reduced to a standard form
\be{standard-form-FT}
     \int \rd^4q\ (q\cdot q)^{-\epsilon}\, \e^{\im q\cdot y} \ ,
\ee
but where the quadratic form $q\cdot q$ has signature $++--$.  For
various choices of quadratic form $q\cdot q$, the Fourier transform~\eqref{standard-form-FT} can be found in the tables at the end of~\cite{Gelfand-Shilov}.  For example, if for $(q\cdot q)^{-\epsilon}$ we take $(q\cdot q +i0)^{-\epsilon}$ (the analytic continuation through the upper-half plane) we obtain 
\be{FT-qqe}
     -\frac{2^{4-2\epsilon}\pi^2\Gamma(2-\epsilon)}{\Gamma(\epsilon)}(y\cdot y-i0)^{-2+\epsilon}\ ,
\ee
where in our context $(y\cdot y-i0)^{-2+\epsilon}=([\mu_1\,\mu_2]-i0)^{-2+\epsilon}$ is understood to be analytically continued through the lower half plane.  Putting this together with the translation, we obtain the final form 
\begin{multline}
\label{1loop-twistor-diverge-explicit}
    -\frac{\mu_{\rm IR}^{2\epsilon}}{\epsilon^2}\frac{2^{4-2\epsilon}\pi^2\epsilon\,\Gamma(2-\epsilon)}{\Gamma(1+\epsilon)}\la12\ra^{-\epsilon}
    \left(\!\left(\mu_1+\mu_3\frac{\la41\ra}{\la34\ra}+\mu_4\frac{\la13\ra}{\la34\ra}\right)\!\cdot\!
    \left(\mu_2+\mu_3\frac{\la42\ra}{\la34\ra}+\mu_4\frac{\la23\ra}{\la34\ra}\right)\!\right)^{-2+\epsilon}\\
    \ \times \frac{\la34\ra^2}{\la12\ra\la23\ra\la34\ra\la41\ra}\,
    \delta^{(0|4)}\left(\chi_1+\chi_3\frac{\la41\ra}{\la34\ra}+\chi_4\frac{\la41\ra}{\la34\ra}\right)\,
    \delta^{(0|4)}\left(\chi_2+\chi_3\frac{\la42\ra}{\la34\ra} + \chi_4\frac{\la23\ra}{\la34\ra}\right)\ 
\end{multline}
for IR divergence in the $s$-channel. The $t$-channel divergence follows by cyclic permutation.

\bigskip

We finish with the rather more straightforward example the all + helicity amplitude.  This amplitude vanishes in the supersymmetric theory, but in the non-supersymmetric case it is non-zero at one loop, and given by the rational expression
\be{all-plus}
     A^{1\, {\rm loop}}_+(1,\ldots,n)=\frac{-\im}{48\pi^2}\delta^{(4)}\left( \sum p_i\right)
     \sum_{1\leq i_1<i_2<i_3<i_4\leq n} \frac{\la i_1\, i_2\ra\la i_3\, i_4\ra [i_2\, i_3][i_4\, i_1]}
     {\la 12\ra\la 23\ra\ldots \la n1\ra}\, .
\ee
This is easily transformed to give 
\be{all-plus-twis}
     \cA^{1\,{\rm loop}}_+(1,\ldots,n)=\frac{-\im}{48\pi^2}\sum_{1\leq i_1<i_2<i_3<i_4\leq n} 
     \la i_1\, i_2\ra\la i_3\, i_4\ra [\del_{i_2}\, \del_{i_3}][\del_{i_4}\, \del_{i_1}]  \cA_{\rm MHV}^{\cN=0}(1,2,\ldots, n)
\ee
where by $\cA_{\rm MHV}^{\cN=0}(1,2,\ldots, n)$ we mean the formula obtained from~\eqref{MHVn} by replacing  $\delta^{(2|4)}$ or $\tilde\delta^{(2|4)}$  by $\delta^{(2|0)}$ or  $\tilde\delta^{(2|0)}$, respectively. The formula therefore has derivative of delta function support along a line as predicted in~\cite{Cachazo:2004zb}.


\section{Twistor Supergravity Amplitudes}
\label{sec:sg}

In this section, we sketch how the recursion rule works for $\cN=8$ supergravity as far as the homogeneous term and its conjugate, and find formulae for the MHV and $\MHVbar$ amplitudes. The BCFW relation itself is the same:
\begin{multline}
\label{BCFWgrav}
    \cM(1,\ldots,n) = \sum\int D^{3|8}\rW\,\cM_L(1,\ldots,\rW)\\
    \ \times\ \sgn\left(\la\rW_1\rW\ra\,\rW_1\!\cdot\!\frac{\del}{\del\rW_n}\,
    \left[\frac{\del}{\del\rW}\frac{\del}{\del\rW_n}\right]\right)\cM_R(\rW,\ldots,\rW_n)\ ,
\end{multline}
except that the sum now runs over all ways of partitioning the external legs over the two sub-amplitudes, with no cyclicity requirement.

To compute the three-point seed amplitudes we start with the momentum space formulae
\be{grav3}
\begin{aligned}
    M_{\rm MHV}(1,2,3)
   &=\frac{\delta^{(4|16)}\left(\sum_{i=1}^3|i\ra\llbracket1\|\right)}{\la12\ra^2\la23\ra^2\la31\ra^2}\\
   M_{\MHVbar}(1,2,3) &=
    \frac{\delta^{(4)}(p_1+p_2+p_3)\,\delta^{(8)}(\eta_1[23]+\eta_2[31]+\eta_3[12])}{[12]^2[23]^2[31]^2}
\end{aligned}
\ee
which are simply the squares of the Yang-Mills three-point amplitudes~\cite{Kawai:1985xq}, provided one strips away the momentum conserving delta functions. We will see a somewhat different structure in twistor space, though the gravitational and Yang-Mills seed amplitudes are still closely related.

Taking the half Fourier transform, the (dual) twistor form of the 3-point MHV amplitude is
\be{MHV3grav1}
    \cM_{\rm MHV}(\rW_1,\rW_2,\rW_3) = 
    \frac{\delta^{(2|8)}\!\left(\mu_1\la23\ra+\mu_2\la31\ra+\mu_3\la12\ra\right)}{\la12\ra^2\la23\ra^2\la31\ra^2}\ .
\ee
This amplitude has homogeneity $+2$ in each of its arguments as
required for on-shell $\cN=8$ supermultiplets.  We can
write~\eqref{MHV3grav1} as 
\be{MHV3-grav-col1} 
    \mathbox{\begin{aligned}
            \cM_{\rm MHV}(\rW_1,\rW_2,\rW_3) 
            &= \left|\la\rW_2\rW_3\ra\,\im\rW_2\!\cdot\!\frac{\del}{\del\rW_1}\,
            \im\rW_3\!\cdot\!\frac{\del}{\del\rW_1}\right|\
          \delta^{(4|8)}(\rW_1)\\
            &= \left(\la\rW_2\rW_3\ra\,\im\rW_2\!\cdot\!\frac{\del}{\del\rW_1}\,
            \im\rW_3\!\cdot\!\frac{\del}{\del\rW_1}\right)\cH^1_{23}\
          \delta^{(4|8)}(\rW_1)
            \end{aligned}}
\ee 
in close analogy to the form~\eqref{MHV3formal} of the SYM amplitude
in terms of the $\cH$-operator; in this formula the sgn-factor in
$\cH$ turns the ordinary differential operator into its formal
modulus. 
Equivalently, in terms of the $\cN=8$
collinear delta function of equation~\eqref{collinear-tilde-delta},
this is 
\begin{equation}
    \cM_{\rm MHV}(\rW_1,\rW_2,\rW_3)
    =|\la\rW_2\rW_3\ra|\,\tilde\delta^{(2|8)}(\rW_1;\rW_2,\rW_3)
\end{equation}
The explicit factor of $|\la\rW_2\rW_3\ra|$ here breaks $\SL(4|8,\R)$ invariance, but $\cN=8$ super-Poincar{\' e} invariance is preserved. Although~\eqref{MHV3-grav-col1} appears to single out state 1, it is clear from~\eqref{MHV3grav1} that the amplitude is really symmetric under exchange of any two states. 

The 3-point $\overline{\rm MHV}$ supergravity amplitude in twistor space is 
\be{MHV-bar-grav}
    \mathbox{
            \begin{aligned}
            \cM_{\overline{\rm MHV}}(\rW_1,\rW_2,\rW_3) &=
            \left|\left[\frac{\del}{\del\rW_2}\frac{\del}{\del\rW_3}\right]\,\im\rW_1\!\cdot\!\frac{\del}{\del\rW_2}\,
            \im\rW_1\!\cdot\!\frac{\del}{\del\rW_3}\right|\delta^{(4|8)}_{1}(\rW_2)\,\delta^{(4|8)}_{1}(\rW_3)
            \\  
& = \left(\left[\frac{\del}{\del\rW_2}\frac{\del}{\del\rW_3}\right]\,\im\rW_1\!\cdot\!\frac{\del}{\del\rW_2}\,
            \im\rW_1\!\cdot\!\frac{\del}{\del\rW_3}\right)
\widetilde \cH^{23}_1 \delta^{(4|8)}(\rW_2)\,\delta^{(4|8)}(\rW_3)
            \end{aligned}}
\ee
in close analogy to~\eqref{MHV3-bar-twist} for SYM (again we have used
$\tilde\cH$ to formally turn a differential operator into its
modulus). Once again, this 
may be written in terms of weighted projective delta functions as 
\be{MHV-bar-grav2}
    \cM_{\MHVbar}(\rW_1,\rW_2,\rW_3)
    =\left|\left[\frac{\del}{\del\rW_2}\frac{\del}{\del\rW_3}\right]\right|\,\tilde\delta^{(3|8)}(\rW_2,\rW_1)\,
    \tilde\delta^{(3|8)}(\rW_3,\rW_1)\ .
\ee


\subsection{The homogeneous term and the  MHV amplitudes}
\label{sec:gravhomog}

The homogeneous term for an N$^k$MHV amplitude again takes the form
\be{gravhomog} \sum_{r\neq1,n}\int
    D^{3|8}\rW\,\cM_k(1,\ldots,\rW)\,\sgn\left(\la1\rW\ra\,\rW_1\!\cdot\!\del_{\rW}\,[\del_\rW\del_n]\right)       \cM_{-1}(\rW,r,n)
\ee
in which the sum over partitions has reduced to a sum over the external state $r$ attached to the three-point $\MHVbar$ sub-amplitude on the right (and hence absent from $\cM_k(1,\ldots,\rW)$). Recalling that 
\be{asdfghk}
\begin{aligned}
    \cM_{-1}(\rW,r,n)
    &=\left|\left[\del_\rW\del_n\right]\,\rW_r\!\cdot\!\del_\rW\,\rW_r\!\cdot\!\del_n\right|
    \delta^{(4|8)}(\rW)\,\delta^{(4|8)}(\rW_n)\ ,
\end{aligned}
\ee
the factor $\sgn [\del_\rW\del_n]$ in the recursion relation combines
with $|[\del_\rW\del_n]|$ in the amplitude to form the standard
differential operator $[\del_\rW\del_n]$. We then integrate by parts
so that $\del_\rW$ acts on $\cM_k(1,\ldots,\rW)$. Since 
\be{zxcvbnm}
    \left|\rW_r\!\cdot\!\frac{\del}{\del\rW}\right|\delta^{(4|8)}(\rW) = \int\frac{\rd t}{t^2}\delta^{(4|8)}(\rW-t\rW_r)
    =\tilde\delta^{(3|8)}(\rW,\rW_r)
\ee
the integral is straightforward and leaves us with
\begin{multline}
\label{gravhomogcontrib}
    -\sum_{r\neq1,n}\frac{\del}{\del\mu^{A'}_{r}}\left(\cM_k(1,\ldots,r)\right)
    \frac{\del}{\del\mu_{nA'}}\left\{\sgn\left(\la1r\ra\,\rW_1\!\cdot\!\frac{\del}{\del\rW_n}\right)
    \left|\rW_r\!\cdot\!\frac{\del}{\del\rW_n}\right|\delta^{(4|8)}(\rW_n)\right\}\\
    \begin{aligned}
    &=-\sum_{r\neq1,n}\frac{\del}{\del\mu^{A'}_r}\left(\cM_k(1,\ldots,r)\right)
    \frac{\del}{\del\mu_{nA'}}\left\{\rW_r\!\cdot\!\frac{\del}{\del\rW_n}\,
    \cH^n_{1r}\,\delta^{(4|8)}(\rW_n)\right\}
    \end{aligned}
\end{multline}
In the second line, we have written
$|\rW_1\!\cdot\!\del_n|=\rW_1\!\cdot\!\del_n\,\sgn(\rW_1\!\cdot\!\del_n)$
to pull out the operator $\cH^n_{r1}$.  
If we introduce the operator
\be{G-def}
    \cG^1_{23}:=|\rW_2\!\cdot \!\del_1|\,\sgn\left(\la23\ra\,\rW_3\!\cdot\! \del_1\right)
    = \rW_2\!\cdot \!\del_1 \, \cH^1_{32}
    \ ,
\ee
then the homogeneous term can be written more compactly as
\be{grav-hom-term}
    -\sum_{r\neq1,n}\cG^n_{r1}\left[{\del_r}\, {\del_n}\right] \,\delta^{(4|8)}(\rW_n) 
    \cM_k(1,\ldots,r)\ . 
\ee

For MHV amplitudes, the homogeneous term is the complete recursion, and iteration of~\eqref{grav-hom-term} gives
\be{MHVn-grav}
    \cM_{\rm MHV}(1,\ldots,n)= \sum_{{\mathcal P}(2,\ldots,n-1)}
    \left(\prod_{r=4}^n\cG^r_{r-1,1} [\del_r,\del_{r-1}]\delta^{(4|8)}(\rW_r)\right)\cM_{\rm MHV}(1,2,3) 
\ee
where $\cP(2,\ldots,n-1)$ denotes the permutations of the labels $2$
to $n$ and, because the terms in the product do not commute, they are
ordered to the left in increasing $r$.


\subsection{The conjugate homogeneous term and
 $\overline{\mbox{MHV}}$   amplitudes}
\label{sec:gravconjhomog}

The conjugate homogeneous term in the decomposition of an N$^{k+1}$MHV
amplitude is  
\be{gravconjhomog}
    \sum_{r\neq1,n}\int
    D^{3|8}\rW\,\cM_0(1,r,\rW)\,\sgn(\la1\rW\ra\,\rW_1\!\cdot\!\del_n\,
    [\del_\rW\del_n])    
    \cM_k(\rW,\ldots,n)
\ee
where
\begin{equation}
\begin{aligned}
    \cM_0(1,r,\rW) &=
\left|\la1\rW\ra\,\rW\!\cdot\!\frac{\del}{\del\rW_r}\,\rW_1\!\cdot\!\frac{\del}{\del\rW_r}\right| 
    \delta^{(4|8)}(\rW_r)\\
    &=\left|\la1\rW\ra\,\rW_1\!\cdot\!\frac{\del}{\del\rW_r}\right|\tilde\delta^{(3|8)}(\rW_r,\rW)\ .
\end{aligned}
\end{equation}
This $\delta$-function again allows us to perform the integral,
setting $\rW=\rW_r$. We are left with the contribution 
\be{zxcvbnmf}
    \sum_{r\neq1,n}\la1r\ra\rW_1\!\cdot\!\frac{\del}{\del\rW_r}\,
    \widetilde \cH^{rn}_1\cM_k(r,\ldots,n)\ .
\ee
Again, defining
\be{G-tilde-def}
    \tilde \cG^{12}_3 =|\rW_3\!\cdot\!\del_2
    |\,\sgn\left([\del_1\,\del_2]\,\rW_3\!\cdot\!\del_1\right) 
    =\rW_3\!\cdot\!\del_2\, \tilde \cH^{12}_3
\ee
we can write
\be{gravconjhomogcontrib}
    \sum_{r\neq1,n}\la1r\ra \widetilde\cG^{nr}_1
\cM_k(r,\ldots,n)\ ,
\ee
and this can be understood as inserting the particle 1.
In particular, this is the complete BCFW decomposition for an
$\MHVbar$ amplitude, so we recursively obtain 
\be{gravMHVbar}
   \cM_{n-4}(1,\ldots,n) = \sum_{{\mathcal P}(2,\ldots,n-1)}
   \left(\prod_{r=2}^{n-2} \la r\,r+1\ra \widetilde
     \cG^{nr}_{r+1}\right)\cM_{-1}(1,2,n) \, ,
\ee
where again the ordering is important in the product which is ordered
to the left with increasing $r$.


\section{An Ambidextrous Approach}
\label{sec:ambi}

For the most part in this paper we have focussed on transforming
amplitudes and their recursion relations from momentum space to dual
twistor space.  We could equally have chosen to transform some
particles to twistor space and others to dual twistor space.  {\it A
priori}, there is no clear rule as to which external particles
should be expressed in terms of twistors and which in terms of dual
twistors\footnote{If one wishes to describe $\cN=4$ SYM using only manifest $\cN=3$
supersymmetry (or explicit $\cN=7$ supersymmetry for $\cN=8$ SG),  there are
naturally two multiplets, one starting from the lowest helicity and
one from the highest. In this case it is natural to transform one multiplet to
twistor space and the other on dual twistor space.  This approach fits
in naturally both with twistor diagrams and with the ambitwistor action~\cite{Mason:2005kn} at the expense of either losing the economy of dealing with the whole spectrum in a single supermultiplet, or having a
redundant representation.}, but a choice was recently discovered by Arkani-Hamed, Cachazo, Cheung \& 
Kaplan~\cite{AHCCK} that leads to significant simplifications for the BCFW recursion. In this section we will first discuss the `twistor transform' that moves from twistor space to dual twistor space, and then use this transform to relate their 
formulae directly to ours.


\subsection{Fourier transforms and twistor transforms}
\label{sec:fourier}

Thus far we have taken the half Fourier transform from functions
$\Phi(\lambda,\tilde \lambda)$ on the split signature light
cone in momentum space to functions on real dual twistor space by
Fourier transforming in the $\tilde\lambda_{A'}$ variable to obtain
\be{half-four-dual-twis}
   f(W):=f(\lambda,\mu)=\int \rd^2\tilde\lambda\,\e^{\im[\mu\tilde\lambda]}\,\Phi(\lambda,\tilde\lambda) 
\ee
as in equation~\eqref{half Fourier-inv}. We could just as easily have Fourier transformed in $\lambda_A$ 
to obtain a function on twistor space
\be{half-four-twis}
   F(Z)=F(\omega,\pi)=\int \rd^2\lambda\,\e^{-\im\la\omega\lambda\ra}\,\Phi(\lambda,\pi)
\ee 
with co-ordinates $Z^\alpha=(\omega^A,\pi_{A'})$ on twistor space (after relabelling $\tilde\lambda_{A'}=\pi_{A'}$).  
Combining these two half Fourier transforms, one obtains a map from functions on twistor space to functions on dual twistor space, given by
\be{four-twis-trans1}
   f(W)=\frac{1}{(2\pi)^2}\int \rd^4Z\, \e^{-\im Z\cdot W} F(Z)\ ,
\ee
where $Z\!\cdot\! W= \omega^A\lambda_A + \pi_{A'}\mu^{A'}$. 

Although~\eqref{four-twis-trans1} is clearly implied by the combined half Fourier transforms, it has some rather puzzling features. The functions $f(W)$ and $F(Z)$ are homogeneous functions on (dual) twistor space, with some well-defined weights $n$ and $-n-4$ respectively. However, if $F(Z)$ has negative weight, then~\eqref{four-twis-trans1} diverges at the origin, while if it has positive weight then~\eqref{four-twis-trans1} diverges at infinity.  So the transform appears not to make sense.

To resolve this issue, we must understand equation~\eqref{four-twis-trans1} as a Fourier transform of
distributions.  To make this explicit and to obtain projective formulae, we co-ordinatise $\R^4$ by $Z=r \Theta$ where $r\in
(-\infty,\infty)$ and $\Theta=\Theta(\theta_i)$ lies on a hemisphere of unit radius on which the $\theta_i$ are
co-ordinates\footnote{These are not quite the usual polar co-ordinates, because $r$ lies on a
complete line, not a half line, and correspondingly, $\theta_i$ lives
on a hemisphere or $\RP^3$ rather than a complete sphere.}. The
$\theta_i$ also provide co-ordinates on $\RP^3$. We will not need to
make the co-ordinatisation of $\RP^3$ explicit and just denote its
volume form by $\D^3\Theta$. This is defined so that  
\be{vol-form}
   \rd^4 Z= |r|^3\rd r \wedge \D^3\Theta \, ,
\ee
where the modulus sign arises from the Jacobian of the co-ordinate
transformation.  Such a choice is not projectively invariant, and two
such choices will differ by some scaling $(r,\Theta)\rightarrow
(r',\Theta')=(ar,a^{-1}\Theta)$ where $a=a(\Theta)$.

The $r$ dependence in~\eqref{four-twis-trans1} can now be made explicit. Assuming $F(Z)$ has homogeneity $-n-4$, one finds 
\be{four-twis-trans}
     f(W)=\frac{1}{(2\pi)^2}\int \sgn(r)\,r^{-n-1}\,\e^{\im r\Theta\cdot W}\,F(\Theta)\,\rd r\,\D^3\Theta\ . 
\ee
The integral for $r$ is a Fourier transform with conjugate variable $\Theta\!\cdot\! W$.  For $n\leq 0$ this integral blows up as $r\to\infty$, and for $n\geq 0$ it blows up at the origin.  These integrals all have a standard regularisation (see {\it e.g.}~\cite{Gelfand-Shilov} for a detailed exposition): For $n< 0$ one obtains
\be{Four-rk-n-neg}
    \int r^{-n-1} \sgn(r)\, \e^{\im r s}\, \rd r =2s^n\im^{-n} (-n-1)!   \ . 
\ee
and for $n\geq 0$ 
\be{Four-rk} 
   \int r^{-n-1} \sgn (r)\, \e^{\im r s}\,\rd r = 2\frac{\im ^{n+1}}{n!} s^n (\log |s|-\alpha_n)
\ee
where $\alpha_n=\Gamma'(1)+\sum_{k=1}^n1/k$. Note that
$\Gamma'(1)=-\gamma$, where $\gamma$ is Euler's constant\footnote{This
gives an alternative derivation of Penrose's `universal bracket
factor'~\cite{Penrose:univbracket}.}. Thus, having
integrated out the scale of $Z$, we obtain the projective formulae 
\be{twis-trans-proj}
   f(W)=
           \begin{cases}\displaystyle
                   \frac {2\im^{-n}(-n-1)!}{(2\pi)^2 } \int D^3Z\, F(Z) (Z\!\cdot\! W)^n \hspace{2cm}  & n\leq-1\vspace{0.2cm}\\
                   \displaystyle
                   \frac {2\im^{n+1}}{(2\pi)^2 n!} \int D^3Z\,F(Z) (Z\!\cdot\! W)^n(\log|Z\!\cdot\! W|-\alpha_n ) & n\geq 0 
           \end{cases}
\ee
where $D^3Z=\varepsilon_{\alpha\beta\gamma\delta}Z^\alpha\rd Z^\beta
\wedge \rd Z^\gamma \wedge \rd Z^\delta$ is the canonical top degree 
form of weight +4 on projective twistor space.

To check the projective invariance we re-scale $Z\rightarrow a(Z) Z$.  Invariance is
obvious when $n<0$, but for $n\geq 0$, the rhs changes by an arbitrary polynomial of degree $n$ in $W$.  While this may seem to violate projective invariance, in fact it is natural to think of a dual twistor function $f(W)$ of homogeneity $n>0$ as being defined modulo polynomials of degree $n$. This is because the X-ray (or Penrose) transform
\begin{equation}
   \phi_{A_1\ldots A_{n+2}}(x) = \int \la\lambda\rd\lambda\ra\, 
   \frac{\del^{n+2}}{\del\lambda_{A_1}\cdots\del\lambda_{A_{n+2}}} f(\lambda_B,x^{CC'}\lambda_C) 
\ee
gives a vanishing space-time field when $f(W)$ is a degree $n$ polynomial. Thus the constant $\alpha_n$ only contributes  such a polynomial to $f(W)$, and is thus irrelevant here. With this proviso, equation~\eqref{twis-trans-proj} now respects the homogeneities and is a clear analogue of the complex version of the twistor transform~\cite{Penrose:1972ia}.

We must also consider twistor and dual twistor functions $\tilde F(Z)$ and $\tilde f(W)$ with `wrong sign' behaviour as in~\eqref{wrongscaling}. In this case, integrating out the scale yields the Fourier transform of $r^{-n-1}$ without the $\sgn\, r$ factor:
\begin{equation}
    \int\rd r\,\e^{\im rs}\, r^{-n-1} =
    \begin{cases}\displaystyle
    \frac{\im^{n+1}\pi s^n}{n!}\sgn(s)\qquad &n\geq 0\vspace{0.2cm}\\
    \displaystyle2\pi(-\im)^{-n-1} \delta^{-n-1}(s) &n\leq -1\ ,
    \end{cases}
\end{equation}
where $\delta^{-n-1}(s)$ denotes the $(|n|-1)^{\rm th}$ derivative of $\delta(s)$. The projective version of the twistor transform for wrong sign functions is thus
\be{twis-trans-proj-odd}
    \tilde f(W)=
    \begin{cases}\displaystyle
            \frac {\im^{n+1}\pi}{(2\pi)^2 n!} \int D^3Z\,\tilde F(Z)  (Z\!\cdot\!W)^n\,\sgn (Z\!\cdot\! W) 
            &\qquad n\geq 0\vspace{0.2cm}\\
            \displaystyle\frac{\im^{n+1}}{2\pi}\int D^3Z\,\tilde F(Z)\,\delta^{-n-1}(Z\!\cdot\! W) &\qquad n\leq -1\, .
    \end{cases}
\ee
For $n\leq -1$ in this odd-odd case, the twistor transform becomes a radon-like transform over planes in twistor space.  This is possible despite the non-orientability of the planes ($\RP^2$s) because the wrong sign behaviour is what is required to define a density on an even-dimensional projective space.


\subsubsection{Supersymmetric twistor transforms}

The positive and negative $n$ formul\ae\ combine to form supersymmetric twistor transforms.  With 
$\cN$ supersymmetries,  the supertwistor transform is
\be{susy-twis-transf}
    f(\rW)=\int D^{3|\cN}\rZ\ F(\rZ)\, (\rZ\!\cdot\!W)^{\cN/2-2}  \log|\rZ\!\cdot\!\rW|
\ee
where $F(\rZ)$ and $f(\rW)$ each have homogeneity $\cN/2-2$ and are of `right sign' type. The `wrong sign' version is
\be{susy-twis-transf-wrong sign}
    \tilde f(\rW)=\int D^{3|\cN}\rZ\ \tilde F(\rZ)\,(\rZ\!\cdot\!\rW)^{\cN/2-2}\, \sgn\left(\rZ\!\cdot\!\rW\right)\ .
\ee
Expanding these transforms in powers of the Grassmann co-ordinates reproduces the 
transforms~\eqref{twis-trans-proj}~\& \eqref{twis-trans-proj-odd}. 


\subsection{The inner product}

In Lorentzian signature, the standard inner product between two massless fields of helicity $h$ is given by multiplying the momentum space wavefunction $\Phi_{2h}(\lambda,\tilde\lambda)$ by the complex conjugate of the wavefunction $\Psi_{2h}(\lambda,\tilde\lambda)$, and then integrating over the momentum light-cone:
\be{mom-inner-prod}
    (\Psi,\Phi) =\frac{\im}{(2\pi)^4}\int \rd^3\ell\ \overline{\Psi}\Phi
\ee
where $\rd^3\ell$ is the standard invariant measure 
$\rd^3\ell=\la\lambda\rd\lambda\ra \rd^2\tilde\lambda -[\tilde\lambda\rd\tilde\lambda]\rd^2\lambda$. 
This Lorentzian inner product is anti-linear in $\Psi$ because of the  complex conjugation; 
in particular $\overline{\Psi}$ has helicity $-h$ because the conjugation exchanges $\lambda$ and $\tilde\lambda$.

On $\R^{2,2}$, complex conjugation does not change particle helicity, so instead of an inner product we simply have a bilinear pairing between fields of helicity $h$ and fields of helicity $-h$, given by
\be{mom-pairing}
    (\Psi_{-2h},\Phi_{2h}) = \frac{1}{(2\pi)^4}\int\rd^3\ell\ \Psi_{-2h}\Phi_{2h}\ .
\ee
The corresponding pairing on twistor space is between a twistor function of weight $n$ and a twistor function of weight $-n-4$. This pairing is again given by multiplying the twistor functions and integrating over the real projective twistor space:
\be{twistor-pairing}
    (F,G) = \int_{\RP^3}\D^3Z\ F_n(Z) G_{-n-4}(Z)\ .
\ee
Unitarity of the half Fourier transform ensures that the momentum space and twistor pairings agree. On twistor space, when $n\geq 0$ or $n\leq -4$ the fact that positive homogeneity twistor functions are defined modulo polynomials is dual to the fact that for~\eqref{twistor-pairing} to be well-defined, the \emph{charge integral}
\be{charge}
    Q^{\alpha\beta\ldots\gamma}:=\int D^3Z\,Z^{\alpha}Z^{\beta}\cdots Z^{\gamma}\, F_n(Z) 
\ee
must vanish, where $F_n(Z)$ has homogeneity $n\leq-4$ and there are $|n|-4$ factors of $Z$ inserted in the integral. When $n=-4$ this is the standard twistor charge integral for a self-dual Maxwell field, and when $n=-6$ it yields the angular-momentum twistor of the corresponding linearised gravitational field~\cite{Penrose-Rindler}.

Combined with the Fourier/Twistor transform described above, we obtain
a pairing between a twistor function $F(Z)$ and a dual twistor function $G(W)$ each of homogeneity degree $-n-4$.  When $F(Z)$ and $G(W)$ both have the `right sign' behaviour this pairing
\be{ambi-inner-prod}
   (F,G)=
   \begin{cases}\displaystyle
           \frac{2\im^{-n}(-n-1)!}{(2\pi)^2}\int D^3Z\wedge D^3W\ F(Z)G(W) (Z\!\cdot\!W)^n \qquad &n\leq-1\vspace{0.2cm} \\
           \displaystyle
           \frac{2\im^{n+1}}{(2\pi)^2n!} \int D^3Z\wedge D^3W\ F(Z)G(W) (Z\!\cdot\!W)^n \log |Z\!\cdot\!W| &n\geq0\ ,
   \end{cases}
\ee
whereas for `wrong sign' functions $\tilde F(Z)$ and $\tilde G(W)$ it is
\be{ambi-inner-prod-odd}
   (\tilde F,\tilde G)=
   \begin{cases}\displaystyle
           \frac{\im^{n+1}\pi}{(2\pi)^2}\int D^3Z\wedge D^3W\ \tilde F(Z)\tilde G(W)\ \delta^{-n-1}(Z\!\cdot\!W)
           \qquad & n\leq-1 \vspace{0.2cm}\\
           \displaystyle\frac{\im^{n+1}\pi}{(2\pi)^2n!}\int D^3Z\wedge D^3W\ \tilde F(Z)\tilde G(W)\, (Z\!\cdot\!W)^n\ 
           \sgn(Z\!\cdot\!W) & n\geq 0\ .
   \end{cases}
\ee
Once again, these formul\ae\ combine into supersymmetric pairings given by
\be{ambi-inner-prod-super}
   \begin{aligned}
   (F,G) &= -\frac{\im^{\frac{\cN}{2}-1}}{(2\pi)^2\left(\frac{\cN}{2}-2\right)!}
           \int D^{3|\cN}\rZ\wedge D^{3|\cN}\rW\ F(\rZ)G(\rW)\,(\rZ\!\cdot\!\rW)^{\frac{\cN}{2}-2}\log|\rZ\!\cdot\!\rW|\\
   (\tilde F,\tilde G) &=\frac{\im^{\frac{\cN}{2}-1}\pi}{(2\pi)^2\left(\frac{\cN}{2}-2\right)!}
           \int D^{3|\cN}\rZ\wedge D^{3|\cN}\rW\ \tilde F(\rZ)\tilde G(\rW)\,(\rZ\!\cdot\!\rW)^{\frac{\cN}{2}-2}\sgn(\rZ\!\cdot\!\rW)
   \end{aligned}
\ee
where all the wavefunctions have homogeneity $\cN/2-2$.


\subsection{BCFW recursion in ambitwistor space}
\label{sec:BCFWambi}

We are now in position to explain the relation of the present paper to that of Arkani-Hamed {\it et al.}~\cite{AHCCK}. The main awkwardness of the twistor space BCFW formula
\begin{multline}
\label{BCFWtwist-again}
   \cM(\rW_1,\ldots,\rW_n) = \sum \int D^{3|\cN}\rW\, \cM_L(\rW_1,\ldots,\rW)\\ \times\ 
   \sgn\left(\la\rW_1\rW\ra\,\rW_1\!\cdot\!\frac{\del}{\del\rW_n}\,\left[\frac{\del}{\del\rW}\frac{\del}{\del\rW_n}\right]\right)
   \cM_R(\rW,\ldots,\rW_n)
\end{multline}
is the presence of the non-local operator $\sgn(\rW_1\!\cdot\!\del_n\,[\del_\rW\,\del_n])$. This can be rendered local by use of the Fourier/Twistor transforms introduced above. To achieve this, first represent $\cM_R(\rW,\ldots,\rW_n)$ in terms of a (non-projective) Fourier transform of $\cM_R(\rZ,\ldots,\rZ_n)$ in its first and last arguments Z and Z$_n$. Then
\begin{multline}
   \sgn\left(\rW_1\!\cdot\!\frac{\del}{\del\rW_n}\,\left[\frac{\del}{\del\rW}\frac{\del}{\del\rW_n}\right]\right)
   \cM_R(\rW,\ldots,\rW_n)\\
   =\frac{1}{(2\pi)^4}\int\rd^{4|\cN}{\rZ}\,\rd^{4|\cN}\rZ_n\ \e^{\im\rZ\cdot\rW}\,\e^{\im\rZ_n\cdot\rW_n}\,
   \sgn\left(\rZ_n\!\cdot\!\rW_1\,[\rZ\,\rZ_n]\right)\cM_R(\rZ,\ldots,\rZ_n)\ ,
\end{multline}
where we abuse notation by not distinguishing $\cM_R$ from its Fourier
transform, and define
$[\rZ\,\rZ_n]:=[\pi\,\pi_n]=[\tilde\lambda\,\tilde\lambda_n]$. (It makes no difference whether or not we similarly transform the
remaining states in $\cM_R$.) The Fourier transform in $\rZ_n$ may be
accommodated by likewise Fourier transforming the lhs of the BCFW
recursion~\eqref{BCFWtwist-again}. Since
$\sgn([\rZ\,\rZ_n])\cM_R(\rZ,\ldots,\rZ_n)$ has wrong sign behaviour
in $\rZ$, we obtain the projective form of ambidextrous BCFW recursion  
\be{BCFWambi}
\mathbox{
   \begin{aligned}
           \cM(\rW_1,\ldots,\rZ_n) &=\sum\int D^{3|\cN}\rZ\wedge D^{3|\cN}W\  \cM_L(\rW_1,\ldots,\rW)\,
           (\rZ\!\cdot\!\rW)^{\cN/2-2}\\
           &\hspace{2cm}\times\ \sgn\left(\la\rW_1\rW\ra\ \rZ_n\!\cdot\!\rW_1\ \rZ\cdot\rW\ [\rZ\,\rZ_n]\right)
           \cM_R(\rZ,\ldots,\rZ_n)
   \end{aligned}}
\ee
in agreement with Arkani-Hamed {\it et. al.}~\cite{AHCCK}. In this form, the non-local sign operators have been replaced by multiplication by the local operator 
$$
   \sgn\left(\la\rW_1\rW\ra\ \rZ_n\!\cdot\!\rW_1\ \rZ\cdot\rW\ [\rZ\,\rZ_n]\right)
$$
in $\cN=4$ SYM or 
$$
   (\rZ\cdot\rW)^2\,\sgn\left(\la\rW_1\rW\ra\ \rZ_n\!\cdot\!\rW_1\ \rZ\cdot\rW\ [\rZ\,\rZ_n]\right)
$$
in $\cN=8$ SG, at the expense of introducing a larger integral for the intermediate state. Comparing the ambitwistorial BCFW recursion~\eqref{BCFWambi} to the supersymmetric pairing $(\tilde F,\tilde G)$ of~\eqref{ambi-inner-prod-super}, we again see the close relation between BCFW recursion and the inner product of the wrong sign wavefunctions
$$
\cM_L(\ldots,\rW)\,\sgn(\la\rW_1\rW\ra)\qquad\hbox{and}\qquad\cM_R(\rZ,\ldots)\,\sgn([\rZ\, \rZ_n])\ .
$$

For some purposes, it may be useful to have a form of recursion relation that is intermediate between the twistorial~\eqref{BCFWtwist-again} and ambidextrous~\eqref{BCFWambi}, and involves the fewest integrals. Such a form may be obtained from~\eqref{BCFWtwist-again} by Fourier transforming $\rW_n\to\rZ_n$, but not Fourier transforming the intermediate state. One finds
\be{BCFWintermediate}
\begin{aligned}
   \cM(\rW_1,\ldots,\rZ_n) &=\sum \int D^{3|\cN}\rW\ \cM_L(\rW_1,\ldots,\rW)\\
   &\hspace{2cm}\times\ 
   \sgn\left(\la\rW_1\rW\ra\ \rZ_n\!\cdot\!\rW_1\ I\!\rZ_n\!\cdot\!\frac{\del}{\del\rW}\right)\cM_R(\rW,\ldots,\rZ_n)\\        
   &=\sum \int  \frac{\rd t}{t}D^{3|\cN}\rW\ \cM_L(\rW_1,\ldots,\rW)\\
   &\hspace{2cm}\times\ 
   \sgn\left(\la\rW_1\rW\ra\ \rZ_n\!\cdot\!\rW_1\right)\cM_R(\rW+tI\!\rZ_n,\ldots,\rZ_n)\, ,
\end{aligned}
\ee
where $I\!\rZ_n:=I_{\alpha\beta}Z^\alpha_n=|n]$. 


\subsection{Ambidextrous form of the 3 point amplitudes}
\label{sec:3ptambi}

To begin the recursion using the ambidextrous form~\eqref{BCFWambi}, we need the three-point amplitudes written in an ambidextrous way. The SYM 3-point amplitudes 
\be{3pt-MHV}
\begin{aligned}
   \cA_{\rm MHV}(\rW_1,\rW_2,\rW_3)&= \sgn \left(\la\rW_2\rW_3\ra\,
   \im\rW_2\!\cdot\!\frac{\del}{\del\rW_1}\,\im\rW_3\!\cdot\!\frac{\del}{\del\rW_1}\right)\,\delta^{(4|4)}(\rW_1)\\
   \cA_{\MHVbar}(\rW_1,\rW_2,\rW_3)&= \sgn\left(\left[\frac{\del}{\del\rW_2}\frac{\del}{\del\rW_3}\right]
   \im\rW_1\!\cdot\!\frac{\del}{\del\rW_2}\,\im\rW_1\!\cdot\!\frac{\del}{\del\rW_3}\right)
   \delta^{(4|4)}(\rW_2)\,\delta^{(4|4)}(\rW_3)
\end{aligned}
\ee
become simply
\be{3pt-ambi}
\begin{aligned}
     \cA_{\rm MHV}(\rZ_1,\rW_2,\rW_3) &=\sgn\left(\la\rW_2\rW_3\ra\,\rZ_1\!\cdot\!\rW_2\,\rZ_1\!\cdot\!\rW_3\right)\\
     \cA_{\MHVbar}(\rW_1,\rZ_2,\rZ_3) & = \sgn\left([\rZ_2\rZ_3]\,\rZ_2\!\cdot\!\rW_1\,\rZ_3\!\cdot\!\rW_1\right)
\end{aligned}
\ee
after representing the delta functions by their Fourier transforms. Similarly, the 3-point gravity amplitudes are
\be{3pt-grav-ambi}
   \begin{aligned}
   \cM_{\rm MHV}(\rZ_1,\rW_2,\rW_3) &= \left|\la\rW_2\rW_3\ra(\rZ_1\!\cdot\!\rW_2)(\rZ_1\!\cdot\!\rW_3)\right|\\
   \cM_{\overline{\rm MHV}}(\rW_1,\rZ_2,\rZ_3) &=\left|(\rW_1\!\cdot\!\rZ_2)(\rW_1\!\cdot\!\rZ_3)[\rZ_2\rZ_3]\right|\ .
   \end{aligned}
\ee

Using~\eqref{3pt-ambi} or~\eqref{3pt-grav-ambi} and the ambidextrous BCFW relation~\eqref{BCFWambi},
one can build up arbitrary tree amplitudes on products of twistor and dual twistor spaces. The amplitudes are, in the first instance, expressed as integrals of these 3-point amplitudes, and make contact with Hodges' twistor diagrams~\cite{Hodges:1980hn,Hodges:2005bf,Hodges:2005aj,Hodges:2006tw}.  In the case of SYM, the integrands are simply products of sgn factors. Further applications of the ambidextrous form of the recursion relations can be found in~\cite{AHCCK}.

Clearly, there is a direct correspondence between the calculations
required to solve the recursion relations on dual twistor space and those
required on ambitwistor space, but we will not expand on this here.  In section~\ref{sec:general-structure}, we showed that on dual twistor space, an arbitrary $n$-point N$^k$MHV tree amplitude involves $2(n-2)$ Hilbert  transforms 
and $(n-2-k)$ delta functions.  Representing each of these $(n-2-k)$ delta functions as a Fourier transform, the Hilbert transforms (and other non-local sign operators) act under the integrals to yield a local function of the $\rZ_i$s (for $i=1,\ldots,n-2-k$) and the remaining $\rW$s. Conversely, it is also possible to transform the result of an ambidextrous calculation onto (multiple copies of) dual twistor space by the reverse procedure.


\section{Conclusions and Outlook}
\label{sec:conclusions}

We have shown that the half Fourier transform provides a practical and
coherent scheme for translating scattering amplitudes for massless
field theories into twistor space.  The BCFW recursion relations can
be reformulated in a useful  form and can in principle be used to
generate the full tree sector of $\cN=4$ super Yang-Mills theory and 
$\cN=8$ supergravity.  In practice, we have show that the twistor
version of BCFW recursion is tractable, and have solved them for small
$k$ N$^k$MHV and googly N$^k$MHV amplitudes. 
In general we have seen that, up to some `signs', a N$^k$MHV amplitude
is expressed on twistor space and an intgegral over $2n-4$ bosonic
parameters (expressed here in the form of Hilbert tranforms) of
$n-k-2$ delta functions $\delta^{4|4}$ of linear combinations of the
$n$ twistors involved. 
As far as loop amplitudes are concerned, in principle there is no
difficulty expressing them on twistor space via the half Fourier
transform, and we showed that for the four-particle amplitude, this is
actually straightforward. It is likely that the generalised unitarity
methods~\cite{Britto:2004nc,Bern:2005iz,Bern:2006ew,Bern:2007ct,Drummond:2008bq} 
that are so successful for constructing loop amplitudes for supersymmetric gauge theories can also be adapted to provide 
a generating principle for loop amplitudes on twistor
space\footnote{See~\cite{ArkaniHamed:2009dn,Mason:2009qx,ArkaniHamed:2009vw,
    Bullimore:2009cb,Kaplan:2009mh} for 
  subsequent developments in this direction}.

Part of the motivation for this work was to express superconformal
invariance more clearly. In fact, the formalism has made transparent
that superconformal invariance is actually broken by factors of $\sgn(\la12\ra\la23\ra\la31\ra)$ and the non-local operator
$\sgn[\del_{\rW_i}\del_{\rW_j}]$. There is complete
cancellation of these factors in even-point MHV and $\MHVbar$
amplitudes (which are therefore conformally invariant), but there is a topological
obstruction to making odd-point MHV amplitudes conformally invariant, even at tree-level. This may simply be reflecting a basic feature of scattering theory: to define an S-matrix, one must
first choose an asymptotic region in which the particles are free.
Indeed, such a choice is necessary to define $\sgn(\la12\ra\la 23\ra\la 31\ra)$, for example.  Choosing a light-cone to 
remove from conformally compactified space-time in order to define the
amplitudes is analogous to choosing a cut of the complex $z$-plane in
order to define $\sqrt z$. It seems also likely that these conformal
symmetry breaking sign factors are an artifact of split signature as
the obstructions to conformal symmetry only apply there (see also the
arguments below).

However, the topological obstruction identified above cannot
apply in Lorentz signature, because twistor lines are then $\CP^1$s
rather than $\RP^1$s, and there is no notion of ordering points
on a sphere. It is thus seems to be the case that the awkward signs and
corresponding  violation of conformal
invariance is a consequence of our use of split
signature. Support for this point of view comes from the fact
that in complex twistor space, space-time fields correspond to
cohomology classes, rather than by functions. Choosing a Dolbeault
representation of the fields, the required antisymmetry of the
kinematic part of the amplitude comes naturally from the wedge product
of forms, as in the holomorphic Chern-Simons vertex 
$$
	\int_{\CP^{3|4}}\hspace{-0.5cm}D^{3|4}\rW\wedge {\rm tr}\left(A\wedge A\wedge A\right)
$$
rather than from an explicit factor of  $\sgn(\la 12\ra\la
23\ra\la31\ra)$. To adapt this Chern-Simons vertex (or indeed the
complete twistor actions
of~\cite{Mason:2005zm,Boels:2006ir,Mason:2008jy}, that also rely on a
Dolbeault description) to the split signature context,  one must
choose Dolbeault representatives defined from the X-ray data, such as
those in~\cite{Bailey:03}. These representatives are conformally
invariant, but one must then choose an open cover to reduce the
integrals from the full complex twistor space to overlaps of open
sets, and eventually down to the real twistor space. These choices of
open sets will break conformal invariance. Conversely, the presence of
conformal breaking factors in the real twistor space amplitudes is
perhaps required in order for them to have a cohomological form which
may be analytically continued to Lorentz signature.

Certainly, in order to make the twistor formulation self-contained,
one should really understand how the twistor recursion relations and seed amplitudes
presented here can be obtained from the twistor actions. Since the
twistor action for Yang-Mills is itself conformally invariant, the
conformal breaking factors must arise either from gauge fixing or from
a choice of {\v C}ech cover to give split signature X-ray
representatives. It remains an open question as to whether one can
introduce a twistor action that is itself naturally adapted either to
a  {\v C}ech description of cohomology, or to the split signature X-ray
transform framework used in this paper\footnote{Similar questions
concern the relation of the twistor diagram approach of
Hodges~\cite{Hodges:2005bf,Hodges:2005aj,Hodges:2006tw} and
Arkani-Hamed {\it et al.}~\cite{AHCCK} to the ambitwistor
action~\cite{Mason:2005kn}. Preliminary calculations show
that~\cite{Hodges:2005bf,Hodges:2005aj,Hodges:2006tw,AHCCK} are
working with the Feynman rules of the ambitwistor action on ``the
momentum space of ambitwistor space'' rather than on ambitwistor
space itself.}.  

However they are viewed, for us the practical consequence of breaking
conformal invariance 
is simply that the factors of $\sgn(\la \rW_i\rW_j\ra)$ and
particularly the non-local operator
$\sgn([\del_{\rW_i}\,\del_{\rW_j}])$ make the twistor BCFW recursion
more complicated. It is tempting to speculate whether there could
perhaps be a sense in which they can be discarded, maybe as a model
for a fully complex BCFW recursion. For example, consider the gluing
rule   
\be{conf-BCFW}
	\cA'(1,\ldots,n) = \sum \int \frac{\rd t}{|t|}D^{3|4}\rW\,\cA'_L(\rW_1,\ldots,\rW)\cA'_R(\rW,\ldots,\rW_n-t\rW_1)
\ee
in real twistor space. Unlike BCFW recursion, this gluing rule is
(manifestly) superconformally invariant as well as projective. If we
seed~\eqref{conf-BCFW} by the three-point objects   
\be{seed-conf}
\begin{aligned}
     \cA'_{\rm MHV}(1,2,3) &= \delta^{(2|4)}(\rW_1,\rW_2,\rW_3)\\
     \cA'_{\MHVbar}(1,2,3) &=
     \delta^{(3|4)}(\rW_1,\rW_2)\,\delta^{(3|4)}(\rW_1,\rW_3) 
\end{aligned}
\end{equation}
then the recursion procedure is manifestly superconformally
invariant\footnote{In split signature, recursion
using~\eqref{conf-BCFW} seeded by~\eqref{seed-conf} corresponds on
momentum space to recursion using the rule  
\be{conf-BCFW-mom} 
A'(1,\ldots,n) = \sum
A'_L(\hat{1},\ldots,-\hat{p})\,\frac{1}{|p^2_L|}\,A'_R(\hat{p},\ldots,\hat{n}) 
\ee
seeded by the three-point objects 
\be{jkhdgjhs}
\begin{aligned}
     A'_{\rm MHV}(1,2,3) &= \frac{\delta^{(4|8)}\left(\sum_{i=1}^3|i\ra\llbracket i\|\right)}{\left|\la12\ra\la23\ra\la31\ra\right|}\\
     A'_{\MHVbar}(1,2,3) &= \frac{\delta^{(4)}\!\left(\sum_{i=1}^3|i\ra[i|\right)\,
     \delta^{(0|4)}\left(\eta_1\la23\ra+\eta_2\la31\ra+\eta_3\la12\ra\right)}{\left|[12][23][31]\right|}
\end{aligned}
\ee
where, compared to the usual BCFW relation and amplitudes, the
propagator and denominators of the three-point amplitudes appear
inside a modulus sign. This is reminiscent of a similar modulus sign
appearing in the calculations of~\cite{Roiban:2004yf} in the connected
prescription of twistor-string theory.}.  The output of this recursion
then seems to produce purely geometric sums of products of delta
functions, just as one might obtain by manually removing the
conformal-breaking signs from amplitudes and replacing the sgn in the
remaining $\sgn(\rW_i\!\cdot\!\del_j)$ operators by logs (see footnote
10 in section~\ref{sec:hilbert}). Although~\eqref{conf-BCFW}
\&~\eqref{seed-conf} have the wrong exchange properties to be
amplitudes, it is perhaps conceivable that the true split signature
amplitudes can be recovered from this simpler recursion rule by
dressing it with conformal-breaking signs once the recursion has been
performed.  One might also speculate that a Dolbeault version of BCFW
recursion in complex twistor space bears more resemblance
to~\eqref{conf-BCFW} than to the actual split signature BCFW rule,
with the exchange properties coming from understanding the (seed)
amplitudes as forms, rather than (delta-)functions. As with the
twistor actions, the conformal-breaking sign factors in the true split
signature BCFW rule might then arise from choosing representatives for
these forms with respect to a cover that is adapted to real twistor
space.

An alternative to making contact with twistor actions is to make
contact with the  MHV formalism form.  It is possible to start from
Risager's shift~\cite{Risager:2005vk} (or its multiline
extensions~\cite{Kiermaier:2008vz,Kiermaier:2009yu}) and proceed as in
this paper, obtaining a twistor representation of the MHV
formalism~\cite{Cachazo:2004kj} in split signature.  This has some
advantages, but the big disadvantage that conformal (and indeed
Lorentz) invariance is expressly broken by the choice of spinor used
in the gauge choice and 
this in the long run leads to more complicated formulae. 

A more
interesting approach might be to work with twistor-string
theory~\cite{Witten:2003nn}, which again 
deals with on-shell amplitudes. The
connected prescription computes an N$^k$MHV super-amplitude by means
of an integral~\cite{Roiban:2004yf} over the moduli space
$\overline{M}_{0,n}(\CP^{3|4},d)$ of $n$-pointed, degree $d$ (stable)
maps  
$$
	f:(\Sigma;p_1,\ldots,p_n)\to\CP^{3|4}\ ,
$$
where $d=k+1$ (see {\it e.g.}~\cite{Fulton:1997} for an introduction
of the bosonic part of this moduli space, and~\cite{Gukov:2004ei} for
a discussion in the context of twistor-strings). This space has
bosonic dimension $4d+n$. If the vertex operators are constructed from
elemental states, they restrict $f$ to take each of the marked
points $p_i$ (insertion points of the vertex operators) on $\Sigma$ to a
specific point $Z_i\in\CP^{3|4}$. This implies three bosonic
constraints per vertex operator, so the space of such maps has virtual
dimension $-2n+4d=-2(n-2k-2)$. The implication of this virtual
dimension being negative is simply that the points $Z_i$ cannot be in
general position in $\CP^3$, and the amplitude only has support on
some algebraic subset. Specifically, there are $2(n-2k-2)$ constraints
on the locations of the points in twistor space. This fits in
precisely with the counting of bosonic delta functions in
section~\ref{sec:general-structure}, where we found that an N$^k$MHV
$4(n-2-k)$ delta-functions, dressed by $2(n-2)$ Hilbert transforms,
each of which soaks up a delta function. It would be interesting to
see how the corresponding parametrisation of the moduli space arises. 

One might then hope to understand the BCFW recursion formula in the
context of the connected prescription by considering the one-parameter
family of amplitudes $\cA(Z_1,\ldots,Z_{n-1},Z_1-tZ_n)$ where the
marked point $p_n$ is only required to be mapped to the line joining
$Z_1$ to $Z_n$. The space of such maps has dimension 1 greater than
the support of $\cA(Z_1,\ldots,Z_{n-1},Z_n)$, and one could seek a
derivation of the twistor BCFW rule by understanding how this curve
interacts with the boundary divisors in
$\overline{M}_{0,n}(\CP^{3|4},d)$, corresponding to degenerations of
$(\Sigma;p_1,\ldots,p_n)$ to a nodal curve. This would make contact
with the work of Vergu~\cite{Vergu:2006np} on the factorisation limits
of the connected prescription\footnote{See \cite{Bullimore:2009cb} for
  the correct explanation of these relationships}.     

A major attraction of reformulating these ideas in connection with twistor-string theory is the possibility of realising Witten's proposal~\cite{Witten:2003nn} that the Yangian symmetry of planar $\cN=4$ super Yang-Mills, seen at both weak~\cite{Drummond:2009fd,Dolan:2004ps} and strong~\cite{Bena:2003wd,Beisert:2008iq} coupling, should arise from the harmonic map equations underlying the twistor-string.


\subsection{Signature signs and loops}\label{sec:signs}

In the discussion of loops, the problems that arise
from working with split signature become sharper.  When one
analytically continues to Euclidean signature, the structure of the
Feynman propagator implies a unique continuation.  However, the
Feynman prescription requires a concept of time ordering that makes no
sense in split signature; in split signature the lightcone is
connected and does not naturally divide into past and future.  
It would therefore be problematic to be restricted to considering scattering
amplitudes in split signature.

Our view is that the split signature methods developed in this paper
(and those of \cite{AHCCK}) represent a clear route to discovering
underlying twistorial 
structures of amplitudes that will also have a direct reformulation
that makes sense in the complex and, in
particular, in Lorentz signature. 
A model for how this is likely to work is provided by twistor diagram
theory which  has never been
restricted to any particular signature.  In this approach, the
integrals are all considered to be contour integrals defined in
complex twistor space.  

Some aspects of the translation procedure into the complex are
straightforward.  
Wave functions that are simply functions of the real twistors should
 be required to have an analytic extension becoming holomorphic
functions on certain open sets.  These would then be re-interpreted as
Cech cocycles as described in \cite{Atiyah:1979,Mason:95,Sparling,
  Woodhouse,Bailey:99, Bailey:01,  
Bailey:03}.   With the reformulation as Cech cocycles, the signs in
the formula would no longer appear explicitly, but be encoded into the
cohomology class containing the information of which overlaps in some
open cover the Cech cocycles are associated with. 
The integrals can then be understood as contour integrals that are
homological to the real slice.
Where, in our formalism we have used a delta
function $\delta(z)$, in the complex, we would replace this with the
pole $1/2\pi i z$ and specify that the topology of the integration
contour should be 
specified so that the integration can be performed by a residue calculation.

The above discussion indicates that in such a complexified reformulation, the
awkward signs should be suppressed from explicit formulae, although
their presence should nevertheless be felt when it comes to specifying
the contour.
Indeed precisely these ingredients, including many details, play a role
in the correspondence between the work of \cite{AHCCK} and Andrew
Hodges' twistor diagrams.

There are of course many technical obtstructions to overcome to make
this programme precise.  Perhaps the most difficult aspect of twistor
diagram theory is the correct cohomological interpretation.  This is
rendered particularly problematic by the infrared divergences that are
present even at tree level.  These mean that one doesnt necessarily
expect the integrals to make good sense even in the simplest examples.
Another related issue is the systematic specification of the contour.

An important role in our formulation of a twistor amplitude is the
concept of an {\em elemental 
  state} ({\it i.e.},  in the real context, this is a twistor 
function on real twistor space that has delta function support at a
point, or, in Fourier transformed form, one of the form $\sgn (Z\cdot
A)$ or
$\log (Z\cdot A)$).  These are essentially the external states that
are being scattered in a
twistor amplitude expression.  Analogues of such states were originally defined
in the complex~\cite{Hodges:1986}, but their cohomological
interpretation remains elusive.
These are used to give us the integral kernel of the amplitude, and
its not so clear that they should have a standard 
interpretation as an external wave-function.  It is reasonable to
expect that kernels of scattering amplitudes should have a different
cohomological status to 
that of wave functions as
one expects them to have 
different analyticity properties; massless
wavefunctions are arbitrary smooth functions on the momentum
light-cone, while tree amplitudes are meromorphic functions on
the complexified lightcone in momentum space.  Under the half Fourier transform,
analyticity in momentum space corresponds to twistor amplitudes that
have restricted support.  In Lorentz signature, crossing symmetry
requires the amplitudes be well-defined when integrated against
external states of both positive and negative frequency. In twistor
space, Lorentzian wavefunctions of positive and negative frequency are
represented by elements of $H^1(\overline{\mathbb{PT}^+}, \cO(-2h-2))$
and $H^1(\overline{\mathbb{PT}^-},\cO(-2h-2))$, respectively  
(see {\it e.g.}~\cite{Eastwood:1981jy}), where
$\overline{\mathbb{PT}^+}$ is the region of complex twistor space for
which the $\SU(2,2)$ inner product $Z\!\cdot\!\overline Z\geq0$, while
$\overline{\mathbb{PT}^-}$ has $Z\!\cdot\!\overline Z\leq0$. In order
to be able to pair a scattering amplitude with either of these, the
amplitude's support must be contained in $\mathbb{PN} =
\overline{\mathbb{PT}^+}\cap\overline{\mathbb{PT}^-}$. It is
suggestive that 
this can be made to be satisfied by our (split signature) formulae if we choose an
$\RP^3$ real slice that lies inside $\mathbb{PN}$ (of which there are
many). Thus the formulae we have obtained may serve perfectly well to
define amplitudes in Lorentz signature at least at tree level.

There have not yet been sufficiently many loop calculations performed
to get a feel for how a complex formulation should work at loops.  The
calculations of \S\ref{sec:loops} show that one can obtain a twistor
amplitude for each choice of analytic continuation from Lorentz
signature through to split signature.  However, there are many choices
for fixing the ambiguity in the continuation of such a loop amplitude
around its branching singularities.
In~\cite{AHCCK} the ambiguity in loop diagrams was
resolved essentially by using a
propagator that was based on the principal value regularization,
rather than the Feynman $i\epsilon$.  They obtained the answer `1' for
this amplitude.  For us, this answer is singular as one must take a Fourier
transform to obtain expressions on twistor space alone.  It was there
proposed that this answer should be related to the physical Lorentzian
signature amplitude by analytic continuation back to Lorentz signature
and being careful to pick up the contributions from the singularities that one
passes en-route.  
According to the ideas outlined above, the final correct answer
will be most directly expressed as a contour integral in the
complex.  However to get a well-defined answer we will also need to
incorporate some satisfactory form of regularization.  Andrew Hodges
has already introduced a regularization procedure in the twistor
diagram formulation that seems to be sufficient for the task at
tree-level.  It
will be interesting to see whether this can used to make sense of loop
amplitudes also.


\vspace{1cm}

{\Large \bf \noindent Acknowledgements}

\medskip
\noindent It is a pleasure to thank N.\ Arkani-Hamed, R.\ Boels, R.\ Britto, F.\ Cachazo,
C.\ Cheung, J.\ Drummond, J.\ Henn, A.\ Hodges, J.\ Kaplan, G.\ Korchemsky,
D.\ Kosower and E.\ Sokatchev for many useful comments and discussions. We would
also like to thank the organisers of the conference ``Hidden
Structures in Field Theory Amplitudes'' at NBIA in September 2008
where the ideas for this paper were sown, as well as the faculty and
staff of both Perimeter Institute and IH{\' E}S where this work was
partially carried out. LM is supported in part by the EU through the
FP6 Marie Curie RTN {\it ENIGMA} (contract number
MRTN--CT--2004--5652) and through the ESF MISGAM network. This work
was financed by EPSRC grant number EP/F016654, see also\\ 
http://gow.epsrc.ac.uk/ViewGrant.aspx?GrantRef=EP/F016654/1.


\appendix

\section{Conventions, Notation and Background}
\label{app:conventions}

For ease of comparison to the scattering theory literature (in which
an MHV amplitude has two \emph{negative} and an arbitrary number of
\emph{positive} helicity states), we will focus on \emph{dual twistor
space}.  We work throughout with \emph{real} (dual) twistors, so
as to use Witten's half Fourier transform, and correspondingly our
space-time signature is $(++--)$. Dual twistor space, $\PT^*$  will therefore be a copy of real
projective 3-space, $\RP^3$ with homogeneous co-ordinates $[W_\alpha] =
[\lambda_A,\mu^{A'}]$, where $A$ and $A'$ denote anti-selfdual and
self-dual two-component spinor indices, respectively. Twistor space
$\PT$ has homogeneous co-ordinates $Z^\alpha=(\omega^A,\pi_{A'})$.

The conformal group in this signature is $PSL(4,R)$ acting linearly on
the homogeneous twistor co-ordinates.  To break the symmetry down to the
Poincar\'e group, we introduce the \emph{infinity twistors}
$I^{\alpha\beta}$ or $I_{\alpha\beta}$ . These are defined so that
for a twistor $Z^\alpha =(\omega^A,\pi_{A'})$ and dual twistor
$W_\alpha=(\lambda_A,\mu^{A'})$ we have  
\be{infinity-twistor}
    I^{\alpha\beta}=\frac12\varepsilon^{\alpha\beta\gamma\delta}I_{\gamma\delta}\, , \qquad 
    Z^\alpha I_{\alpha\beta}=(0,\pi^{A'})\, \qquad I^{\alpha\beta}W_\beta =(\lambda^A,0)\, .  
\ee 
The equation $I^{\alpha\beta}W_\beta=0$ gives the line in dual twistor
space corresponding to the point at infinity in Minkowski space, and
the scale of $I^{\alpha\beta}$ fixes the space-time metric via the
spinor inner products below. 

We often use the spinor helicity notation
\be{spinor-products} \la \alpha\beta\ra:=\alpha_A\beta^A\ ,\qquad
[\tilde\alpha\tilde\beta]:=\tilde\alpha^{A'}\tilde\beta_{A'}
\qquad\hbox{and}\qquad \alpha_AU^{AA'}\tilde\beta_{A'}:=\langle \alpha
|U|\tilde\beta] \ee for inner products of spinors. On twistor space,
the Poincar\'e invariant inner products are \be{inf-bracket}
\begin{aligned} 
    \la W_1W_2\ra&:= I^{\alpha\beta} W_{1\alpha}W_{2\beta}= \lambda_{1A}\lambda^A_2 = \la12\ra\\
   [Z_1Z_2]&:= I_{\alpha\beta} Z_1^\alpha Z_2^\beta=\pi_1^{A'}\pi_{2A'} =[12] 
\end{aligned}
\ee

The null cone $p^2=0$ in momentum space may be co-ordinatised by
$p_{AA'}=\lambda_A\tilde\lambda_{A'}$. In split signature $\lambda_A$
and $\tilde\lambda_{A'}$ are each real, independent, two-component
spinors defined up to the scaling
$(\lambda_A,\tilde\lambda_{A'})\to(r\lambda_A,r^{-1}\tilde\lambda_{A'})$
for $r$ any non-zero real number.  A wavefunction
$\Phi(\lambda_A,\tilde\lambda_{A'})$ on the light-cone in momentum
space can be related to a (dual) twistor function $f(W)$ by Witten's
`half Fourier transform' 
\begin{equation} 
    f(W) = \int\rd^2\tilde\lambda\ \e^{\im\mu^{A'}\tilde\lambda_{A'}}\,\Phi(\lambda,\tilde\lambda)
    \qquad;\qquad 
    \Phi(\lambda,\tilde\lambda) = \frac{1}{(2\pi)^2}\int\rd^2\mu\ \e^{-\im\mu^{A'}\tilde\lambda_{A'}}\,f(\mu,\lambda) 
\ee 
which makes sense only when $\tilde\lambda_{A'}$ and $\mu^{A'}$ are real.

In discussing supersymmetric theories, we use on-shell supermultiplets such as the $\cN=4$ Yang-Mills supermultiplet in the $\eta$-representation
\begin{equation}
    \Phi(\lambda,\tilde\lambda,\eta) = G^+(\lambda,\tilde\lambda) +\eta^a\Gamma_a(\lambda,\tilde\lambda) + \cdots +
    \frac{1}{4!}\epsilon_{abcd}\eta^a\eta^b\eta^c\eta^dG^-(\lambda,\tilde\lambda)\ .
\ee 
Here $G^\pm$ are the on-shell momentum space wavefunctions of the helicity $\pm 1$ parts
of the multiplet {\it etc.}, and $\eta^i$ are Grassmann co-ordinates on the on-shell momentum superspace. A supersymmetric half Fourier transform
\be{susy-halfFT}
    f(\lambda,\mu,\chi) = \int\rd^2\tilde\lambda\rd^\cN\eta\,\e^{\im(\mu^{A'}\tilde\lambda_{A'} + \chi_i\eta^i)}\,
    \Phi(\lambda,\tilde\lambda,\eta)
\ee
relates this to a supertwistor multiplet. We often write
\begin{equation}
     \llbracket \mu\tilde\lambda\rrbracket := \mu^{A'}\tilde\lambda_{A'}+\chi_a\eta^a\ ,\qquad
     \rd^{2|\cN}\tilde\lambda:=\rd^2\tilde\lambda\rd^\cN\eta\ ,\qquad\rd^{2|\cN}\mu :=\rd^2\mu\rd^\cN\chi
\end{equation}
for these commonly occurring supersymmetric combinations. We denote the homogeneous co-ordinates on (dual) supertwistor space ${\RP^{3|\cN}}^*$ by $\rW = (W_\alpha,\chi_a) = (\lambda_A,\mu^{A'},\chi_a)$.


\section{The X-ray and Half Fourier Transforms}
\label{app:halfFT}

In this appendix we will examine the relation of Witten's half Fourier transform~\cite{Witten:2003nn}
\be{half-Four}
    f(W)=\int \rd^2\tilde\lambda\ \e^{\im\mu^{A'}\tilde\lambda_{A'}}\,\Phi(\lambda,\tilde\lambda)\, .  
\ee 
to the usual Penrose transform~\cite{Penrose:1972ia}
\be{Penrosetrans}
    \phi(x) = \left.\oint\la\lambda\rd\lambda\ra f(W)\right|_{\mu^{A'} = -x^{AA'}\lambda_A}
\ee
and its generalisations to other helicities.

\bigskip

In much of the twistor literature, space-time has either the physical,
Lorentzian signature, or has Euclidean signature, or else is treated as
complex. For (conformally) flat space-times, twistor space is then complex projective three-space $\CP^3$.  The half Fourier transform does not apply in these contexts. Instead, massless fields are canonically related to \emph{cohomology
classes} (either {\v C}ech or Dolbeault) in twistor space via the abstract Penrose
transform~\cite{Eastwood:1981jy}. The concrete Penrose integral formula~\eqref{Penrosetrans} involves a representative $f$ of this cohomology classes. However, there is gauge freedom inherent in picking such a representative, and this shows itself in~\eqref{Penrosetrans} because (in the {\v Cech} picture) one can add to $f(W)$ any function whose singularities are not separated by the contour without changing the space-time field $\phi(x)$.  In the Dolbeault picture $f$ should be thought of as a $(0,1)$-form and~\eqref{Penrosetrans} interpreted as an integral over the full Riemann sphere. Again, adding any $\bar\del$-exact piece to $f$ does not change $\phi(x)$. Furthermore, to even pick a Dolbeault representatives in the first place one needs to solve a $\bar\del$-equation, whilst to pick a {\v C}ech representative one must first specify a covering of twistor space.  Thus, in Lorentzian or Euclidean space-time, the twistor representation is rather subtle.

However, in $(++--)$ signature space-time, the twistor representation becomes much simpler: Twistor space is now \emph{real} projective three-space and fields are represented in terms of straightforward \emph{functions}. This simplicity has been our main reason for working with split signature in this paper. See {\it e.g.}~\cite{Mason:2005qu} for a detailed discussion of twistor theory in $(++--)$ signature space-time. 

As pointed out by Atiyah in~\cite{Atiyah:1979}, in $(++--)$ signature,
the Penrose transform~\eqref{Penrosetrans} in its {\v C}ech format can be reinterpreted as the
`X-ray transform' of Fritz John~\cite{John:1938}. The X-ray transform\footnote{It
gets its name from X-ray imaging, where $f$ is taken to be the density
of the body to be X-rayed, and the integral of $f$ along lines gives
the attenuation of the X-ray as it passes through.} takes a
function $f$ on $\R^3$ and transforms it to a function $\phi$ on the
four-dimensional space of directed lines in $\R^3$ given by
integration along corresponding line in $\R^3$.  It naturally extends
to a map from functions on $\PT=\RP^3$ to the space $\widetilde M$ of
oriented lines in $\PT$. On dual twistor space (to fit with the scattering theory community) it can be expressed on
${\PT'}^*=\PT^*-\{\lambda_A=0\}$ by
\begin{equation}
    \phi(x) := \oint_{L_x}\, \la\lambda\rd\lambda\ra\, f(W)\, ,
\label{Pen0}
\end{equation}
where $L_x$ is the line in twistor space given by the incidence relation
\be{incidence}
    \mu^{A'} = -x^{AA'}\lambda_A
\ee
for $x$ in (2,2) signature space-time $\M$, and $f$ is an
arbitrary smooth function on real twistor space ${\PT}$.  Fritz John
showed that $\phi$ satisfies the wave equation and that, under
suitable assumptions, the X-ray transform is an isomorphism.  There
has by now been much work over the years to understand how the X-ray
transform and its relatives fit into the Penrose
transform~\cite{Mason:95,Sparling, Woodhouse,Bailey:99, Bailey:01, 
Bailey:03} 
and its non-linear analogues~\cite{LeBrun:2002,
LeBrun:2005qf,Mason:2005zm,LeBrun:2008ch}.  Early proofs of
invertibility followed by expanding $f(Z)$ in spherical harmonics,
where the X-ray transform integral may be done explicitly, and then
using the completeness relations of the spherical
harmonics. Subsequent proofs used complex analysis and adapted
versions of the Penrose transform.  We now give an alternative proof
that also gives a proof of Witten's half Fourier transform.

\subsection{Scalar fields}
\label{app:scalars}

For ease of comparison to the scattering theory literature, we will
focus on the transform from dual twistor space $\PT^*=\RP^{3*}$ with
real co-ordinates $W_\alpha=(\lambda_A,\mu^{A'})$. The dual twistor X-ray transform for the scalar wave equation is
\be{dual-pen}
    \phi(x)=\int \langle \lambda \rd\lambda\rangle\, f(\lambda,-x^{AA'}\lambda_A)\, ,
\ee
where $f(W)$ is an arbitrary function of homogeneity $-2$ (so $f(rW)=r^{-2}f(W)$).  Differentiating under the integral sign, it is easy to check that any scalar field constructed via~\eqref{dual-pen} automatically obeys the massless field equation $\Box\phi(x)=0$. In fact, it will follow from the argument below that \emph{all} such fields may be constructed this way. Hence, the Fourier transform $\tilde\phi(p)$ satisfies $p^2\,\tilde\phi(p) = 0$,
and so 
\begin{equation}
    \tilde\phi(p)=\delta(p^2)\Phi(\lambda,\tilde\lambda)
\end{equation}
for some function $\Phi(\lambda,\tilde\lambda)$ defined on the null cone in
momentum space, where $p_{AA'}=\lambda_A\tilde\lambda_{A'}$. 

Combining the X-ray and Fourier transforms, we have
\be{combine-F-X} 
    \delta(p^2)\Phi(\lambda,\tilde\lambda) = \int\rd^4x\ \e^{-\im p\cdot x} \phi(x) = \int\rd^4x\ \e^{-\im p\cdot x}
    \left\{\int\,\la{\lambda'}\rd{\lambda'}\ra\,f(\lambda'_A,-x^{AA'}\lambda'_A)\right\}
\end{equation}
where $|\lambda'\ra$ is a dummy spinor variable. Now, because
$f$ depends on $x$ only through the combination $x^{AA'}\lambda'_A$,
half of the $x$-integrals may be performed directly. To do this,
choose\footnote{The choice will soon be seen to drop out.} a constant
spinor $|\alpha\ra$ with $\la\alpha\lambda'\ra\neq0$ and decompose $x$
as \be{foliation} x^{AA'} =\frac{\mu^{A'}\alpha^{A} -
\chi^{A'}{\lambda'}^A}{\la\alpha\lambda'\ra}\ , \ee 
where $\mu^{A'} = -x^{AA'}\lambda'_A$ and $\chi^{A'}:=x^{AA'}\alpha_A$. (Equation~\eqref{foliation} is easily checked by contracting both sides with either $|\alpha\ra$ or
$|\lambda'\ra$.) The measure $\rd^4x$ decomposes as \be{d4xfoliation}
\rd^4x = \frac{\rd^2\mu\,\rd^2\chi}{\la\alpha\,\lambda'\ra^2}\ .
\ee We now integrate out $|\chi\ra$ to obtain 
\be{fourier-X-ray}
\begin{aligned}
    \delta(p^2)\Phi(\lambda,\tilde\lambda)
    &= \int\rd^4x\,\la\lambda'\rd\lambda'\ra\, \e^{-\im p\cdot x} f(\lambda',-x|\lambda'\ra) \\
    &=\int \frac{\rd^2\mu\,\rd^2\chi\,\la\lambda'\rd\lambda'\ra}{\la\alpha\,\lambda'\ra^2}\, 
    \exp\left(-\im\frac{[\mu|p|\alpha\ra-[\chi|p|\lambda'\ra}{\la\alpha\lambda'\ra}\right)\,f(\lambda',\mu)\\
    & = \int\frac{\rd^2\mu\,\,\la\lambda'\rd\lambda'\ra}{\la\alpha\,\lambda'\ra^2}
    \delta^2\left(\frac{p_{AA'}{\lambda'}^A}{\la\alpha\,\lambda'\ra}\right)
    \e^{-\im\frac{[\mu|p|\alpha\ra}{\la\alpha\,\lambda'\ra}}\,f(\lambda',\mu)\        . 
\end{aligned}
\ee
The $\delta$-functions inside the integral may be converted into
$\delta$-functions involving the integration variable  
$\lambda'$ at the expense of a Jacobian
\begin{equation}
    \delta^2\left(\frac{p_{AA'}{\lambda'}^A}{\la\alpha\,\lambda'\ra}\right)
    = \left|\la\alpha\,\lambda'\ra\la\alpha\,\lambda\ra\right|\,\delta(p^2)\delta(\la\lambda\,\lambda'\ra)\ ,
\ee
where $p = \lambda\tilde\lambda$ on the support of $\delta(p^2)$. The $\rd\lambda'$ integral may now be performed, yielding finally
\be{Fourier-xray-fin}
\begin{aligned}
    \delta(p^2)\Phi(\lambda,\tilde\lambda) &= \delta(p^2)\int\rd^2\mu \,\la\lambda'\rd\lambda'\ra \,  
    \left|\frac{\la\alpha\lambda\ra}{\la\alpha\lambda'\ra}\right|
    \delta(\la\lambda\lambda'\ra)\ \e^{-\im[\mu\tilde\lambda]\frac{\la\alpha\,\lambda\ra}{\la\alpha\,\lambda'\ra}}\,
    f(\lambda',\mu)\\   
    &= \delta(p^2)\int\rd^2\mu\,\e^{-\im[\mu\tilde\lambda]} f(\lambda,\mu)
\end{aligned}
\ee which is precisely Witten's half Fourier transform~\cite{Witten:2003nn}. Provided
$\Phi$ is sufficiently well-behaved, \eqref{Fourier-xray-fin} can be
inverted by standard Fourier analysis to give 
\be{half Fourier-inv}
    \mathbox{ 
            f(\lambda,\mu) = \int\rd^2\tilde\lambda\,
            \e^{\im[\mu\tilde\lambda]}\,\Phi(\lambda,\tilde\lambda) } 
\ee
as is used throughout the text.


\subsection{Generalisation to other helicities}
\label{app:otherhelicities}

Like the Penrose transform, the X-ray  transform may be generalised to fields of helicity $h$. 
Specifically, on-shell massless fields of helicity $h$ are represented on real (dual) twistor space by functions of homogeneity $2h-2$ and the X-ray transform is either
\begin{equation}
    \phi_{AB\cdots D}(x)
    =\left.\int\la\lambda\,\rd\lambda\ra\, \lambda_A\lambda_{B}\cdots \lambda_{D}\,f_{2h-2}(W)\right|_{\mu=-x\lambda}
\label{Pen-}
\end{equation}
when $h<0$, or
\begin{equation}
    \phi_{A'B'\cdots D'}(x)
    =\int\left.\la\lambda\,\rd\lambda\ra\, \frac{\del^{2h}}{\del \mu^{A'}\del\mu^{B'}\cdots \del\mu^{D'}} 
    f_{2h-2}(W)\right|_{\mu=-x\lambda}
\label{Pen+}
\end{equation}
when $h>0$. In particular, when $|h|=1$ they
represent the anti-self-dual and self-dual parts of the linearized
Yang-Mills field strength 
\be{YMfieldstrength} 
    F_{ABA'B'} =\epsilon_{A'B'}\phi_{AB}(x) +\epsilon_{AB}\phi_{A'B'}(x) 
\ee 
and when $|h|=2$ they likewise represent the self-dual and anti-self-dual
parts of the linearized Weyl curvature 
\be{Weylfieldstrength}
W_{ABCDA'B'C'D'}(x) =
\epsilon_{A'B'}\epsilon_{C'D'}\phi_{ABCD}(x)+\epsilon_{AB}\epsilon_{CD}\phi_{A'B'C'D'}(x)\
. 
\ee
Again,  differentiating under the integral sign in~\eqref{Pen-} and~\eqref{Pen+}, one verifies that 
that these fields automatically satisfy the linearized Yang-Mills or Einstein
equations, which in spinor form are 
\be{eom-h}
    \nabla^{AA'}\phi_{AB\cdots D}(x)=0
    \quad\hbox{and}\quad
    \nabla^{AA'}\phi_{A'B'\cdots D'}(x)=0\ .
\ee 
Since we are dealing with linearized curvatures, these formul\ae\ are
all gauge invariant.

For general helicity, the field equations~\eqref{eom-h} imply that the
Fourier transformed fields obey 
\be{F-wave-eq-h}
    p^{AA'}\tilde\phi_{AB\cdots D}(p) = 0\ , \qquad
    p^{AA'}\tilde\phi_{A'B'\cdots D'}(p) = 0.
\ee
Away from $p^2 = 0$,  $p^{AA'}$ is invertible. Hence the Fourier transforms take the form\footnote{The factor of 
$\im^{2h}$ is included for later convenience.}
\be{fourier}
\begin{aligned}
    \tilde\phi_{AB\cdots D}(p) 
    &= \im^{2h}\delta(p^2)\, \lambda_A\lambda_B\cdots\lambda_D\, \Phi_{2h}(\lambda,\tilde\lambda)\\
    \tilde\phi_{A'B'\cdots D'}(p) 
    &=  \delta(p^2)\,\tilde\lambda_{A'}\tilde\lambda_{B'}\cdots\tilde\lambda_{D'}\,\Phi_{2h}(\lambda,\tilde\lambda)
\end{aligned}
\ee where as before $p_{AA'}=\lambda_A\tilde\lambda_{A'}$ and
$\Phi_{2h}(\lambda,\tilde\lambda)$ must scale under $(\lambda,\tilde\lambda) \to (r\lambda,r^{-1}\tilde\lambda)$ as
\be{Phiscaling}
    \Phi_{2h}(r\lambda,r^{-1}\tilde\lambda) = r^{2h}\Phi_{2h}(\lambda,\tilde\lambda)
\end{equation}
so as to balance the scaling
of the spinor pre-factors (recall that $h<0$ for $\tilde\phi_{AB\cdots
D}$). Equations~\eqref{fourier} are fixed purely by kinematics, and hold irrespective of the particular wavefunction of the
on-shell state. The wavefunction itself is determined by a choice of particular
function $\Phi_{2h}(\lambda,\tilde\lambda)$.

The half Fourier transform for non-zero helicity has the same relation
to the X-ray transform as it does for the scalar field
treated above. When $h<0$, the same steps as in
equations~\eqref{combine-F-X}-\eqref{Fourier-xray-fin} give 
\be{FXh-}
    \tilde\phi_{AB\cdots D}(p) =
    \delta(p^2)\int\rd^2\mu\,\e^{-\im[\mu\tilde\lambda]}\,\lambda_{A}\lambda_{B}\cdots\lambda_{D}\,f_{2h-2}(W)\ , 
\ee 
whereas when $h>0$, the $\rd^2\mu$ integral is done by parts to
find similarly \be{FXh+}
\begin{aligned}
    \tilde\phi_{A'B'\cdots D'}(p) &= \delta(p^2)\int\rd^2\mu\,\e^{-\im[\mu\tilde\lambda]}\,
    \frac{\del^{2h}}{\del\mu^{A'}\del\mu^{B'}\cdots\del\mu^{D'}}f_{2h-2}(W)\\
    &=\im^{2h} \delta(p^2)\int\rd^2\mu\,\e^{-\im[\mu\tilde\lambda]}\,
    \tilde\lambda_{A'}\tilde\lambda_{B'}\cdots\tilde\lambda_{D'}\,f_{2h-2}(W)\ ,
\end{aligned}
\ee
Both of these are captured by the simple half Fourier transforms
\be{FX}
\mathbox{
\begin{aligned}
    \Phi_{2h}(\lambda,\tilde\lambda) &= \int\rd^2\mu\,\e^{-\im[\mu\tilde\lambda]}\,f_{2h-2}(W)\\
    f_{2h-2}(W) &= \frac{1}{(2\pi)^2}\int\rd^2\tilde\lambda\,\e^{\im[\mu\tilde\lambda]}\,\Phi_{2h}(\lambda,\tilde\lambda)
\end{aligned}}\ .
\ee
It is easy to see that~\eqref{Phiscaling} implies that $f$ has homogeneity $2h-2$ under $W_\alpha\rightarrow r W_\alpha$. This agrees with the well-known fact that massless fields of helicity $h$ on space-time or momentum space correspond to homogeneous functions on dual twistor space of weight $2h-2$. 


\subsection{Supersymmetry}
\label{app:halfsuperFT}

The Penrose transform naturally extends to supersymmetric theories by
adjoining $\cN$ anti-commuting variables to the non-projective twistor
space. Supertwistor space is then the projectivisation of $\R^{4|\cN}$
(or $\C^{4|\cN}$, in the complex case~\cite{Ferber:1977qx}). The space of
$\RP^1$s inside projective supertwistor space is the (conformal
compactification of) anti-chiral split signature superspace with
co-ordinates $(x_-^{AA'},\tilde\theta^{aA'})$, while the space of
$\RP^1$s inside dual projective supertwistor space is chiral
superspace\footnote{Chiral superspace can also be realised as the
space of $\RP^{1|\cN}$s inside supertwistor space, and similarly
anti-chiral superspace is also the space of $\RP^{1|\cN}$s in dual
supertwistor space, but we will not use this fact here.} with
co-ordinates $(x_+^{AA'},\theta^A_a)$. The incidence relation on dual supertwistor space is
\be{superincidence2} 
    \mu^{A'} = -x_+^{AA'}\lambda_{A}\ , \qquad \chi_a = \theta^A_a\lambda_{A} \ .
\ee

Supertwistor space and its dual carry a natural action of the (4:1
cover of the) space-time superconformal group
$SL(4|\cN;\R)$. Concentrating on the dual twistor space, this action
is generated by the vector fields 
\be{twistorsuperconfgen}
    \left(W_\alpha\frac{\del}{\del W_\beta}-\frac{1}{4-\cN}\delta_\alpha^{\ \beta}\Upsilon\right)\,,
    \quad W_\alpha\frac{\del}{\del\chi_b}\,,
    \quad \chi_a\frac{\del}{\del W_\beta}\,,
    \quad \left(\chi_a\frac{\del}{\del\chi_b}-\frac{1}{4-\cN}\delta_a^{\ b}\Upsilon \right) 
\ee 
where $(W_\alpha,\chi_a)$ are homogeneous
co-ordinates on the superspace and 
\be{upsilondef}
    \Upsilon:=W_\gamma \frac{\del}{\del W_\gamma}+\chi_c\frac{\del}{\del\chi_c}
\ee
is the Euler homogeneity operator\footnote{For $\cN=4$, $\SL(4|4)$ needs to be defined slightly
differently as being generated by linear transformations of $\R^{4|4}$ that have vanishing supertrace -- we can no longer use $\Upsilon$ to remove the trace because ${\rm str} (\mathbb{I})=0$ when $\cN=4$. Instead, $\Upsilon$ may either be considered as a {\it bona fide} generator of the group itself, or else may be omitted, in which case one is dealing with ${\rm PSL}(4|4)$.}.


\subsubsection{Superfields in twistor space and on-shell momentum space}

When $\cN=4$, we can construct a twistor supermultiplet representing
an on-shell SYM multiplet by taking a (Lie algebra-valued) function
$A(W,\chi)$ of the supertwistors, homogeneous of degree 0 as
in~\cite{Ferber:1977qx}. Its component expansion is 
\be{symtwist} 
    A(W,\chi) = a(W) + \chi_a\psi^a(W) + \cdots +\frac{1}{4!}\epsilon^{abcd}\chi_a\chi_b\chi_c\chi_d\,g(W) 
\ee
where the component fields $a,\psi^a,\ldots,g$ have homogeneities
$0,-1,\ldots,-4$ corresponding to on-shell space-time fields of
helicities $1,\half,\ldots,-1$ and transform in the appropriate
representation of the $SL(4;\R)$ R-symmetry group. We want to understand the relation between this representation and the on-shell momentum supermultiplet 
\be{symmom}
    \Phi(\lambda,\tilde\lambda,\eta) = G^+(\lambda,\tilde\lambda) +\eta^a\Gamma_a(\lambda,\tilde\lambda) + \cdots +
    \frac{1}{4!}\epsilon_{abcd}\eta^a\eta^b\eta^c\eta^dG^-(\lambda,\tilde\lambda) 
\ee 
used by {\it e.g.}~\cite{Drummond:2008vq,Brandhuber:2008pf,Drummond:2008cr,Drummond:2008bq,ArkaniHamed:2008gz} in the supersymmetric BCFW Yang-Mills recursion relations.

Taking the X-ray transform of the complete
supermultiplet~\eqref{symtwist} gives a chiral superfield\footnote{We
henceforth drop the subscript on the chiral co-ordinate $x_+$.}
\be{superPen}
    \phi_{A'B'}(x,\theta) = \int\la\lambda\rd\lambda\ra\frac{\del^2}{\del\mu^{A'}\del\mu^{B'}}
    A(\lambda_A,-x^{AA'}\lambda_A,\theta_a^A\lambda_A)\ .
\ee 
$\phi_{A'B'}(x,\theta)$ is independent of the anti-chiral $\tilde\theta$s and can be defined without auxiliary fields precisely because the multiplet is on-shell (see {\it e.g.}~\cite{Buchbinder} for a full discussion). By differentiating under the integral sign, we see that
\be{supereom}
    \nabla^{AA'}\phi_{A'B'} =0\qquad\hbox{and}\qquad\nabla^{CC'}\frac{\del}{\del\theta^C_a}\phi_{A'B'} = 0\ .
\ee

We now take the super-Fourier transform of $\phi_{A'B'}$, {\it i.e.}
\be{superFourier}
    \tilde\phi_{A'B'}(p,\xi) = \int\rd^{4|8}x\,\exp\left(-\im x^{CC'}p_{CC'} -\im\theta^C_a\xi^a_C\right)\, 
    \phi_{A'B'}(x,\theta)\ , 
\ee 
where $\rd^{4|8}x=\rd^4x\, \rd^8\theta$ is the usual chiral superspace
measure and $\xi^a_A$ is conjugate to $\theta^A_a$. The superfield equations~\eqref{supereom} imply that
$\tilde\phi_{A'B'}(p,\xi)$ obeys
\be{constraints-susy}
    p^{AA'}\tilde\phi_{A'B'}(p,\xi) = 0\qquad\hbox{and}\qquad p^{CC'}\xi^a_C\tilde\phi_{A'B'}(p,\xi) = 0\ .
\end{equation}
As before, the first of these implies that 
\begin{equation}
    \tilde\phi_{A'B'}(p,\xi)
    =\delta(p^2)\tilde\lambda_{A'}\tilde\lambda_{B'}\Phi(\lambda,\tilde\lambda,\xi) 
\end{equation}
with $p_{AA'}=\lambda_A\tilde\lambda_{A'}$ on-shell. The second
equation in~\eqref{constraints-susy} means that $\Phi(\lambda,\tilde\lambda,\xi)$ vanishes when multiplied by
$\lambda^{A}\xi^a_A$ for any choice of the R-symmetry index. Thus the super-Fourier transform is a multiple of 
\be{mom-super-delta}
    \delta^{(0|4)}(\xi^a_A\lambda^{A}) := \frac{1}{4!}\epsilon_{abcd}\xi^a_A\xi^b_B\xi^c_C\xi^d_D\,
    \lambda^{A}\lambda^{B}\lambda^{C}\lambda^{D}\, .  
\ee 
and the factor $\delta(p^2)\delta^{(0|4)}(\xi^a_A\lambda^{A})$ restricts the support of the super Fourier transform to the `super light-cone' $p^2=0=\xi^a_A\lambda^A$ in momentum superspace.  On
restriction to this super light-cone, $\xi_A^a\lambda^A\delta^{(0|4)}(\xi_B^b\lambda^B) =0$, so we have
\be{super-light-cone}
    \xi^a_A\delta^{(0|4)}(\xi_B^b\lambda^B)=:\eta^a\lambda_A\,\delta^{(0|4)}(\xi_B^b\lambda^B) 
\ee 
for some odd co-ordinates $\eta^a$ with opposite weight to $\lambda_A$ (the
same weight as $\tilde\lambda_{A'}$). Thus we co-ordinatise the momentum space
super-light-cone by $(\lambda_A,\tilde\lambda_{A'},\eta^a)$ defined up to the
scaling 
\be{susyscaling}
    (\lambda_A,\tilde\lambda_{A'},\eta^a)\sim (r \lambda_A,r^{-1}\tilde\lambda_{A'},r^{-1}\eta^a) \, , 
    \qquad r\in \R\ .
\ee
On this super light-cone $(p_{AA'},\xi_A^a)= (\lambda_A\tilde\lambda_{A'},\lambda_A\eta^a)$ and a wavefunction may be written as
\begin{equation}
\label{mom-superspace}
    \tilde\phi_{A'B'}(p,\xi) = \delta(p^2)\,\delta^{(0|4)}(\xi^a_A\lambda^{A})\,\tilde\lambda_{A'}\tilde\lambda_{B'}\,
    \Phi_{-2}(\lambda,\tilde\lambda,\eta)\ .
\end{equation}
Combining equations~\eqref{superPen} \&~\eqref{superFourier}, and
following the same argument as before, but now performing the integral over both the $\chi^{A'}$ of~\eqref{foliation}
and the $\zeta^a$ of $\theta_a^A = (\nu_a\alpha^A+\zeta_a{\lambda'}^A)/\la\alpha\lambda\ra$ yields the
extra delta function $\delta^{(0|4)}(\xi^a_A\lambda^A)$.  Omitting the details, we obtain the formula 
\be{symrel} 
    \mathbox{
            \Phi_{-2}(\lambda,\tilde\lambda,\eta) 
            =\frac{1}{(2\pi)^2}\int\rd^{2|4}\mu\ \e^{-\im\llbracket\mu\tilde\lambda\rrbracket}\,A(W,\chi) 
            }
\ee
where we use the supersymmetric notation 
\be{supershorthand} 
    \llbracket\mu\tilde\lambda\rrbracket:= [\mu\tilde\lambda] + \chi_a\eta^a\ . 
\ee 
Thus the on-shell momentum space SYM multiplet~\eqref{symmom} is simply the half Fourier
transform (taken over both $\mu^{A'}$ and $\chi_a$) of the twistor supermultiplet.

\medskip

The same arguments applies for all $\cN$, and in particular for an
$\cN=8$ supergravity multiplet:  According to the X-ray or Penrose transform, a
linearized graviton of helicity $+2$ may be represented  on
dual twistor space by a twistor function $h_2(W)$ of homogeneity
$+2$, while a graviton of helicity $-2$ is represented by a
twistor function $h_{-6}(W)$ of weight $-6$. For $\cN=8$
supertwistor space, both graviton helicities are contained in the
single twistor supermultiplet 
\be{sugratwist} 
    H(W,\chi) = h_2(W) +\chi_a\gamma^a(W) + \cdots + (\chi)^8h_{-6}(W)\ .  
\ee
The above extends straightforwardly to a proof that 
this multiplet corresponds to an on-shell $\cN=8$
supergravity multiplet in momentum space via the half Fourier
transform 
\be{sugrarel}
    \mathbox{
            \Phi_{-4}(\lambda,\tilde\lambda,\eta) = \frac{1}{(2\pi)^2}\int\rd^{2|8}\mu\ 
            \e^{-\im\llbracket\mu\tilde\lambda\rrbracket}\,H(W,\chi) 
            }
\ee
and again $\Phi_{-4}(\lambda,\tilde\lambda,\eta)$ is the supermultiplet
used in the BCFW rules for $\cN=8$ SG~\cite{ArkaniHamed:2008gz,Drummond:2009ge}.

\bigskip

We finally remark that the half Fourier transform yields the 
supersymmetric substitutions 
\be{superreplace} 
\begin{aligned}
    \tilde\lambda_{A'} &\to \im\frac{\del}{\del\mu^{A'}}\,\qquad\qquad\frac{\del}{\del\tilde\lambda_{A'}} &\to-\im\mu^{A'}\\
    \eta^a &\to \im \frac{\del}{\del\chi_a}\ , \qquad\qquad \frac{\del}{\del\eta^a} &\to -\im\chi_a 
\end{aligned}
\ee
This relates the dual twistor space superconformal generators of~\eqref{twistorsuperconfgen} to
\be{supmom}
\begin{aligned}
    Q_{Aa} &= \lambda_A\frac{\del}{\del\eta^a} &\qquad \tilde Q_{A'}^b &= \tilde\lambda_{A'}\eta^b\\
    S^{Ab} &= \eta^b\frac{\del}{\del\lambda_A} &  \tilde S^{A'}_a &= \frac{\del^2}{\del\eta^a\del\tilde\lambda_{A'}}
\end{aligned}
\ee
together with the R-symmetry generator
\begin{equation}
    R^a_{\ b} =\eta^a\frac{\del}{\del\eta^b} - \frac{1}{\cN}\delta^a_{\ b}\eta^c\frac{\del}{\del\eta^c}\ ,
\end{equation}
on the momentum super light-cone.


\section{The Half Fourier Transform of Seed Amplitudes}
\label{app:amplitudes}

In this appendix we explicitly perform the half Fourier transform of the $3$-particle MHV and $\overline{\rm MHV}$, and
the $n$-particle MHV SYM amplitudes to (dual) twistor space. 


\subsection{The 3-point MHV amplitude}
\label{app:N=4amp-3MHV}

The three-particle MHV amplitude in on-shell momentum superspace is
\begin{equation}
\label{app:MHV3}
    A_{\rm {MHV}}(p_1,p_2,p_3)
    =\frac{\delta^{(4|8)}\left(\sum_{i=1}^3|i\ra\llbracket i\|\right)}
    {\la12\ra\la23\ra\la31\ra}
\ee 
where again $\llbracket i\| = ([i|,\eta_i)$. We will re-express the momentum $\delta$-functions as spinor
$\delta$-functions to expedite the half Fourier transforms. For the three-point MHV amplitude we may assume 
$\la12\ra\neq0$ and so can expand unprimed spinors in the basis $\{|1\ra,|2\ra\}$.  This gives
\be{mom-basis2} 
    \sum_{i=1}^3 |i\ra\llbracket i\|=|1\ra\left(\llbracket1\|+\frac{\la32\ra}{\la12\ra}\llbracket3\|\right)
     +|2\ra\left(\llbracket2\|+  \frac{\la13\ra}{\la12\ra}\llbracket3\|\right) 
\ee
and since $|1\ra$ and $|2\ra$ are linearly independent, the $\delta$-function becomes
\be{mom-delta-2}
    \delta^{(4|8)}\!\left(\sum_{i=1}^3 |i\ra\llbracket i\|\right)
    =\la12\ra^2\ \delta^{(2|4)}\!\left(\llbracket1\|+\frac{\la32\ra}{\la12\ra}\llbracket3\|\right)\, 
     \delta^{(2|4)}\!\left(\llbracket2\|+  \frac{\la13\ra}{\la12\ra}\llbracket3\|\right) 
\ee
The half Fourier transform of states 1 and 2 can be done straightforwardly using these
$\delta$-functions, yielding 
\be{app:MHV-twist}
    \cA_{\rm {MHV}}({\rW}_1,{\rW}_2,\rW_3)
    =\frac{\delta^{(2|4)}\!\left(\mu_1\la23\ra+\mu_2\la31\ra+\mu_3\la12\ra\right)}{\la12\ra\la23\ra\la31\ra}\, ,
\ee 
as in equation~\eqref{MHV-twist-explicit}. This has homogeneity degree zero in each of the three supertwistors, as required.  It has support precisely where W$_1$, W$_2$ and W$_3$ are collinear and is appropriately antisymmetric under permutations of $1,2,3$. 

We can elucidate the behaviour of~\eqref{app:MHV-twist} under conformal transformation by relating it to the superconformally invariant $\tilde\delta$-function
\begin{equation}
    \tilde\delta^{(2|4)}(\rW_1;\rW_2,\rW_3) = \int\frac{\rd s}{s}\frac{\rd t}{t}\,\delta^{(4|4)}(\rW_1-s\rW_2-t\rW_3)
\ee
Notice that this $\tilde\delta$-function is manifestly symmetric under the exchange $2\leftrightarrow3$. By using the $|\lambda\ra$ spinor components to perform the integrals we obtain
\begin{equation}
    \int \frac{\rd s}{s}\frac{\rd t}{t} \delta^{(2)}(\lambda_1-s\lambda_2- t\lambda_3)
    = \frac{|\la23\ra|}{\la12\ra\la31\ra}\ ;\qquad 
    s=\frac{\la31\ra}{\la21\ra}\, , \quad t= \frac{\la21\ra}{\la32\ra}\ ,
\ee 
where the modulus sign arises from a Jacobian in the delta functions, generalising the standard scaling $\delta (a x )= |a|^{-1}\delta(x)$. Combining this with the $\|\mu\rrbracket$ components of $\tilde\delta^{(2|4)}(\rW_1,\rW_2,\rW_3)$ yields 
\begin{equation}
    \tilde\delta^{(2|4)}(\rW_1,\rW_2,\rW_3)= \sgn(\la23\ra)
    \frac{\delta^{(2|4)}\!\left(\mu_1\la23\ra +  \mu_2\la31\ra+\mu_3\la12\ra\right)}{\la12\ra\la23\ra\la31\ra}\ . 
\ee 
Thus we see that
\be{app:MHV3-twist}
\mathbox{
    \cA_{\rm MHV}(\rW_1,\rW_2,\rW_3)=\sgn(\la\rW_2\rW_3\ra)\, \tilde\delta^{(2|4)}(\rW_1;\rW_2,\rW_3)
                        }
\ee
as in equation~\eqref{MHV3-twist}.


\subsection{The 3-point $\MHVbar$ amplitude}
\label{app:MHV-bar}

The three-particle $\MHVbar$ amplitude for
$\cN=4$ SYM in on-shell momentum superspace is given by~\cite{Brandhuber:2008pf}
\be{MHV-bar3}
    A_{\overline{\mathrm{MHV}}}(p_1,p_2,p_3)
    =\frac{\delta^{(4)}\!\left(p_1+p_2+p_3\right)\delta^{(4)}\!\left(\eta_1[23]+\eta_2[31]+\eta_3[12]\right)}{[12][23][31]}\, . 
\ee 
Momentum conservation here implies proportionality of the unprimed spinors, so to construct a basis
for this space we must introduce an arbitrary independent spinor $|\alpha\ra$ with $\la1\alpha\ra\neq0$. We then express $|2\ra$ and $|3\ra$, say, in the $\{|1\ra,|\alpha\ra\}$ basis, leading to
\be{MHV-bar-mom}
\begin{aligned}
    \delta^{(4)}(p_1+p_2+p_3)&=\frac{1}{\la1\alpha\ra^2}
    \delta^{(2)}\! \left(|1]+|2]\frac{\la2\alpha\ra}{\la1\alpha\ra}  +|3]\frac{\la3\alpha\ra}{\la1\alpha\ra}\right)\,
    \delta^{(2)}\!\left(|2]\frac{\la21\ra}{\la\alpha1\ra}+|3]\frac{\la31\ra}{\la\alpha1\ra}\right)\\
    &=\frac{1}{|[23]|}
    \delta^{(2)}\! \left(|1]+| 2]\frac{\la2\alpha\ra}{\la1\alpha\ra}  +| 3]\frac{\la3\alpha\ra}{\la1\alpha\ra}\right)\,
    \delta(\la12\ra)\,\delta(\la13\ra)\ ,
\end{aligned}
\ee 
The second line follows from the fact that $|2]$ and $|3]$ are
linearly independent in the $\overline{\rm MHV}$ amplitude and the
pre-factors $1/\la1\alpha\ra^2$ in the first line and $1/|[23]|$ in the second come from Jacobians.  The
modulus sign in the second Jacobian again arises from the scaling relation $\delta (a x )= |a|^{-1}\delta(x)$ for the standard 1-dimensional $\delta$-function. On the support of the $\delta^{(2)}$-functions for $|1]$, the remaining unprimed spinor factors in~\eqref{MHV-bar3} become
\be{MHV-bar-rest}
    \frac{\delta^{(4)}\!\left(\eta_1[23]+\eta_2[31]+\eta_3[12]\right)}{[12][23][31]}
    =\frac{[23]\la1\alpha\ra^2}{\la2\alpha\ra\la3\alpha\ra}
    \delta^{(4)}\!\left(\eta_1+\eta_2\frac{\la2\alpha\ra}{\la1\alpha\ra}+\eta_3\frac{\la3\alpha\ra}{\la1\alpha\ra}\right)\ .
\ee
Notice that the factor of $[23]$ in the numerator here cancels the overall Jacobian in~\eqref{MHV-bar-mom} only up to its sign.

The twistor ${\overline{\rm MHV}}$ amplitude is the Fourier transform
\begin{multline}
    \cA_{\overline{\rm MHV}}(\rW_1,\rW_2,\rW_3) 
    = \delta(\la12\ra)\,\delta(\la13\ra)\frac{\la1\alpha\ra^2}{\la2\alpha\ra\la3\alpha\ra}\\
    \times\int\prod_{j=1}^3 \rd^{2|4}\tilde\lambda_j\  \e^{\im \llbracket\mu_j\tilde\lambda_j\rrbracket}\ {\rm sgn}([23])\, 
    \delta^{(2|4)}\!\left(\llbracket1\|+\llbracket2\|\frac{\la2\alpha\ra}{\la1\alpha\ra}+\llbracket3\|
    \frac{\la3\alpha\ra}{\la1\alpha\ra}\right)
\end{multline}
The $\rd^{2|4}\tilde\lambda_1$ integral may be done immediately using the delta functions, yielding
\begin{multline}
\label{MHV-bar-int}
    \int \prod_{j=2,3}\left\{ \rd^{2|4}\tilde\lambda_j\ 
    \e^{\im\llbracket\mu_j\tilde\lambda_j\rrbracket - 
    \im\frac{\la j\alpha\ra}{\la1\alpha\ra}\llbracket\mu_1\tilde\lambda_j\rrbracket}\ 
    \delta(\la1j\ra) \frac{\la1\alpha\ra}{\la j\alpha\ra}\right\}\sgn[23]\\
    =\int\prod_{j=2,3}\left\{\rd^{2|4}\mu'_j\,\rd^{2|4}\tilde\lambda_j\ 
    \e^{\im\llbracket\mu_i\tilde\lambda_i\rrbracket-\im\llbracket\mu'_i\tilde\lambda_i\rrbracket}\ 
    \delta^{(2|4)}\!\left(\mu_1-\mu'_j\frac{\la1\alpha\ra}{\la j\alpha\ra}\right)\frac{\la j\alpha\ra}{\la1\alpha\ra}
    \delta(\la1j\ra)\right\} \sgn[23]\ .
\end{multline}
In the second line here, we have introduced integrals over dummy variables $\mu'_j$ together with weighted $\delta$ functions which re-enforce $\mu'_j=\frac{\la j\alpha\ra}{\la 1\alpha\ra}\mu_1$. The virtue of this step is that, since we can also write
\begin{equation}
     \frac{\la j\alpha\ra}{\la1\alpha\ra}\delta(\la1j\ra) =
     \int_{-\infty}^\infty\frac{\rd t}{t}\,\delta^{(2)}\!\left(|1\ra-t|j\ra\right)\ ,
\ee
the $\delta$-functions combine into our superconformally invariant, but tilded $\delta$-functions
\begin{equation}
    \delta^{(2|4)}\!\left(\mu_1-\mu'_j\frac{\la1\alpha\ra}{\la j\alpha\ra}\right)\frac{\la j\alpha\ra}{\la1\alpha\ra}\delta(\la1j\ra)
    =\int \frac{\rd t}{t}\, \delta^{(4|4)}(\rW_1-t\rW'_j) =: \tilde\delta^{(3|4)}(\rW_1,\rW'_j)\ ,
\ee
where $\rW'_j$ is the supertwistor $(\lambda_j,\mu'_j,\chi'_j)$ for $j=2,3$.

As in the BCFW recursion relations, we now replace the factor $\sgn[23]$ by the formal operator
\be{Hilbert3}
    \sgn[23] = \sgn\left[\frac{\del}{\del\mu'_2}\frac{\del}{\del\mu'_3}\right]
     = \sgn\left[\frac{\del}{\del\rW'_2}\frac{\del}{\del\rW'_3}\right]
\ee
acting inside the $\rd^{2|4}\tilde\lambda_j\rd^{2|4}\mu'_j$ integrals. Performing these integrals then simply replaces $\rW'_j$ by $\rW_j$. Overall, the dual twistor form of the three-point $\overline{\rm MHV}$ super-amplitude in $\cN=4$ SYM is
\begin{equation}
    \mathbox{
            \cA_{\overline{\rm MHV}}(\rW_1,\rW_2,\rW_3) = 
            \sgn\left(\left[\frac{\del}{\del\rW_2}\frac{\del}{\del\rW_3}\right]\right)
            \tilde\delta^{(3|4)}(\rW_1,\rW_2)\,\tilde\delta^{(3|4)}(\rW_1,\rW_3)
            }
\ee
as in equation~\eqref{MHV3-bar-twist}. In fact, it is easy to show that $\cA_{\overline{\rm MHV}}(\rW_1,\rW_2,\rW_3)$ is given by the explicit formula
\begin{multline}
    \cA_{\overline{\rm MHV}}(\rW_1,\rW_2,\rW_3)
    = \frac{\lambda_1^2}{\lambda_2\,\lambda_3}\,\delta(\la12\ra)\,\delta(\la13\ra)\ \times\\
    \delta^\prime\!\left(\!\!\left(\mu_2^{A'}-\frac{\lambda_2}{\lambda_1}\mu_1^{A'}\right)
    \!\!\left(\mu_{3A'}-\frac{\lambda_3}{\lambda_1}\mu_{1A'}\right)\!\!\right)
    \delta^{(0|4)}\left(\chi_2-\frac{\lambda_2}{\lambda_1}\chi_1\right)
    \delta^{(0|4)}\left(\chi_3-\frac{\lambda_3}{\lambda_1}\chi_1\right)\ ,
\end{multline}
where the ratios $\lambda_2/\lambda_1$ and $\lambda_3/\lambda_1$ are meaningful on the support of $\delta(\la12\ra)\,\delta(\la13\ra)$.


\subsection{The $n$-point MHV amplitude}
\label{app:MHVSYM}

The $n$-point Parke-Taylor super-amplitude is
\be{app:Parke-Taylor}
    A_{\rm MHV}(p_1,\ldots,p_n) 
    = \frac{\delta^{(4|8)}\!\left(\sum_{i=1}^n |i\ra\llbracket i\|\right)}{\la12\ra\la23\ra\cdots\la n1\ra} \, .
\end{equation}
To transform~\eqref{app:Parke-Taylor} to dual twistor space, we use a straightforward extension of~(\ref{mom-delta-2}) for the momentum $\delta$-function, using the unprimed spin basis $\{|1\ra,|2\ra\}$ to rewrite it as
\be{mom-delta-MHVn} 
    \delta^{(4|8)}\left(\sum_{i=1}^n |i\ra\llbracket i\|\right)=\la12\ra^2\, 
    \delta^{(2|4)}\!\left(\llbracket1\| +\sum_{i=3}^n \frac{\la i2\ra}{\la12\ra}\llbracket i\|\right)
    \delta^{(2|4)}\!\left(\llbracket2\|+\sum_{i=3}^n \frac{\la i1\ra}{\la21\ra}\llbracket i\|\right)  \, .  
\ee 
Insert this into the the half Fourier transform and immediately performing the $\llbracket1\|$ and $\llbracket2\|$ integrals one obtains
\be{MHVn-twis}
\begin{aligned}
    \cA_{\rm MHV}(\rW_1,\ldots,\rW_n)&= 
    \frac{\la12\ra}{\la n1\ra}\prod_{i=3}^n \frac1{\la i-1\, i\ra}
    \delta^{(2|4)} \left(\mu_i+\frac{\la i1\ra}{\la12\ra}\mu_2 -\frac{\la i2\ra}{\la12\ra}\mu_1 \right) \\
    &= \frac{\la12\ra}{\la n1\ra}\prod_{i=3}^n \frac1{\la i-1\, i\ra\la12\ra^2} 
    \delta^{(2|4)} \left(\mu_1\la2i\ra +\mu_2\la i1\ra +\mu_i\la12\ra \right)
\end{aligned}
\ee
Using the identity 
\begin{multline}
\label{collinear-delta-id} 
    \delta^{(2|4)}\left(\mu_1\la2i\ra +\mu_2\la i1\ra +\mu_i\la12\ra\right)
    \delta^{(2|4)}\left(\mu_1\la2j\ra +\mu_2\la j1\ra +\mu_j\la12\ra \right)\\
    =
    \frac{\la12\ra^2}{\la1i\ra^2}
    \delta^{(2|4)}\left(\mu_1\la2i\ra +\mu_2\la i1\ra +\mu_i\la12\ra\right)
    \delta^{(2|4)}\left(\mu_1\la ij\ra + \mu_i\la j1\ra +\mu_j\la1i\ra\right)\ ,
\end{multline}
one can show
\be{app:MHVn}
    \mathbox{
    \cA_{\rm MHV}(\rW_1,\ldots,\rW_n) = (-1)^{n-3}\prod_{i=3}^n \cA_{\rm MHV}(\rW_1,\rW_{i-1},\rW_i)
    }
\ee
as in equation~\eqref{MHVn}.


\end{document}